\documentclass[twocolumn,10pt]{article}
\usepackage{layout}
\usepackage{epsfig}
\usepackage{multirow}
\usepackage{array}
\usepackage{graphicx,xcolor,textpos}
\usepackage{bm}
\usepackage{geometry}
\title{\textbf{Review on the progress in nuclear fission}}
\author{Karl-Heinz Schmidt\footnote{e-mail: schmidt-erzhausen@t-online.de},
		Beatriz Jurado\footnote{e-mail: jurado@cenbg.in2p3.fr}\\ ~ \\
		CENBG, CNRS/IN2 P3, Chemin du Solarium B.P. 120,\\ 
	   F-33175 Gradignan, France\\~}
\begin{document}

\maketitle

\tableofcontents{}

\newpage

\paragraph{Abstract:}
An overview is given on some of the main advances in experimental methods, experimental results and theoretical models and ideas of the last years in the field of nuclear fission. 
New approaches extended the availability of fissioning systems for experimental studies of nuclear fission considerably and provided a full identification of all fission products in $A$ and $Z$ for the first time. 
In particular, the transition from symmetric to asymmetric fission around 
$^{226}$Th and some unexpected structure in the mass distributions in the fission of systems around $Z$ = 80 to 84 as well as an extended systematics of the odd-even effect in fission fragment $Z$ distributions have been measured [A. N. Andreyev et al., Rep. Progr. Phys. 81 (2018) 016301].
Three classes of model descriptions of fission presently appear to be the most promising or the most successful ones:
Self-consistent quantum-mechanical models fully consider the quantum-mechanical features of the fission process. Intense efforts are presently made to develop suitable theoretical tools [N. Schunck, L. M. Robledo, Rep. Prog. Phys. 79 (2016) 116301] for modeling the non-equilibrium, large-amplitude collective motion leading to fission.  
Stochastic models provide a fully developed technical framework. The main features of the fission-fragment mass distribution are well reproduced from mercury to fermium and beyond [P. M\"oller, J. Randrup, Phys. Rev. C 91 (2015) 044316]. However, the limited computer resources still impose restrictions, for example on the number of collective coordinates and on an elaborate description of the fission dynamics.
In an alternative semi-empirical approach [K.-H. Schmidt et al., Nucl. Data Sheets 131 (2016) 107], considerable progress in describing the fission observables has been achieved by combining several theoretical ideas, which are essentially well known.
This approach exploits 
(i) the topological properties of a continuous function in multidimensional space, 
(ii) the separability of the influences of fragment shells and macroscopic properties of the compound nucleus, 
(iii) the properties of a quantum oscillator coupled to the heat bath of other nuclear degrees of freedom, 
(iv) an early freeze-out of collective motion, and 
(v) the application of statistical mechanics for describing the thermalization of intrinsic excitations in the nascent fragments. 
This new approach reveals a high degree of regularity and allows calculating
high-quality data that are relevant for nuclear technology without 
specific adjustment to empirical data of individual systems.

\newpage

\section{Introduction} \label{1}

The discovery of nuclear fission revealed that the heaviest
nuclei are barely bound in their ground state. 
An excitation energy in the order of a few 
per mille of their total binding energy is
sufficient to induce the disintegration into two large
pieces, 
%in a collective shape evolution that
%resembles the division of living cells, 
releasing 
a huge amount of energy of about 200 MeV. 
Thus, the specific energy content of nuclear fuel is 
about 10$^8$ times larger compared to fossil 
fuels like coal, mineral oil or natural gas. This
explains the importance of fission in nuclear technology.

The energy stored
in heavy nuclei, and even the synthesis of 
an appreciable portion of matter in the
Universe has its origin in the astrophysical r-process, 
a process of consecutive neutron capture and beta
decay in an environment with a very high neutron 
flux in some astrophysical site, which
had not fully been identified for long \cite{Arnould07,Thielemann17,Kajino17}.
The recent observation of the merging of a binary neutron-star system \cite{Abbott17}
affirmed the importance of this scenario for the r-process nucleosynthesis.

Fission is believed to play an important
role in the r-process itself by fission cycling that 
limits the mass range of the r-process path and has an influence on
the associated nuclide abundances \cite{Goriely15,Eichler15}.
The r-process nucleo-synthesis
cannot be fully understood without a good knowledge 
of the fission properties of very
neutron-rich isotopes of the heaviest elements, which 
are presently not accessible to
direct measurements \cite{Jesus15}.
Therefore, additional interest for a
better understanding of the fission properties of nuclei
far from stability comes from astrophysics.

In a general sense,
nuclear fission offers a rich laboratory for a broad 
variety of scientific research on
nuclear properties and general physics. 
The relatively flat potential 
energy of fissile nuclei reaching to very large
deformations, if compared to lighter nuclei, allows studying nuclear properties like shell 
effects in super- and hyper-deformed 
shapes \cite{Delaroche06}. Phenomena connected with the decay 
of the quasi-bound nuclear
system beyond the fission barrier yield information on 
nuclear transport properties like
nuclear viscosity \cite{Jacquet09,Vardaci15} and heat transfer between the 
nascent fragments \cite{Schmidt10}. They even
offer a valuable test ground of general importance 
for non-equilibrium processes in isolated
mesoscopic systems, where quantum mechanics and 
statistical mechanics play an important role \cite{Kadmensky07}.

During the last years, there has been a considerable
activity in the field of nuclear fission, both 
experimental and theoretical. Several detailed and 
some comprehensive papers have been written on the 
development of theoretical approaches and formalisms with an impetus
on fission dynamics and its signatures. This is illustrated by a
representative list of
papers that were published during the last seven years 
\cite{Baran11,Bernard11,Mirea11,Randrup11,Randrup11a,Robledo11,Sadhukhan11,Schmidt11a,Younes11,
Andreev12,Carjan12,Moeller12,Panebianco12,Warda12,Warda12a,
Andreev13,Giuliani13,Ivanyuk13,Randrup13,Sadhukhan13,Staszczak13,
Aritomo14,Giuliani14,Ivanyuk14,McDonnell14,Mirea14,Moeller14,Nadtochy14,Naderi14,
Rodriguez14a,Rodriguez14b,Sadhukhan14,Schunck14,Simenel14,Soheyli14,Aritomo15,
Carjan15,Goddard15,Jurado15,Lemaitre15,Mazurek15,Moeller15,Moeller15a,Pahlavani15,Radionov15,Scamps15,Scamps15a,
Schunck15,Tanimura15,Vardaci15,Vonta15,Warda15,Zhao15,
Andreev16,Balasubramaniam16,Bulgac16,Capote16a,Eremenko16,Goddard16,Maslyuk16,Nadtochy16,
Pahlavani16,Pasca16,Poenaru16,Regnier16,Rodriguez16a,Sadhukhan16,Santhosh16,Schunck16,Singh16,Usang16,Zhao16,
Denisov17,Eslamizadeh17,Ishizuka17,Kumar17,Moeller17,Pahlavani17,Pomorski17,Sadhukhan17,
Sierk17,Usang17,Tanimura17,Tao17,Ward17,Zdeb17}. 
The description of fission cross sections and fission probabilities in induced fission that
involves the entrance channel, the transmission through the fission
barrier and the competition with other exit channels, is a field of continuous activity
\cite{Bjornholm80,JEFF18,Younes03,Sin06,
Capote09,Goriely09,Kawano09,Goriely11,Bouland13,Romain16,Sin16}. 
In this domain, the coupled-channels model, the dispersive optical potential, and an heuristic model for resonant transmission are of eminent importance.
On the experimental side there have been several publications 
on refinements of existing \cite{Martin14,Blanc15,Jandel15,Oberstedt15,Bhike17,Gaudefroy17} or the development
of novel techniques \cite{Boutoux13,Rodriguez14} as well as on new experimental findings
\cite{Bail11,Goeoek11,Naik11,Nishinaka11,Noda11,Ressler11,Ryzhov11,Sarmento11,
Adili12,Calviani12,Chyzh12,Enqvist12,Hughes12,Kessedjian12,Mueller12,Mukhopadhyay12,Naik12,
Naik12a,
Andreyev13b,Billnert13,Caamano13,Chyzh13,Czeszumska13,Desai13,Desai13a,Elseviers13,Lane13,
Liberati13,Oberstedt13,Salvador13,Ullmann13,
Ayyad14,Casperson14,Chatillon14a,Chyzh14,Desai14,Ghys14,Goeoek14,Hughes14,Ishkhanov14,
Kalaninova14,Karadimos14,Mueller14,Oberstedt14,Yanez14,
Belyshev15,Bhatia15,Caamano15,Deppman15,Desai15,Farget15,Kessedjian15,Lebois15,Martin15,
Oberstedt15a,Pal15,Paradela15,Prasad15,Rodriguez15,Salvador15,Salvador15a,
Adili16,Diakaki16,Duke16,Gagarski16,Leguillon16,Meierbachtol16,Naik16,Oberstedt16,Penttilae16,
Truesdale16,Gatera17,Gupta17,Hirose17,Pellereau17,Sen17,Wang17,Wilson17,Naik18}. 
Usually, the technical development of the specific theoretical or
experimental approach, its challenges and achievements, are in the focus of these papers.
Theoretical papers often intend to demonstrate the quality of a specific approach
by showing its ability for reproducing some distinct data and to present computational  
algorithms that provide suitable approximate descriptions when exact solutions are out of reach,
which is often the case.

Experiments and evaluation\footnote{Evaluation assesses measured data and their uncertainties, reconciles discrepant experimental data and fills in missing data by exploiting systematic trends of the measured data in order to provide reliable data primarily for applications in nuclear technology. Evaluation work is organized, and resulting nuclear-data tables are disseminated by several Nuclear Data Centres under the auspices of the International Atomic Energy Agency \cite{NRDC}.}
are often driven by the interest for reliable nuclear data that are required for 
technical applications, mostly in reactor technology. 
In this context, we mention for example (i)
measurements of prompt-neutron spectra at the LANSCE accelerator facility at the Neutron Science Centre in Los Alamos, New Mexico, USA \cite{Haight14}, (ii)
fission-fragment yield, cross-section and prompt-neutron and gamma-emission data from actinide isotopes at the Joint Research Centre in Geel (previously called IRMM), Belgium \cite{Hambsch14},
(iii) a coordinated research project (CRP) on the evaluation of prompt-fission-neutron spectra of actinides, organized by the IAEA, Vienna, Austria \cite{Capote16},
(iv) measurements of the energy dependence of fission-product yields from $^{235,238}$U and $^{239}$Pu at TUML, Durham, North Carolina, USA \cite{Gooden16},
%\cite{Lestone16},
(v) MCNP studies on multiple scattering contributions, over-corrected background, and inconsistent deconvolution methods used in evaluated prompt-fission-neutron spectra of $^{239}$Pu(n,f) \cite{Neudecker16}, (vi)
high-precision measurement of the prompt-neutron spectrum of $^{252}$Cf at the Rensselaer Polytechnic Institute, Troy, New York, USA \cite{Blain17}, (vii)
a measurement of the energy and multiplicity distributions of neutrons from the photo-fission of $^{235}$U, again at the Los Alamos Neutron Science Center \cite{Clarke17}, and
(viii) measurements of beta decay of fission products with the Modular Total-Absorption Spectrometer at the Holifield Radioactive Ion Beam Facility, Oak Ridge, Tennessee, USA  \cite{Fijalkowska17,Rasco17}.
%, and (ix)
%measurements of the prompt-gamma-ray emission characteristics as a function of excitation energy at %the Oslo Cyclotron Laboratory, Oslo, Norway \cite{Rose17}.
We would also like to mention that 
there has been a European initiative in the last years to measure the 
neutron-induced fission cross section of $^{242}$Pu using very different 
experimental methods in order to provide several completely independent 
measurements of this cross section. This is the only means to strongly 
reduce the final systematic uncertainties in this cross section and 
comply with the very demanding uncertainty required by reactor 
applications \cite{Salvatores08}. Measurements of the $^{242}$Pu(n,f) cross section were 
recently carried out for example at the JRC Geel \cite{Salvador15}, at the National 
Physical Laboratory in Teddington, UK \cite{Matei17}, at the Van de Graaff 
accelerator of Bruyères le Châtel, France \cite{Marini17}, at the 
Physikalisch-Technische Bundesanstalt in Braunschweig, Germany \cite{Belloni17} and 
at the nELBE facility of the Helmholtz-Zentrum Dresden - Rossendorf, 
Germany \cite{Kloeger17}. 

The present review article has a different goal. 
It aims at promoting an improved
understanding of the nuclear-fission process
by establishing a synopsis
of different theoretical approaches and of
empirical knowledge on a general level.
Its impetus lies in tracing back experimental
findings to the underlying physics on different 
levels, reaching from
microscopic descriptions to 
statistical mechanics while covering essentially all
fission quantities.
%, excluding those connected with the
%entrance channel, for example the fission cross sections. 

We carefully consider the fission barrier,
because it is a key quantity for establishing the probability of fission to occur in
competition with other decay channels. In particular, it is important 
for determining the individual
contributions when modeling multi-chance fission, that is fission after
the emission of one or more neutrons. 
While direct information on the
evolution of the system between saddle and scission is scarce, the 
fission-fragment yields, the total kinetic energies and the prompt-neutron multiplicities
are the most important observables that result from this
evolution by evidencing the division of the protons and 
neutrons between the fragments and the magnitude and the division of the 
excitation energy between the fragments.
The energy spectra of prompt neutrons, \cite{Capote16} and references therein, and prompt gammas 
\cite{Chyzh12,Billnert13,Chyzh13,Oberstedt13,Ullmann13,Chyzh14,Oberstedt14,Oberstedt16} are
calculable in a rather straightforward manner within the statistical model, once the conditions
at scission (fission-fragment
distributions in mass and atomic number, excitation energy and angular momentum) 
are imposed.
This is demonstrated by the rather successful attempts \cite{Chen12,Becker13,Vogt13,Stetcu14,
Talou14,Talou14a,Vogt14,Litaize15,Verbekea15,Talou16,Madland17,Vogt17}, 
in cases, where the fragment properties mentioned above are empirically known or can be estimated. 
The latter work is motivated by the interest of nuclear technology to better
reproduce the measured energy spectra in order to
control the irradiation load and the heat production in fission .
The GEF model \cite{Schmidt16} (see section \ref{4-3}) also gives a rather accurate reproduction of the measured
spectra and other characteristics of the prompt neutrons and prompt gammas. In this case, 
the conditions at scission are provided by the model itself.
Therefore, the de-excitation process does not carry so much information on the fission 
process itself,
if we disregard the quest for scission neutrons, which are neutrons of non-statistical
nature emitted at scission \cite{Capote16}, whose existence is controversially discussed \cite{Lestone16}, 
and the efforts to better understand the generation of angular momentum of the fission 
fragments, which is still not well understood \cite{Kadmensky07}.
 
We will concentrate on low- and medium-energy fission, where binary fission, ending up in two heavy fragments, is the dominant decay channel. 
Ternary fission with its very specific features is
not included. A compact but rather exhaustive record on the most 
relevant experimental and theoretical work in the field of ternary fission is given 
in the introduction of Ref.~\cite{Santosh15}.
The decay of highly excited
nuclei, where the phase space favors 
multi-fragmentation \cite{Bondorf82,Borderie08}, that is the 
simultaneous decay into more than two fragments,
and quasi-fission \cite{Rietz13} after heavy-ion reactions that 
preserve a memory on the entrance channel are not 
covered neither.  

The present article is structured as follows.
After a short reminder on the former status of knowledge in section \ref{2}, we
will give a review on the recent innovations in experimental and
theoretical work in sections \ref{3} and \ref{4}, respectively.
In detail, major steps in experimental fission research will be reported that 
were made during the last years by the application of inverse kinematics and by technical innovations 
in the application of beta-delayed fission.
On the theoretical side, applications and further developments of fully dynamical descriptions
of the fission process in quantum-mechanical and classical models will be presented
in sections \ref{4-1} and \ref{4-2}.
In addition, the application of a number of general concepts in a semi-empirical model,
described in section \ref{4-3} and \ref{4-4}, delivered very interesting explanations or opened well targeted
questions for some prominent and some very peculiar observations that stayed
unexplained for long time or that emerged from the results of recent experiments. 
Models on specific aspects of fission are presented in section \ref{4-5}.
A general discussion of current problems that covers 
experimental results and different theoretical models and ideas is 
provided in section \ref{5}, followed by a discussion on the expected evolution of fission
theory and experiments to satisfy the main identified needs.
A full scenario for fission is then proposed, and the main
associated theoretical achievements are highlighted. Finally, a summary is found in section \ref{6}. 

In the interest of the comprehensive and consistent discussion 
in section \ref{5}, section \ref{2} does not include former experimental 
results that were interpreted only recently in the framework of
new theoretical ideas. In a similar way, section \ref{4} presents
predominantly the basic and the technical aspects of the different
models.

\section{Former status of knowledge} \label{2}

Since the discovery of nuclear fission \cite{Hahn39,Hahn39a}, the bulk of the experimental 
results has been obtained in neutron-induced fission of available 
and manageable target nuclei. Limitations arose from the 
small number of primordial or long-lived heavy target nuclides 
and from the technical difficulties of experiments with mono-energetic
neutrons of arbitrary energy. Therefore, and due to some other 
practical and technological reasons, the majority of
the experiments on fission yields has been made with thermalized reactor
neutrons and, to a lesser extent, with fast neutrons (that denotes typical
energies of unmoderated or slightly moderated fission neutrons) and neutrons 
of 14 MeV, e.g. from the $^2$H + $^3$H $\to$ n + $^4$He reaction. 
Moreover, in all experiments performed in direct
kinematics, the kinetic energies of the fission products are
hardly sufficient for obtaining an unambiguous identification
of the fission products in $A$ $and$ $Z$ in kinematical measurements,
which implies that complete fission-product distributions on the $N$-$Z$ plane 
could not be obtained.
%are unable to provide a complete fission-product distribution in A and Z. !!! (Beatriz)

An exhaustive overview of the experimental results on 
neutron-induced fission is given in a recent review by
G\"onnenwein \cite{Goennenwein14}. It essentially covers
fission cross sections,
the fission-product mass distributions, 
%and the underlying fission modes, 
kinetic and excitation energies, nuclear-charge 
distributions in the group of the lighter fission products with an emphasis on the odd-even staggering, 
and the emission of prompt neutrons and gammas.

Also many experiments on charged-particle-induced fission \cite{Kailas14}
and photon-induced fission (bremsstrahlung or mono-energetic 
photons), e.g. Ref.~\cite{Belyshev15} and references therein, 
have been and are still being performed. 
The use of charged-particle or heavy-ion projectiles
made a number of additional nuclei available for fission experiments
by transfer reactions \cite{Gavron76} or fusion (e.g. 
refs.~\cite{Hulet89,Hoffman95,Hessberger17}). The easy accessibility of
higher excitation energies and the inevitable population of 
larger angular momenta allow to study other aspects
of the fission process. They are described in the recent review by
Kailas and Mahata \cite{Kailas14}. These are not covered in this work.

The prominent theories developed in parallel with these observations 
provided potential-energy surfaces of the
fissioning systems in macroscopic-microscopic \cite{Mosel71} or in self-consistent
microscopic approaches \cite{Berger84}. 
Brosa et al.~\cite{Brosa90} associated gross structures that manifest experimentally
by clear structures in the mass and nuclear-charge yields, prompt neutron yields
and fragment total kinetic energies (TKE) with valleys in the multi-dimensional potential,
which imply particularly stable combinations of mass splits.
The nomenclature introduced in \cite{Brosa90}, which distinguishes the "super-long" (SL) 
symmetric fission mode and the asymmetric "standard" fission modes in sequence of
increasing mass asymmetry (S1,S2), is still used. Often, they are labeled as fission channels.
Fine structures in the fission-fragment mass and nuclear-charge distributions 
were analyzed with the combinatorial model 
of Nifenecker et al.~\cite{Nifenecker82}. However, both models \cite{Brosa90,Nifenecker82} rather
remained on the level of empirical parametrizations.
After first qualitative considerations on the dynamical evolution
of the fissioning system at low excitation energies,  
e.~g.~by the S-matrix formulation of N\"orenberg \cite{Noerenberg74},
two types of microscopic
quantum-dynamic calculations have been performed rather early. 
Already in 1978, the time-dependent Hartree-Fock
method \cite{Negele78} was applied to fission. Later, time-dependent
calculations based on the generator-coordinate method using
Hartree-Fock-Bogoliubov states were performed, and the most
probable fission configuration of $^{240}$Pu was analyzed \cite{Berger84}.
Wilkins et al.~\cite{Wilkins76} performed quantitative 
calculations of fission quantities, for example fission-fragment mass
distributions, charge polarization (that leads to different $N/Z$ ratios of the two complementary fragments), total kinetic energies and 
prompt-neutron multiplicities, with a static statistical 
scission-point model, including the influence of shell effects
and pairing correlations.\footnote{See section \ref{4} for detailed information about 
the different theoretical approaches.}
Although this model disregards any
influence of the dynamics, which prevents for example obtaining any information
on dissipation or fission times, successors of this model are still being
developed \cite{Ivanyuk14,Carjan15,Lemaitre15}, often achieving a
good reproduction of measured mass distributions and other quantities, for example
for thermal-neutron-induced fission of $^{232}$Th, $^{235}$U, $^{239}$Pu, and $^{245}$Cm \cite{Ivanyuk14},
and for spontaneous fission of nuclei around $^{258}$Fm and above \cite{Carjan15}.

The status of experimental and theoretical research described in this section, which
we denote as former knowledge,
has been achieved around the 50th anniversary of the discovery of fission in 1989, 
when the ``conventional" technical possibilities of experimental equipment and computer power, 
respectively, had been exploited to a large extent.
Considerable progress has been achieved in the following years up to present times, mostly 
due to novel experimental methods and the exponentially developing possibilities for performing appreciably more 
complex theoretical calculations, as described in the following sections.

\section{Experimental innovations} \label{3}

There has been a continuous progress in the quality of experimental
equipment by the development in technology on many fields. 
This allowed to improve the quality and to extend the quantity
of experimental results in many aspects. As a direct consequence,
the data basis for applications in nuclear technology 
has considerably improved.
In the present section,
we give a concise overview on only a few major developments that gave a 
considerably improved insight into the physics of the fission process.
A comprehensive and detailed overview on technological developments and 
new experimental results is presented in a dedicated review
that appeared very recently in the same journal \cite{Andreyev18}.

\subsection{Accessible fissionable nuclei} \label{3-1}

The progress in the understanding of fission heavily relied and still relies on the
development of advanced experimental methods. A severe restriction is still the
availability of fissionable nuclei as target material. Therefore, the traditional use of
neutrons for inducing fission offers only a rather limited choice of fissioning systems.
These limitations were more and more overcome by alternative methods: For instance, 
spontaneously fissioning heavy nuclei are being produced by fusion reactions \cite{Ghiorso67}
since many years, as already mentioned. These experiments provide fission-fragment 
mass distributions for very heavy nuclei, finally
limited by statistical uncertainties due to the low number of nuclei being produced
\cite{Hessberger17}.

Recently, very neutron-deficient nuclei, e.g. in the $Z = 80$ region, were
produced in spallation reactions at ISOLDE \cite{ISOLDE}, CERN, which undergo beta-delayed fission \cite{Andreyev13}.
This experiment profited from an unambiguous identification of the fissioning nuclei, 
mass selection by ISOLDE, and $Z$ selection by the resonance-ionization laser
ion source RILIS \cite{RILIS}. 
A pronounced double-humped mass distribution was found for the 
fission fragments of the compound nucleus $^{180}$Hg, formed as the daughter nucleus 
after beta decay of $^{180}$Tl, which has similarities with the 
double-humped mass distribution observed previously by Itkis et al.~\cite{Itkis88} for the fission of
excited $^{201}$Tl that is situated close to beta stability, in an alpha-induced reaction.
Fission-fragment mass distributions were also studied in the beta-delayed fission of
the daughter nuclei $^{178}$Hg and $^{194,196}$Po at ISOLDE and $^{202}$Rn
at SHIP (GSI) \cite{Andreyev13}, and asymmetric fission, respectively, complex shapes were observed. 
These unexpected observations triggered several experiments with different techniques (heavy-ion fusion-fission reactions and electromagnetic-induced fission in inverse kinematics) and a number of theoretical works with self-consistent and stochastic models.
They will be described in detail in the following.
Again, by the study of beta-delayed fission at ISOLDE, Ghys et al.~\cite{Ghys14} found
a single-humped fission-fragment mass distributions in the beta-delayed 
fission of the daughter nucleus $^{194}$Po and
indications for triple-humped distributions for $^{196}$Po and $^{202}$Rn.
Asymmetric fission was observed following the population of $^{182}$Hg at an excitation energy of
22.8 MeV above the saddle point, while symmetric fission was observed for $^{195}$Hg nuclei at
a similar excitation energy above the saddle point in heavy-ion fusion-fission reactions \cite{Prasad15}. 
In this case, the situation is complicated by the possibility of multi-chance fission, that means
contributions from fission of neighboring nuclei at lower excitation energy after pre-fission
emission of neutrons and, in the case of very neutron-deficient systems, also protons. 
Multi-chance fission should also be considered when interpreting
the measured mass distributions of $^{180}$Hg and $^{190}$Hg, formed in
$^{36}$Ar-induced reactions \cite{Nishio15}, where asymmetric fission was found, 
and for the measured mass distributions of $^{179}$Au and $^{189}$Au, formed in $^{35}$Cl-induced reactions \cite{Tripati15}, where symmetric fission was observed for the lighter system, while the heavier system showed indications for an asymmetric component.  
These experiments revealed the complexity in the fission-fragment mass distributions in the
lead region, but they are still too fragmentary to establish a full systematics of the variations
with the fissioning system. Although beta-delayed fission is especially suited for studying low-energy fission, it is very restricted to nuclei with beta Q values above or not much below the fission threshold and high fission probabilities. These are mostly odd-odd, very neutron-deficient nuclei. 
Alternative techniques are required to establish a more complete systematics of low-energy fission in the
lead region. The most promising one is electromagnetic fission of secondary beams at relativistic energies that will be mentioned below.
% \cite{Ghys14,Prasad15 8. Juni,Tripati15 12. August,Nishio15 30.Juni}

Advanced experimental studies on light-charged-particle-induced fission 
probabilities of systems 
that are not accessible by neutron-induced fission are being performed 
systematically.  
These surrogate-reaction studies focus on the ability of these alternative reactions to
simulate neutron-induced reactions
\cite{Escher12}. 
%\cite{Kessedjian15}.
Also recently, the use of heavier ions, for example $^{16}$O, in transfer 
reactions allowed to appreciably extend 
the range of fissionable nuclei available for fission studies and
for measuring the fission-fragment mass distributions over an extended range of
excitation energy \cite{Hirose17} at the JAEA tandem facility at Tokai.
Moreover, comprehensive studies on fission of transfer products of 
$^{238}$U projectiles, impinging on a $^{12}$C target, have been performed at GANIL, Caen, in 
inverse kinematics, covering fission probabilities \cite{Rodriguez14} and 
fission-fragment properties like yields of all produced nuclides that are identified in $Z$ and $A$, as well as fission-fragment kinetic energies \cite{Farget15}.

The most remarkable enlargement in the number of systems, being accessible for low-energy
fission studies was provided by exploiting the fragmentation of relativistic $^{238}$U projectiles at GSI, Darmstadt. 
From the results obtained up to now, one can estimate that
more than 100 mostly neutron-deficient projectile fragments with $A \le 238$ are accessible for
low-energy fission experiments in inverse kinematics by electromagnetic 
excitations \cite{Schmidt00,Boutoux13,Martin15,Pellereau17}. 
When fission events after
nuclear interaction are suppressed, the induced excitation-energy distribution  
centers at about 14 MeV above the ground state with a FWHM of about 7 MeV \cite{Schmidt00}.
First results provided a systematic mapping of the transition from asymmetric fission around $^{235}$U to symmetric fission around $^{217}$At by measuring the fission-fragment $Z$ distributions \cite{Schmidt00}.
Recently, investigations with relativistic secondary beams
were performed also at GSI by the SOFIA collaboration (Studies Of Fission with Aladin) with an improved set-up that
provided an unambiguous identification of the fission fragment in $Z$ $and$ $A$
\cite{Boutoux13,Martin15,Pellereau17}.
An exploratory study revealed that this method is also
applicable for ligher nuclei, extending to the neutron-deficient 
lead region \cite{Martin15}.
 
\subsection{Boosting the fission-fragment kinetic energies} \label{3-2}

The identification of fission products poses a severe problem. First experiments
were based on radiochemical methods \cite{Crough77,Laurec10}. Although this approach
provides unambiguous nuclide identification (in $A$ and $Z$) of the fission fragments, 
it is not fast enough 
for determining the yields of short-lived fragments, and it suffers from normalization 
problems, e.g.~by uncertainties of the gamma-spectroscopic properties.
Identification with kinematical methods by double time-of-flight 
\cite{Stein57,Milton58} 
and double-energy measurements \cite{Schmitt65}
provides complete mass distributions, however, with limited resolution of typically 4 mass units FWHM \cite{Gaudefroy17} and problems
in the mass calibrations due to uncertainties in the correction for prompt-neutron emission 
\cite{Adili12}. 
At the expense of a very small detection efficiency,
the COSI-FAN-TUTTE set-up \cite{Oed84,Sicre86} had some success in measuring mass and 
nuclear charge of fission fragments at high total kinetic energies in the light group 
in thermal-neutron-induced fission of some suitable target nuclei
by combining double time-of-flight, double-energy and energy-loss measurements. The LOHENGRIN
separator brought big progress in identifying the fission products in mass
and nuclear charge \cite{Moll75}, although the $Z$ identification was also limited to 
the light fission-product group. This technique was applied to thermal-neutron-induced
fission of a number of suitable targets that were mounted at the ILL high-flux
reactor, see Ref.~\cite{Goennenwein14}. Recent attempts for developing COSI-FAN-TUTTE - like detector assemblies with 
higher detection efficiency and better resolution are presently being made, but showed 
only limited success up to now 
\cite{Wang13,Meierbachtol15,Fregeau15,Tsekhanovich07,Panebianco14}. 
In particular, the $Z$ resolution is severely impeded by straggling phenomena.

Full nuclide identification (in $Z$ and $A$) of $all$ fission
products has only been achieved by boosting the energies of the products in 
inverse-kinematics experiments and by using powerful magnetic spectrometers
\cite{Schmidt00,Boutoux13,Martin15,Pellereau17,Caamano13,Navin14}.

\subsection{Results} \label{3-3}

Some of the most prominent new results have been obtained in fission 
experiments performed in inverse kinematics 
on electromagnetic-induced fission at relativistic energies \cite{Schmidt00,Martin15,Pellereau17} 
and on transfer-induced fission at energies slightly above
the Coulomb barrier \cite{Farget15} at the VAMOS spectrometer of the GANIL facility. 

In Ref.~\cite{Schmidt00}, the fission-fragment 
$Z$ distributions of 70 fissionable nuclides from $^{205}$At to $^{234}$U were measured,  
using beams of projectile fragments produced from a 1 $A$ GeV $^{238}$U primary beam 
and identified by the fragment separator of GSI, Darmstadt. 
The measured $Z$ distributions show a gradual transition from single-humped 
to double-humped distributions with increasing mass,
with triple-humped distributions for fissioning nuclei in the intermediate region 
around $A = 226$. The position of the heavy component of asymmetric fission
could be followed over long isotopic chains and turned out to be very close to 
$Z = 54$ for all systems investigated. In a refined analysis, it was 
shown that the mean $Z$ values of the contributions to the heavy component 
from the two most prominent asymmetric fission channels are nearly the same for all 
actinides \cite{Boeckstiegel08}. 
Moreover, the odd-even structure in the 
$Z$ yields was found to systematically increase with asymmetry and to have similar
magnitudes for even-$Z$ and for odd-$Z$ fissioning nuclei at large asymmetry
\cite{Steinhaeuser98,Caamano11}. 
The importance of these findings for the theoretical understanding of the
fission process is further discussed in sections \ref{4-3-2} and \ref{4-3-3}.   

The SOFIA experiment \cite{Boutoux13,Martin15,Pellereau17} that used a refined and extended set-up 
compared to the one used in Ref.~\cite{Schmidt00}
%The refined and extended set-up of the SOFIA experiment \cite{Boutoux13}
made possible to fully identify unambiguously event by event
all fission products in $Z$ $and$ $A$ from
electromagnetic-induced fission of relativistic $^{238}$U beam and its fragmentation residues.
The experiment profited also from the higher available beam intensity, which allowed to
extend the range of nuclei to be investigated and to reduce the statistical
uncertainties.

\begin{figure} [h]  
\begin{center}
\includegraphics[width=0.5\textwidth]{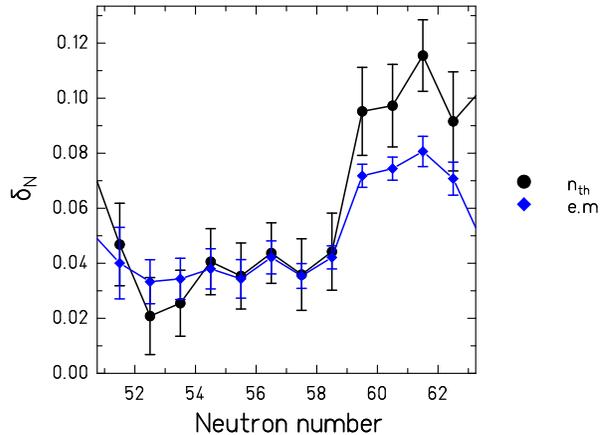}
\caption{(Color online)
Logarithmic four-point differences $\delta_n$ \protect\cite{Tracy72}
in the fission-fragment neutron-number distribution
for electromagnetic-induced fission of $^{238}$U
measured in the SOFIA experiment \cite{Pellereau13} and in thermal-neutron-induced fission
of $^{235}$U \cite{Lang80}. 
%In addition, this quantity is also shown for
%the fission of $^{250}$Cf, produced with an initial excitation energy of 45 MeV 
%in the fusion of $^{238}$U and $^{12}$C \cite{Caamano13}.
The figure shows the most relevant data in the range that contains 
about 90\% of the $N$ distribution of the system $^{235}$U(n$_{th}$,f)
in the light fission-fragment group.
}
\label{N-EO}
\end{center}
\end{figure}

\begin{figure*}[h]  % top
\begin{center}
\includegraphics[width=1.3\textwidth,angle=90]{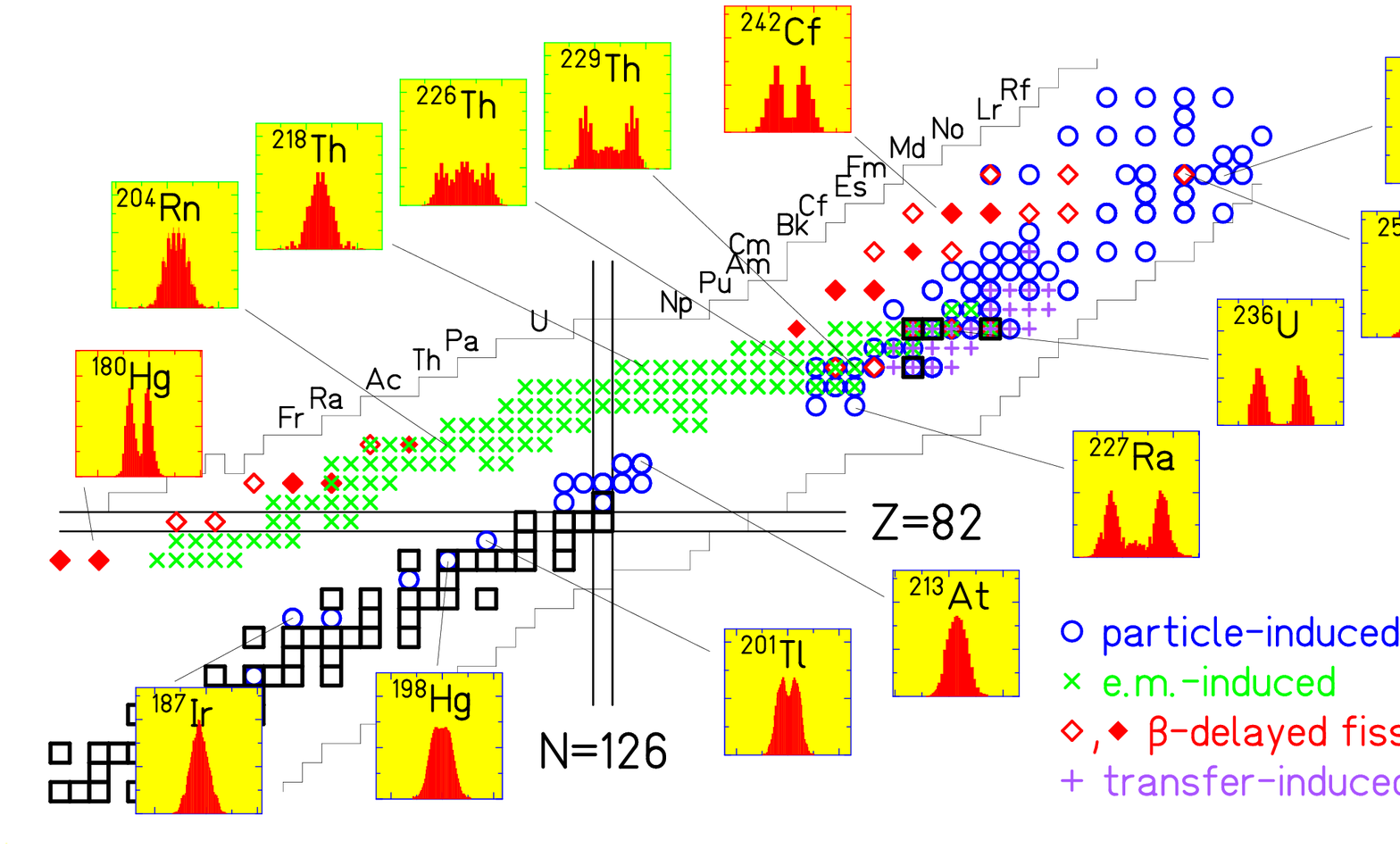}
\caption{(Color online) Updated overview of fissioning systems
investigated up to $\sim$2016 in low-energy fission with excitation energies
up to 10 MeV above the fission barrier.
In addition to the systems for which fission-fragment mass distributions (FFMDs) have previously been obtained in
particle-induced and spontaneous fission ($\circ$), the nuclei
for which fission-fragment $Z$ distributions after electromagnetic excitations
were measured in
the 1996 experiment \protect\cite{Schmidt00} and in the recent SOFIA experiment \protect\cite{Martin15,Pellereau17} in inverse kinematics
at the FRS at GSI ($\times$) and
the fissioning daughter nuclei studied in $\beta$-delayed fission
($\diamond$) are shown. Full diamonds mark systems for which FFMD were measured,  the data are from \protect\cite{Andreyev13,Ghys14} and references therein. Furthermore, 25 nuclei are marked ($+$), including FFMDs obtained from multi-nucleon-transfer-induced fission with $^{18}$O+$^{232}$Th protect\cite{Leguillon16} and $^{18}$O+$^{238}$U target \protect\cite{Nishio15}.
Several examples of measured FFMDs are
shown, data from Refs.~\protect\cite{Itkis88,Weber76,Lang80,Britt84}.
For orientation, the primordial isotopes are indicated by squares.
The figure is a modification of figure 1 in Ref.~\protect\cite{Schmidt00}, of figure 1 in Ref.~\protect\cite{Andreyev13}, and of figure 5 in Ref.~\protect\cite{Andreyev18}.
}
\label{Fig_KHS}
\end{center}
\end{figure*}

As one of the most prominent results, 
this experiment showed for the first time that the fine structure in the
fission-product $N$ distribution depends only weakly on the excitation energy
of the fissioning system, in contrast to the odd-even staggering in the $Z$ distribution.
This can be seen in figure \ref{N-EO}, where the logarithmic four-point 
differences\footnote{The logarithmic four-point difference \cite{Tracy72} quantifies the deviations 
of a distribution
from a Gaussian function, see section \ref{4-3-3}. It is used to determine the local odd-even staggering 
\cite{Gonnenwein92}, but
it contains also other contributions, for example from fine structures due to shell 
effects and from a kerf between different fission channels.},
%\newline
$\delta_n(N + 3/2) = 1/8 (-1)^{N+1} ( \ln Y(N+3) - \ln Y(N) - 3 [\ln Y(N+2) - \ln Y(N+1) ] ) $,
%\newline
of the fission-fragment $N$ distribution from
the SOFIA results for electromagnetic-induced
fission of $^{238}$U are compared with those obtained for thermal-neutron-induced
fission of $^{235}$U.\footnote{Unfortunately, there are no data available yet that allow comparing the 
$N$ distributions for the same fissioning nucleus. However, the neutron separation energies, which are
believed to be at the origin of the odd-even staggering in neutron number, 
see Ref.~\cite{Ricciardi11} and section \ref{4-3-3}, 
vary only little along the slight displacement in the fission-fragment $Z$ distributions for a fixed $N$. 
Moreover, a direct comparison is anyhow not possible due to the presence of multi-chance fission
in the SOFIA experiment. }
Around the maximum of the $N$ distribution at $N \approx 56$,
the $\delta_N$ values are almost identical, and they are fairly close below and above
$N = 56$.
In contrast, the odd-even staggering in 
the $Z$ distribution decreases by about 50\% \cite{Caamano11} when comparing
electromagnetic-induced with thermal-neutron-induced fission. 
See section \ref{4-3-3}
for further discussion of the fine structure in the fission-fragment yields.

Also in the VAMOS experiment on transfer-induced fission and fusion-fission around the Coulomb barrier
in inverse kinematics, a separation in $Z$ and $A$ 
of all fission products was obtained \cite{Caamano13}, although the peaks showed 
some overlap, preventing an unambiguous event-by-event identification.
This experiment provided for the first time complete fission-product nuclide 
distributions after the formation of a $^{250}$Cf compound nucleus at an excitation
energy as high as 45 MeV, produced in the fusion of $^{238}$U projectiles with $^{12}$C 
\cite{Caamano13} and a number of transfer products \cite{Ramos16}. 
This enabled, for the first time, to systematically study the dependence 
of the fission-fragment $N/Z$ degree of freedom (charge polarization and fluctuations) on excitation 
energy \cite{Caamano13,Caamano15,Ramos16}, regarding that the full nuclide identification had 
previously been obtained 
for the light fission products from thermal-neutron-induced fission of a small
number of fissioning systems, only.

The observation of a double-humped mass distribution in the fission of the very
neutron-deficient $^{180}$Hg nucleus in beta-delayed fission and different kind 
of structure in the 
mass distributions from fission of other nuclei in this region in different
experiments \cite{Andreyev13} demonstrated that 
complex structural effects are a rather general phenomenon in low-energy fission,
not restricted to asymmetric fission in the actinides and multi-modal fission 
around $^{258}$Fm.
This result demonstrates that, contrary to the symmetric fission in neutron-rich 
Fm isotopes, which is explained by the simultaneous formation of two fragments
close to the doubly-magic $^{132}$Sn, the production of two semi-magic $^{90}$Zr
fragments is not favored in the case of $^{180}$Hg.   
A large variety of neutron-deficient nuclei reaching down even below mercury is 
also accessible to fission studies at energies close to the fission barrier with 
the SOFIA experiment. A few exploratory measurements have already been made
\cite{Chatillon14}.

The experimental observations in the lead region were related to peculiarities due to shell effects in the
potential-energy landscape, calculated with the macroscopic-microscopic approach \cite{Andreev12,Andreev13,Ghys14} or
with microscopic self-consistent methods \cite{Panebianco12,Warda12,McDonnell14,Ghys14}, partly
applied in full calculations of the mass distributions with the scission-point model \cite{Andreev12,Panebianco12,Andreev13,Andreev16} or in
systematic calculations with the stochastic random-walk formalism \cite{Andreyev13,Ghys14, Moeller15a}.
% \cite{Andreev13,McDonnell14,Ghys14,Panebianco12,Warda12}, 

A glance on the present status of experimental fission research is exhibited in figure \ref{Fig_KHS}
that gives an overview on the observed fission-fragment mass and nuclear-charge distributions.
These are among the most prominent signatures of nuclear structure in low-energy fission. 
Although the coverage on the chart of the nuclides is still very incomplete, the systematic variations
of the fission-fragment distributions
with mass and nuclear charge of the fissioning system are already rather well evidenced. 
The experiments performed during the last two decades account for a large share of this achievement.   

\clearpage

\section{Theoretical innovations} \label{4}

In the following, we will give a survey on the ability of different presently used theoretical approaches and ideas to closely reproduce experimental fission observables and to reveal the physics behind. 
Because the dynamics of the fission process and the influence of shell effects and pairing are considered to be essential assets for the understanding of low-energy fission, static and purely macroscopic approaches are not included. 
The survey comprises microscopic self-consistent approaches, stochastic models and a recently developed semi-empirical approach. 
In contrast to the self-consistent and the stochastic approaches, which are presently restricted to the description of a rather small subset of fission observables, this semi-empirical model \cite{Schmidt16} covers essentially all fission quantities and their mutual correlations. 
This is why the physics background and the abilities of this model will be described in more detail in this review.

The {\it{self-consistent microscopic theory}} aims at describing the fission process in the quantum-mechanical framework under the influence of an effective nuclear force, see section \ref{4-1}. This is the most ambitious and the most fundamental approach, but it still faces considerable technical and computational difficulties.
Also several formal or conceptual issues are not fully solved. 
%To quote some of them: separation of relevant and non-relevant degrees of freedom, fragments recognition and definition of the neck in a quantum liquid, quantum tunneling in many-body evolving system, and ssymmetry breaking and their restauration.
A review on the present status of this type of models, in particular of the implementation, solution methods and results of microscopic self-consistent fission modeling based on the nuclear-density-functional formalism is given by Schunck and Robledo \cite{Schunck16}. 

{\it{Stochastic models}} describe the fission process by a transport equation, whereby the concepts of statistical mechanics are applied to describe the irreversible approach towards statistical equilibrium, see section \ref{4-2}.
In recent applications, the stochastic processes are introduced by a random force in the numerical solution of the Langevin equations or by a random-walk approach.
These stochastic models are essentially classical, while quantum-mechanical features only enter partially, e.g. by the macroscopic-microscopic description of the potential-energy landscape. 
Detailed considerations on the validity of transport theories and a survey of the respective theoretical approaches can be found in 
Refs.~\cite{Weidenmueller80,Abe96}. 
Also stochastic models are computationally demanding and restricted to the description of a limited number of degrees of freedom.
The reader, who is interested in an exhaustive overview on self-consistent and stochastic fission theories can find it, among other topics, in the textbook of 
Krappe and Pomorski \cite{Krappe12}. 

A recently developed {\it {semi-empirical model description}} \cite{Schmidt16}
by-passed the great complexity that is encountered when striving for an ab-initio modeling of 
the fission process by concentrating on the essential features of the process that mostly influence the observables, see section \ref{4-3}.
Qualitative and semi-quantitative ideas in this direction have been developed long time ago, see for example Ref.~\cite{Jensen74}.
The ability to obtain rather accurate consistent quantitative results for practically all fission observables was achieved by linking the observables for different conditions (for example for fissioning systems with different $A$ and $Z$ and for different initial excitation energies and angular momenta) by a tailored theoretical frame. 
Thus, the experimental data were traced back to a rather limited number of about 100 empirical model parameters, from which about 50 are decisive for the actual fission process. These simultaneously describe a large variety of fissioning systems with a unique set of parameter values.

\subsection{Microscopic self-consistent approaches} \label{4-1}

\subsubsection{Basic considerations}

A number of dynamical self-consistent quantum transport theories have been developed for handling nuclear reactions ranging from ab initio to self-consistent mean-field approaches. A description of the different methods can be found for example in the habilitation thesis of D. Lacroix 
\cite{Lacroix10}. 
The application to nuclear fission poses considerable challenges on suitable algorithms and computation resources and is presently an active field of development. 
Up to now, only theories based on the mean-field approximation have been applied to fission.
%\textcolor{red}{Due to the tremendous number of possible final configurations, the fissioning system must be treated as an open system, where only a sub-class of the degrees of freedom associated to the system is explicitly considered, because it is technically impossible to simultaneously treat all (collective and intrinsic) degrees of freedom explicitly
%in a quantum-mechanical framework \cite{Negele89,Tanimura15}. 
%{\bf{(Is this also true for the Bulgac approach? May be, this is connected to the semi-classical character of TDDFT? If not, I would keep the references, but with a different formulation: Negele and Tanimura assumed ...)}} 
%Indeed, often only few collective degrees are treated, and in all cases microscopic calculations ignore many of the single-particle excited states that arise during the fission process because of the difficulties to correctly include nuclear dissipation. 
% The preceding section was replaced by theh following text (proposition B. Jurado), same for the next sections.
Due to the tremendous number of possible final configurations, the fissioning
system must be treated as an open system, where only a sub-class of the degrees of freedom associated to the
system is explicitly considered, because it is technically
impossible to simultaneously  treat all (collective and intrinsic) degrees
of freedom explicitly in a $fully$ quantum-mechanical framework \cite{Negele89,Tanimura15}.
Either only few collective degrees of freedom are treated or the collective
degrees of freedom result from the assumption of non-interacting nucleons
in a mean field. 
%Techniques have been proposed to model quantum dissipation in collective motion as an interaction with a heat bath that represents all intrinsic degrees of freedom, e.g. \cite{Koch08,Lacroix08,Hupin10}, but also this simplified scenario has not yet been used in modeling fission.
%Current microscopic fission models, which will be described in the following, either assume adiabaticity (decoupling of collective and intrinsic degrees of freedom), or they only include one-body dissipation and the lowest quasi-particle excitations.
Moreover, microscopic theories suffer also from an incomplete treatment of
nuclear dissipation. Techniques have been proposed to model quantum dissipation in collective motion as an interaction with a heat bath that represents all intrinsic degrees of freedom, e.g. \cite{Koch08,Lacroix08,Hupin10}, but also this simplified scenario has not yet been used in modeling fission.
$In$ $general$, current microscopic fission models, which will be described in the
following, either assume adiabaticity (decoupling of collective and intrinsic
degrees of freedom), or they only include one-body dissipation.
Two-body dissipation is only partially treated by including pairing correlations
in the calculations that consider one-body dissipation.

%Usually, these are some of the collective degrees of freedom. 
%The other degrees of freedom are attributed to an environment and not explicitly followed, because it is technically impossible to simultaneously treat all collective and single-particle degrees of freedom explicitly in a quantum-mechanical framework \cite{Negele89,Tanimura15}.  Moreover, fission is a dynamical process and it should be treated as such.}

\subsubsection{Density-functional theory}

As said before, all the existing microscopic models of fission are based on teh mean-field approximation.
Unfortunately, there is a significant ambiguity in the names given to the different approaches in literature. 
Therefore, for the sake of clarity, here we adopt the notation given in the recent review article on the microscopic theory of fission by Schunck and Robledo \cite{Schunck16}. 

As shown in \cite{Schunck16}, all the microscopic models that have been applied to fission so far make use of the density-functional theory (DFT), where the term DFT includes self-consistent mean-field theory and extensions beyond mean field. 
Within this framework, the many-body nucleon interactions are approximated by a mean field, where the nucleons evolve independently, the mean field being itself determined by the ensemble of nucleons. 
The energy of the nucleus is a functional of the one-body density matrix, which is determined by solving the Hartree-Fock (HF) or the Hartree-Fock-Bogoliubov (HFB) equation, although in some cases pairing is introduced by using the BCS approximation to the HFB equation. 
Introducing an energy-density functional (EDF) \cite{Kohn65} offers considerable conceptual and practical simplifications to the solution of the many-body problem. 
In general, the EDF is derived from an effective two-body nuclear potential, typically the Skryme or the Gogny effective interactions. 
Note that the parameters of these effective interactions are adjusted to experimental data. 
For attempts to deduce the nuclear force from QCD see Refs.~\cite{Machleidt11,Baldo16}.

Fission dynamics can only be treated approximately either by assuming adiabaticity, i.e. no coupling between collective and intrinsic degrees of freedom, or by representing the wave function at each time by the solution of a HF or, only very recently, a HFB mean field. We refer to \cite{Schunck16} for a detailed description of the DFT theory applied to fission and a complete list of the appropriate references. Here, we only intend to give an overview on the achievements of this kind of approach in contributing to a better understanding of the fission process, on the variety of sophistication and the challenges self-consistent microscopic fission theory is facing.

\subsubsection{Application to fission}

Table \ref{FISSION-Tab1} lists the main approaches based on the DFT that have been applied to fission, together with their main assumptions and the fission quantities and fissioning systems that have been considered. 
Notice that the study of odd-mass fissioning nuclei poses additional difficulties with respect to even-even nuclei and, thus, they are only rarely considered in DFT. 

\begin{figure} [h]  
\begin{center}
\includegraphics[width=0.5\textwidth]{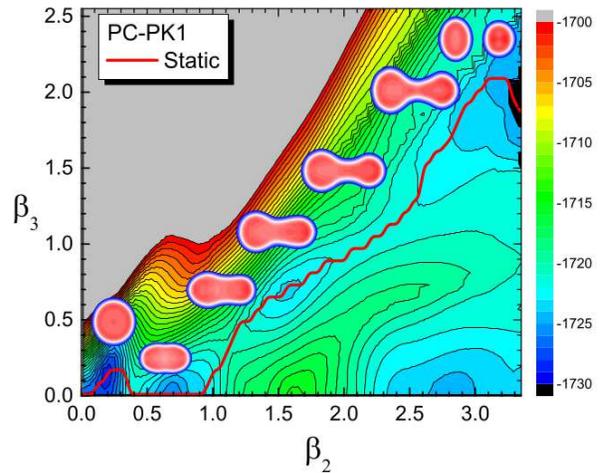}
\caption{(Color online)
Self-consistent quadrupole- and octupole-constrained deformation-energy surface (in MeV) of $^{226}$Th in the plane spanned by the quadrupole and octupole deformation parameters
$\beta_2$ and $\beta_3$. (PC-PK1 denotes the used energy-density functional.) 
The red curve is the static fission path. The density distributions for selected deformations along the fission path are also shown. 
Figure taken from \protect\cite{Tao17}.
}
\label{Tao-Epot}
\end{center}
\end{figure}

The two first approaches listed in Table \ref{FISSION-Tab1} have in common a first static step, where the potential-energy surface (PES) as a function of few selected collective coordinates and the collective inertia, which determines the response of the fissioning nucleus to changes in the collective variables, are calculated by solving the HFB equation.
For computing the PES, the energy of the system is determined under the constraint of the coordinates of the relevant degrees of freedom on a grid of points. 
In each point, the shape of the system is optimized in a self-consistent way with respect to its energy while respecting the imposed constraints.
Figure \ref{Tao-Epot} shows the potential-energy surface of $^{226}$Th, calculated in a self-consistent mean-field approach, as an example. 

The choice of the collective variables is a key point in these approaches. 
Reasonable choices of the relevant collective variables (often coming from studies performed with the semi-classical models described in section \ref{4-2}) are usually made, but there is always some degree of arbitrariness in the selection. 
The addition of collective variables or changes in the definition of the relevant collective degrees of freedom can significantly reduce discontinuities in the PES
\cite{Dubray12} and strongly impact the results. 
We further discuss the issues of the number of collective coordinates and the PES discontinuities below and in section \ref{5-1}. 
Fission dynamics is considered in a second step. 

%\begin{figure} [h]  
%\begin{center}
%\includegraphics[width=0.5\textwidth]{FIG-Zdeb-Epot.eps}
%\caption{(Color online)
%The potential energy surface of $^{252}$Cf on the $Q_2$ - $Q_3$ plane, calculated %with the HFB model. The scission line is depicted in white.
%Figure taken from \cite{Zdeb17}.
%}
%\label{Zdeb-Epot}
%\end{center}
%\end{figure}

The first approach listed in Table 1 is mainly used
in spontaneous fission, where the evolution of the nucleus from the ground state to scission proceeds by tunneling through the fission barrier. 
Spontaneous fission is generally characterized by the fission half-live, which is inversely proportional to the transmission coefficient through the multidimensional PES. 
The transmission coefficient is computed semi-classically with the Wentzel-Kramers-Brillouin (WKB) approximation \cite{Brack72}. 
From the calculated PES, it is possible to deduce a one-dimensional fission barrier, which can be compared to the one-dimensional barriers deduced from experimental data on fission cross sections or probabilities; this will be discussed in section \ref{4-3-1}. 

The second approach in Table 1 is generally applied to induced fission at low energies. 
The evolution of the fissioning nucleus is computed using the adiabaticity approximation. 
As said above, this implies that the collective coordinates are completely decoupled from the intrinsic degrees of freedom, and, consequently, the system does not experience any intrinsic excitation during the course of the reaction. 
The dynamical evolution up to scission is obtained with the time-dependent generator coordinate method TDGCM \cite{Berger84,Goutte05,Regnier16} assuming the Gaussian overlap approximation GOA \cite{Hill53}. 
A step towards a full dynamical microscopic description of the fission process that goes beyond the adiabatic approximation is the generalization of the TDGCM approach by including two-quasi-particle excitations on the whole fission path in the Schr\"odinger Collective Intrinsic Model by Bernard et al.~\cite{Bernard11}. 
To our knowledge, this approach has not yet been applied to infer measurable fission quantities.

Within the adiabatic approximation, there is no scission mechanism, and there is again some part of arbitrariness in the definition of the scission configuration. 
The definitions of scission are either geometrical (e.g. the neck size) or dynamical (ratio between the Coulomb and the nuclear forces). 
These definitions ignore quantum-mechanical effects in the neck region. 
These effects were investigated by Younes and Gogny \cite{Younes11}, who observed that the fission-fragment density distributions have large tails that extend into the other fragment, even when the size of the neck is very small. 
The authors extended the quantum-localization method used in molecular physics to fission by introducing a localization indicator to sort the quasi-particles into the two pre-fragments. 
They showed that the application of quantum localization yields more realistic fission-fragment properties at scission. 
The quantum-localization approach was extended to finite temperature (i.e. excitation energy above zero) by Schunck et al.~\cite{Schunck15}.
The latter studies of scission are static in nature. 
We will show below how the third type of approaches can contribute to a better understanding of the dynamics of scission. 

In \cite{Schunck16}, the third type of approaches listed in Table \ref{FISSION-Tab1} is grouped in the so-called time-dependent density-functional theory (TDDFT). 
Although, in literature, these approaches are denoted in different ways: time-dependent Hartree-Fock TDHF, time-dependent Hartree-Fock-Bogoliuvov TDHFB or time-dependent energy-density functional TDEDF. 
In TDDFT, the real-time evolution of the nuclear system is simulated from an initial condition up to scission by calculating the nuclear wave function at each time with a mean field that results from solving the HF or the HFB equation. 
%TDDFT treats the evolution of single-particle states quantum-mechanically and deduces the behavior of any collective one-body degree of freedom from the single-particle wave functions. 
%However, the evolution of the collective degrees of freedom is treated only semi-classically. {\bf{(Reference?)}} 
%As a consequence, fluctuations of the collective degrees of freedom are strongly underestimated. 
%This prevents this type of approaches from predicting realistic fission-fragment distributions and generally only average fission-fragment properties can be calculated. 
TDDFT performs the evolution of single-particle states self-consistently under
the assumption that the system is described by a state of independent particles
or a Slater determinant at all times. 
From this evolution, one can infer the
behavior of any collective one-body degree of freedom (as e.g. 
multipole deformation, neck formation, fragment kinetic energies, etc.). 
However, the assumption of independent particles leads to a strong underestimation
of the fluctuations of the collective degrees of freedom. In this sense, one can
consider that in TDDFT the evolution of the collective degrees of freedom is treated semi-classically \cite{Flocard78}. As explained in \cite{Schunck16},
this prevents this type of approaches from predicting realistic fission-fragment
distributions, and, generally, only average fission-fragment properties can be
calculated.
In addition, tunneling through the barrier is not included, which implies that TDDFT calculations of low-energy or spontaneous fission have to start beyond the fission barrier. The starting point or the initial condition of the calculations is usually obtained following the same procedure as in the static step of the two first approaches of Table \ref{FISSION-Tab1}. 
The choice of variables used in this static part can have a significant impact in the subsequent dynamical calculations since in principle TDDFT cannot spontaneously break symmetry.  
For example, octupole deformation, which is necessary to form fragments
with different masses, does not emerge spontaneously but has to be imposed already
at the starting point of the dynamical evolution.
TDDFT has the advantage that it is not based on the hypothesis of adiabaticity. 
Indeed, it automatically includes one-body-dissipation effects: as the nucleus changes its shape, single-particle excitations (or quasi-particle excitations when pairing correlations are taken into account) are included, which slows down the collective motion. 
%TDDFT calculations {\bf{(References?)}} based on HFB or HF-BCS include also the Landau-Zener effect [267, 268], which is crucial to properly describe dissipation when single-particle levels with different occupation numbers cross. 
TDDFT calculations that include pairing correlations \cite{Blocki76} account
partly for the Landau-Zener effect \cite{Landau32,Zener32}. Therefore, in HFB or HF-BCS calculations two-body dissipation is partially considered. 
In the TDDFT, scission occurs naturally at some time as a result of the competition between nuclear and Coulomb forces. 
Because the fragments can experience a rapid change in shape at scission, the TDDFT is better suited to describe scission dynamics than the adiabatic approach previously described. 

\begin{figure} [h]  
\begin{center}
\includegraphics[width=0.5\textwidth]{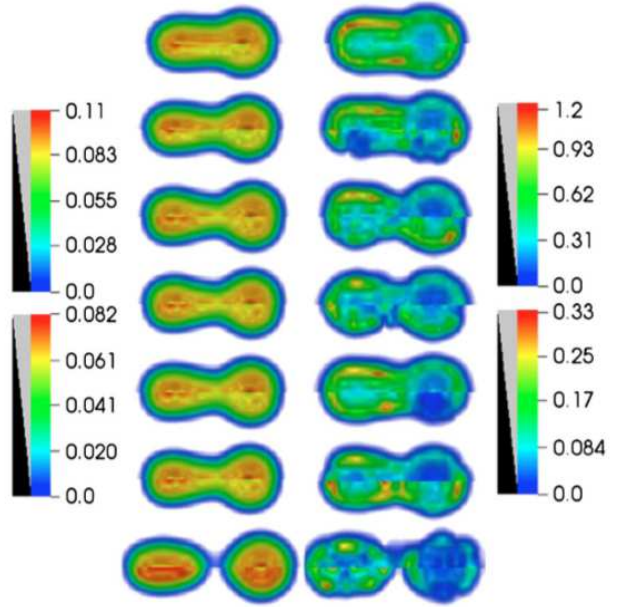}
\caption{(Color online)
The left column shows the neutron (proton) densities $\rho _n,_p(\bm{r})$ (per fm$^{3}$) in
the top (bottom) half of each frame. In the right column the
pairing field $\Delta _{n,p}(\bm{r})$ (in MeV) for the neutron (proton) systems are displayed in the
top (bottom) of each frame. The time difference between frames
is $\Delta t$ = 1600 fm/$c$. The scale of the values is given by the color bars on the left (right) for
densities (pairing gaps), and with upper (lower) ones for neutrons (protons).
Figure taken from \protect\cite{Bulgac16}.
}
\label{Bulgac}
\end{center}
\end{figure}

In most calculations based on the TDDFT it is required that the starting point is located well beyond the outer barrier (e.g.~\cite{Goddard15,Scamps15,Tanimura15}) or that it has an additional initial collective energy or boost (e.g.~\cite{Tanimura15,Goddard16}) for the system to evolve to scission. 
The appearance of such a dynamical threshold was recently attributed by Bulgac et al.~\cite{Bulgac16} to restrictions for transitions of a 
Cooper-pair state with opposite angular momenta into other Cooper-pair states with
different angular momenta. This prevents the system to adopt the most strongly bound states. 
Bulgac et al.~have been the first to perform a full TDHFB calculation of induced fission, which they refer to as time-dependent superfluid local-density approximation TDSLDA. 
Solving the TDHFB equation is substantially more involved numerically than solving the TDHF equation. 
In \cite{Bulgac16}, the evolution from a point slightly beyond the barrier up to scission was achieved by allowing transitions between magnetic sub-states by means of a complex pairing field that varies in time and in space during the evolution. 
The evolution of the proton and neutron densities and of the corresponding pairing fields for the fission of $^{240}$Pu at an exciation energy of 8.05 MeV are shown in figure \ref{Bulgac}. 

More recently, Tanimura et al.~\cite{Tanimura17} proposed a novel method to describe quantum fluctuations and spontaneous symmetry breaking together with the possibility to obtain fully microscopically fragment mass and TKE distributions in spontaneous fission.
The method consists of simulating quantum and thermal effects by a sampling of initial conditions, followed by quasiclassical evolutions with a TDHF + BCS approach.
A significant increase of the fluctuations in the collective degrees of freedom by the sampling of initial conditions was observed.
Moreover, with the exploration of the initial phase space (even without inclusion of pairing effects) no dynamical threshold was observed.
In total, few hundred trajectories starting from different initial densities were propagated in time. This sampling of trajectories was possible because TDHF+BCS is much more affordable numerically than TDHFB \cite{Schunck16}. 
%{\bf{May be, we should discuss this also with Bulgac. Of course, we should review and not judge, but I am confused about the claims of Tanimura et al.}}

%{\bf{The following section is not present any more in your DOC file (intentionally?):}}
%As explained in \cite{Schunck16}, TDDFT can be seen as the quantum-mechanical analog of the Langevin equation described in next section. 
%Unfortunately, because of the huge demand of computing power, calculations based on the TDDFT can only simulate single fission events. 
%However, predicting fission-fragment distributions in this framework would probably require sampling thousands of such events, which is currently inapplicable. 
%Therefore, TDDFT provides only most probable fission-fragment properties such as kinetic and excitation energies. 
%{\bf{For a realistic distribution, also different initial conditions and stochastic effects need to included in a realistic way: Monte-Carlo TDDFT?, otherwise all calculations give the same result. $<$KHS: This is the inclusion of additional, may be heuristic, fluctuations in the dynamical calculation with a Monte-Carlo technique. I do not remember in the moment, where this was mentioned. Probably one has to move this sentence on another place and give a reference.$>$}} 

\begin{table*}[h] 
\begin{center}
\small{
\caption{Synthetic description of the main microscopic approaches that have been applied to fission.}
\label{FISSION-Tab1}
%\begin{tabular}{|p{2.7cm}|p{2.8cm}|p{2.7cm}|p{2.7cm}|} \hline \hline 

\begin{tabular}{@{}m{2.4cm}p{5.2cm}p{3.2cm}p{1.8cm}@{}} \hline \hline 
 Theory & Main assumptions & Quantities and systems described & Selected references \\ \hline \hline
 \multirow{3}{2.7cm} {~\\~\\~\\~\\~\\
 Density-Functional Theory with Energy-Density Functionals based on Skyrme or Gogny effective potential.}  
 & Description of the potential-energy surface with few selected collective degrees of freedom.
  
 Tunnel transmission coefficient obtained with the WKB approximation.
 
 ~
 & Spontaneous-fission half-lives of even-even fissioning nuclei. 
 
 ~ 
 & \cite{Sadhukhan14},\cite{Staszczak09}, \cite{Staszczak13}
 
 ~ \\
 & Description of the potential-energy surface with few (presently two) selected collective degrees of freedom.
 
   The scission configuration must be explicitly defined.
   
   Dynamical evolution from the ground state to scission obtained with TDGCM-GOA assuming adiabaticity, i.e. no coupling between collective and intrinsic degrees of freedom.
   
 ~   
 & Fission-fragment mass and charge distributions of few even-even fissioning nuclei, mainly
   $^{226}$Th, $^{236,238}$U, $^{240}$Pu, and $^{252}$Cf at low energies.
 & \cite{Goutte05},\cite{Younes12}, \cite{Regnier16},\cite{Tao17},\cite{Zdeb17} \\
 & Dynamical evolution from a point beyond the fission barrier with 
TDHF, TDHF-BCS or TDHFB. This implies:

-Semi-classical evolution of one-body collective degrees of freedom. 
No tunneling.
 
-One-body dissipation is included. TDHFB and TDHF-BCS include also partly the
Landau-Zener effect.
   
 & Most probable kinetic and excitation energies of the fission fragments for a few even-even
   fissioning nuclei, mainly $^{240}$Pu at low energy and $^{258,264}$Fm, spontaneous fission. 
   Only recently, fairly realistic kinetic-energy and fragment-mass distributions have been obtained
for $^{258}$Fm.
 & \cite{Simenel14},\cite{Goddard15},\cite{Scamps15}, \cite{Tanimura15},\cite{Bulgac16},\cite{Goddard16}, \cite{Tanimura17} \\ 
                 
 \hline

\end{tabular}
}
\end{center}
\small{Note:
All approaches are based on the Density-Functional Theory, as described in \cite{Schunck16}. The second column lists
the main assumptions underlying the different approaches, the third column gives the fission
quantities and fissioning systems that have been computed, and the last column gives some selected
references for each approach.}
\end{table*}

\clearpage

\subsubsection{Selected results}

In the following, we will discuss some selected results obtained with the three approaches listed in Table \ref{FISSION-Tab1}. 

Staszczak et al. \cite{Staszczak09} studied the multi-modal spontaneous fission of isotopes from californium to hassium with the DFT solving the HF-BCS equation in two-dimensional collective spaces with the Skyrme effective interaction.
They calculated a two-dimensional PES involving the quadrupole and octupole moments and another two-dimensional PES involving the quadrupole and hexadecupole degrees of freedom. 
Observed fission characteristics in this region were traced back to topologies in the multidimensional collective space by searching for the optimum collective trajectory. 
The authors predicted tri-modal fission for several rutherfordium, seaborgium, and hassium isotopes where the compact symmetric (consisting of two spherical fragments of the same mass), the elongated symmetric and the elongated asymmetric modes co-exist. 
The spontaneous-fission half-lives were calculated with the WKB approximation for the double-humped potential barrier, assuming a one-dimensional tunneling path along the elongation degree of freedom.

In a later work, Staszczak et al. \cite{Staszczak13} solved the HFB equation with the Skyrme effective interaction and computed the spontaneous-fission half-lives along the entire Fm isotopic chain reproducing within 1-2 orders of magnitude the experimental trend of the lifetime as a function of neutron number of even-even isotopes.
Geometrical degrees of freedom, typically multipole moments, are very important for a realistic description of fission. 
However, recent work showed that fission can also be very sensitive to non-geometric collective variables such as pairing correlations.
For example, Sadhukhan et al.~\cite{Sadhukhan14} demonstrated that the inclusion of pairing collective degrees of freedom has a huge impact on the spontaneous-fission half-lives of $^{264}$Fm and $^{240}$Pu.

\begin{figure} [h]  % top
\begin{center}
\includegraphics[width=0.45\textwidth]{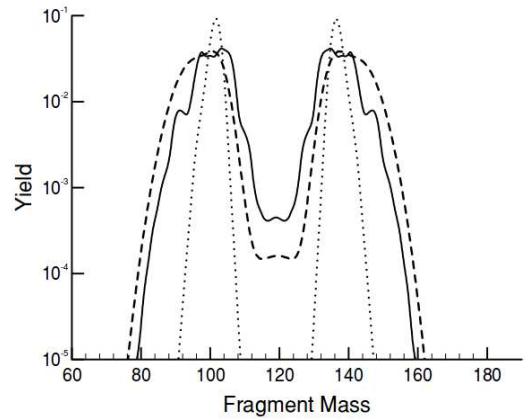}
\caption{ 
Comparison between the calculated one-dimensional mass distribution 
(dotted line), the mass distribution of $^{238}$U resulting from the
dynamic calculation (solid line) with the initial state located 2.4 MeV
above the barrier, and the Wahl evaluation (dashed line) \protect\cite{Wahl02}.
The calculation has been performed with the TDGCM-GOA approach \protect\cite{Goutte05}.
The figure is taken from Ref.~\protect\cite{Goutte05}.}
\label{FIG-Goutte}
\end{center}
\end{figure}

The first realistic calculations of fission-fragment kinetic-energy and mass distributions obtained with the second approach were performed by Goutte et al.~\cite{Goutte05}.
The authors solved the TDGCM equations using the GOA and the D1S Gogny force for $^{238}$U at excitation energies slightly above the fission barrier. 
The quadrupole and octupole moments were considered to be the two relevant collective parameters of the fissioning system. 
Figure \ref{FIG-Goutte} shows the  evaluation \cite{Wahl02} for the fission-fragment mass distribution of $^{238}$U in comparison with the results of a static one-dimensional calculation
of collective stationary vibrations along the sole
mass-asymmetry degree of freedom (left-right asymmetry) for nuclear configurations
just before scission
and with a full time-dependent calculation involving the two-dimensional PES. 
The maxima of the one-dimensional mass distribution lie close to the Wahl evaluation indicating that the most probable fragmentation is essentially due to the properties of the potential-energy surface at scission, i.e. to shell effects in the nascent fragments. 
However, it is clear that reproducing the width of the mass distribution requires including dynamical effects in the descent from saddle to scission of the two-dimensional PES.

The same approach as Goutte et al. was used by Younes and Gogny \cite{Younes12} to compute fission-fragment mass distributions of $^{236}$U and $^{240}$Pu at different neutron incident energies ranging from 0 to 5 MeV.
Instead of the commonly used multipole moments, Younes and Gogny used two collective variables similar to the ones used in macroscopic-microscopic approaches: the distance between the fragments and the mass asymmetry. 
As a result of this choice, the discontinuities in the PES were significantly reduced, particularly near scission. 
Figure 30 in \cite{Schunck16} shows that the resulting mass distributions of $^{240}$Pu agree fairly well with the data. 

\begin{figure} [h]  
\begin{center}
\includegraphics[width=0.45\textwidth]{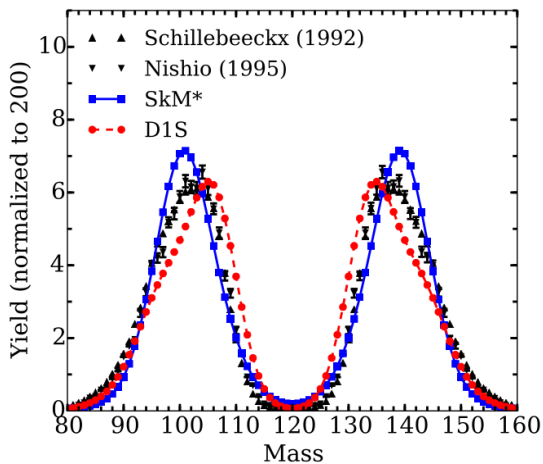}
\caption{(Color online)
Pre-neutron mass yields for $^{239}$Pu(n,f). 
The calculations by Regnier et al.~\protect\cite{Regnier16} obtained with the Skyrme (SKM*) and Gogny (D1S) effective nuclear potentials are compared with the experimental datasets by Schillebeeckx and Nishio. Figure taken from \protect\cite{Regnier16}.
}
\label{Regnier}
\end{center}
\end{figure}

Regnier et al.~\cite{Regnier16} performed an elaborate study of the influence of using different effective forces and different initial conditions on the pre-neutron mass and $Z$ distribution in the neutron-induced fission of $^{239}$Pu at $E^*$ = 1 MeV above the fission barrier with the adiabatic TDGCM-GOA method. 
They found that the qualitative features of the fission-fragment mass distributions are rather robust with respect to the effective force (see figure \ref{Regnier}) and/or the various ingredients of the model. The calculated mass distributions agree rather well with experimental data, although there remain ambiguities in deducing the widths of the dominating asymmetric components from the calculation due to difficulties in the modeling of the fluctuations.

Calculations for the fission of $^{252}$Cf were performed by Zdeb et al.~\cite{Zdeb17} using the TDGCM-GOA approximation with the aim
to examine, how the initial conditions affect the obtained fission-fragment mass yield, especially how
this quantity depends on the excitation energy and the parity of the initial state.
The dependence of the mass distribution from the initial conditions was found to be rather weak up to
initial excitation energies 4 MeV above the fission barrier. 
Possible modifications of the model are discussed in order to enhance the yields at large asymmetry, which are
considerably underestimated.

\begin{figure} [h]  
\begin{center}
\includegraphics[width=0.45\textwidth]{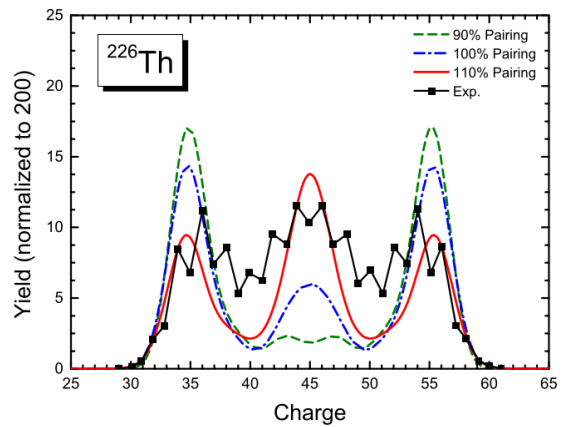}
\caption{(Color online)
Nuclear-charge yields for electromagnetic-induced
fission of $^{226}$Th. The results of calculations for three different values
of the pairing strength are compared to the data from \protect\cite{Schmidt00}.
Figure taken from \protect\cite{Tao17}.
}
\label{Tao-Th226}
\end{center}
\end{figure}

\begin{figure} [h]  
\begin{center}
\includegraphics[width=0.45\textwidth]{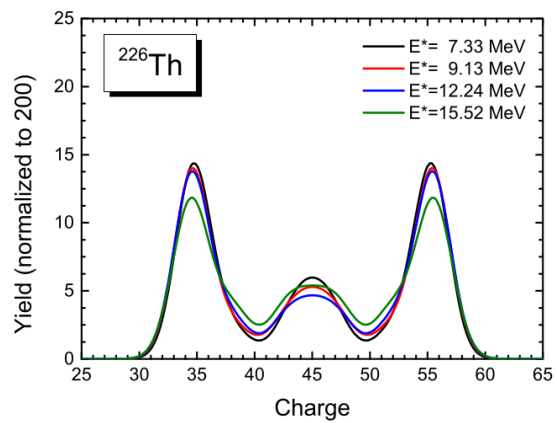}
\caption{(Color online)
Nuclear-charge distribution of fission fragments for different
excitation energies.
Figure taken from \protect\cite{Tao17}.
}
\label{Tao-Th226-Edependence}
\end{center}
\end{figure}

Recently, induced fission of $^{226}$Th was investigated with the TDGCM-GOA using a relativistic energy-density functional by Tao et al.~\cite{Tao17}. 
The overall topography of the calculated PES and the TKE values were found to be consistent with previous studies based on the Gogny density functional \cite{Dubray08}. The calculated charge distribution reproduces the main characteristics of the measured distribution, which presents a triple-humped structure with symmetric and asymmetric peaks. 
The influence of the strength of the pairing interaction and of the initial excitation energy were studied. 
Some prominent results are shown in figures \ref{Tao-Th226} and \ref{Tao-Th226-Edependence}. 
Note that, by its nature, the model cannot describe the odd-even staggering.
The dependence on the excitation energy appears to be rather weak, compared to the experimental systematics \cite{Tsang72,Specht74,Schmidt00}.

%\clearpage

As said above, the approaches based on the adiabatic approximation require a criterion to define the scission configuration. 
Moreover, close to scission an acceleration of the fragments may occur, inducing nonadiabatic effects. 
Thus, the adiabatic approximation is expected to break down in the latter stage of fission. 
These effects are crucial to properly describe properties of the fragments such as their kinetic and excitation energy. The TDDFT is more appropriate for calculations of these properties since there is no need to characterize scission, and one-body dissipation is included. 

Simenel and Umar \cite{Simenel14} studied dynamical effects in the vicinity of scission on the kinetic and excitation energies of the fragments in symmetric fission of $^{264}$Fm with time-dependent-Hartree-Fock calculations. 
They considered as starting point the position after the barrier where the properties of the emerging double-magic $^{132}$Sn fragments are well defined. 
Simenel and Umar showed that a significant fraction of the final excitation energy of the fragments is acquired right before scission and that it is at least partly stored in low-energy collective vibrational states of the fragments. 
However, Simenel and Umar neglected pairing correlations, which may be reasonable when studying $^{264}$Fm but not to investigate fission dynamics across the nuclear chart. 
Their approach was extended by including pairing correlations within the BCS approximation by Scamps et al.~\cite{Scamps15} and applied to investigate multi-modal fission of $^{258}$Fm. 
The resulting TKE of the fragments were compared with experimental data showing good agreement. 
Their result for the TKE of the compact symmetric mode was located near the main peak of the TKE distribution in agreement with previous interpretations of the experimental data. 
The measured lower TKE tail was mostly attributed to the elongated asymmetric mode, although these calculations predict that the elongated symmetric mode leads to the lowest values of the TKE where only a few events have been observed experimentally. 
They showed that the time associated with the non-adiabatic descent of the potential to scission for the compact symmetric mode is shorter than for the two other modes, reflecting that magic fragments are difficult to excite and deform and, thus, fission occurs faster as less dissipation is involved. 

Tanimura et al. \cite{Tanimura15} proposed a method to obtain the collective momentum and
mass associated to any collective observable from TDDFT. They applied this method to the fission of $^{258}$Fm and showed, how the use of different amounts of collective energy to induce a boost for overcoming the dynamical threshold can impact the results.
They also showed that in the non-adiabatic case the fragments stick together for a longer time than in the adiabatic case, indicating that the scission point derived in the non-adiabatic case occurs at a larger distance between the fragments. 

In the first ever application of TDHFB to fission, Bulgac et al.~\cite{Bulgac16} presented rather accurate calculations with the time-dependent superfluid local-density approximation (TDSLDA) of some specific average fission quantities (fragment excitation energies, kinetic energies, saddle-to-scission time) in thermal-neutron-induced fission of $^{239}$Pu. 
The threshold anomaly (the necessity to impose an initial boost in order to bring the system to fisssion that was observed in previous studies) is solved by the treatment of dynamical pairing.
Long fission times were obtained that were attributed to the excitation of a large number of collective degrees of freedom (confirming early qualitative results of N\"orenberg \cite{Noerenberg74}) and not to a particularly large viscosity, i.e. coupling between collective and intrinsic degrees of freedom.

\begin{figure} [h]  
\begin{center}
\includegraphics[width=0.5\textwidth]{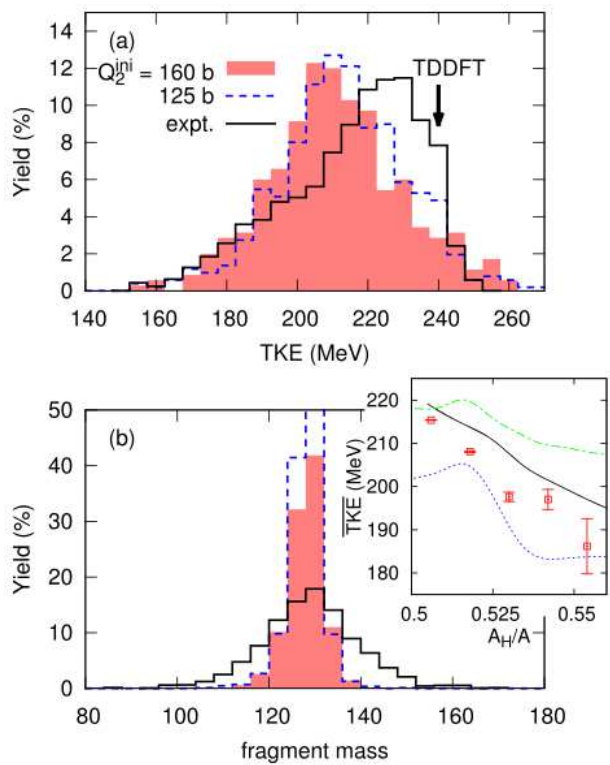}
\caption{(Color online)
TKE (a) and fragment mass (b) distributions of $^{258}$Fm obtained with two different initial conditions. The solid line represents the experimental data. In (a) the arrow indicates the mean TKE obtained in Ref.~\protect\cite{Scamps15}. In the inset, the correlation between the average TKE and the heaviest fragment mass (red squares) are shown in comparison with results of a scission-point model \protect\cite{Wilkins76,Balagna71} in dotted and dot-dashed lines. The $^{257}$Fm data are shown by a solid line.
Figure taken from \protect\cite{Tanimura17}.
}
\label{Tanimura}
\end{center}
\end{figure}

%With their method described above, to get the collective momentum and mass associated to any collective observable directly from TDDFT, Tanimura et al.~\cite{Tanimura17} studied the non-adiabatic effects and dissipation.
Tanimura et al.~\cite{Tanimura17} studied the fission-fragment mass and TKE distributions
of $^{258}$Fm spontaneous fission with TDHF-BCS. 
They simulate quantum and thermal effects by a sampling of initial conditions,
followed by the quasiclassical evolution of the collective motion with TDDFT.
As said above,
the authors observed that this exploration of the initial phase space 
(even without inclusion of dynamical pairing effects) also solves 
the dynamical-threshold anomaly.
The calculated TKE distribution is in rather good agreement with the experimental data, whereas the mass distribution is still too narrow, see figure \ref{Tanimura}. 
From their model they infer information on the fission process. 
For instance, they show that at scission the fragments are deformed and that the deformation relaxes as the two fragments move apart from each other after scission. 
In their calculation, the fragments are excited before reaching scission, which leads to a non-negligible probability for pre-scission neutron emission.

TDDFT appears to be the most ambitious self-consistent microscopic approach for the description of the fission process. 
Therefore, it is expected that it will gain importance and that it is going to play a leading role in fission research in the future. 
As described above, two implementations of this approach have been developed, either
using the BCS approximation (e.g. by Tanimura et al.~\cite{Tanimura17}) or
the Hartree-Fock-Bogoliubov approximation (by Bulgac et al.~\cite{Bulgac16}).
Apparently, some important results are not compatible. Table \ref{TDDFT} summarizes some prominent differences in the ingredients, in the application and in the results of the two approaches.
The scientific discussion that aims for solving this problem is in progress.
A crucial point is the approximate treatment of the pairing correlations within the
TDBCS approximation in ref.~\cite{Tanimura17}, which, according to ref.~\cite{Bulgac16},
violates the continuity equation.

 % \cite{Bulgac17}.

%\clearpage

\begin{table*}[t] 
\begin{center}
\caption{Comparison of the different TDDFT approaches}
\label{TDDFT}
%\begin{tabular}{|p{3cm}|p{5cm}|p{5cm}  |c|} \hline \hline 
\begin{tabular} {@{}m{3.4cm}p{5.5cm}p{5.5cm}@{}}  
           Approach  & TDHF-BCS, Ref.~\cite{Tanimura17} &  TDSLDA, Ref.~\cite{Bulgac16} \\  \hline \hline
  Pairing model      &   BCS approximation             &  Bogoliubov theory    \\  \hline
  Starting point of the    &   Between saddle and scission   &  Near outer saddle    \\ 
  dynamic calculation  &   with thermal fluctuations             &                   \\  \hline
  Main influence on    & Dissipation                     &  Excitations of \\
  fission time         & (intrinsic excitations)         &  collective shape          \\  
                       &                                 &  and pairing modes
 \\ \hline
  Typical fission time &  1500 fm/c                   &       10000 fm/c         \\ \hline
  Disappearance of     & Fluctuations of the         &  Dynamical pairing     \\   
  threshold anomaly    & fission path                &        \\  
  mainly attributed to        &          &        \\  \hline
\end{tabular}
\end{center}
\small{
}
\end{table*}

As we have seen, at present, the importance of microscopic self-consistent models for the description of nuclear fission lies in the qualitative understanding of several fundamental aspects, while the achievements in completeness and accuracy are in a vivid process of development. 
Only a limited variety of fission observables (mostly spontaneous-fission half-lives, fission-fragment mass or nuclear-charge distributions and fragment kinetic energies) for a few fissioning nuclei are treated up to now. 
The practical restrictions due to the very high demand on computing power are still severe. 
Even more importantly, there are several formal or conceptual issues that are not fully solved, as already mentioned. They will be further discussed in section \ref{5}. 
%To quote some of them: (i) separation of relevant and non-relevant degrees of freedom, (ii) fragments recognition and definition of the neck in a quantum liquid, (iii) quantum tunneling in many-body evolving system, and (iv) symmetry breaking and its restauration.
In the long term, it is hoped that many of the approximations described above will not be needed any more and microscopic self-consistent models will be particularly strong in providing reliable predictions for exotic systems that are not accessible to experimental studies.

\subsection{Stochastic approaches} \label{4-2}

\subsubsection{Basic considerations}

Early dynamical studies of the nuclear-fission process have 
been performed 
%on the basis of the liquid-drop model 
%in the framework of classical statistical mechanics, for example
by Nix \cite{Nix69} with a non-viscous irrotational liquid-drop model and
by Davies et al. \cite{Davies76} with the inclusion of two-body dissipation. 
Later,
%with static, dynamical and statistical methods to the Hamiltonian
%of the fissioning system and 
Adeev et al.~\cite{Adeev88} studied the influence of one-body and
two-body dissipation on the fission dynamics
with a transport equation of the Fokker-Planck type. Also in this case,
like in the microscopic self-consistent approaches, 
only a limited part of the large number of degrees of freedom
was explicitly treated. In Ref.~\cite{Adeev88},
the evolution of the probability-density
distribution in a space defined by the restricted number of degrees
of freedom that are considered to be relevant
is described under the influence of a driving force
and friction, including the associated statistical fluctuations. 
Driving force and friction represent the interactions 
with the other degrees of freedom, which are treated as a heat bath and, thus, are
not explicitly considered.
However, the solution of the Fokker-Planck equation was limited
to simple cases or subject to strong approximations. 
Abe et al.~\cite{Abe86} replaced the Fokker-Planck equation by
the equivalent Langevin equations that can be solved numerically 
in the application to fission.
Monte-Carlo sampling of individual trajectories in the space of
the relevant degrees of freedom proved to be a more practicable 
way for obtaining more accurate solutions in complex cases. 
However, this method requires considerable computing resources. 

\subsubsection{Stochastic transport equations}

The Langevin equations in their discretized form for the evolution 
of the system in the time interval $\Delta t$  between the time step $i$ and $i+1$, read

\begin{equation}
  q_{i+1} = q_i + {{p_i} \over {m} } \Delta t
\label{Eq-Langevin-1}  
\end{equation}
and
\begin{equation}
 p_{i+1} = p_i - T {{dS}  \over {dq}} \Delta t - \beta p_i \Delta t + 
  \sqrt{\beta m T \Delta t} \cdot \Gamma .
\label{Eq-Langevin-2}  
\end{equation}

$q$ is the coordinate in the space of the relevant degrees of freedom,
$p$ the corresponding momentum. $m$ and $\beta$ are the mass parameter and the 
dissipation coefficient, respectively. 
$- T {{dS}  \over {dq}}$ is the driving force, $- \beta p_i$ the friction force, 
and  $\sqrt{\beta m T \Delta t} \cdot \Gamma $ the fluctuating term that expresses
the stochastic transfer of thermal energy from the heat bath to the collective coordinates. 
$T$ is the temperature and $S$ is the entropy. Both are related
to the local level density\footnote{Strictly speaking, 
the degeneracy of the magnetic sub-states should be considered.}
$\rho$, which has to be provided by an appropriate model,
see for example \cite{Senkov16}.

\begin{equation}
T = ( {{d \mbox{ln}(\rho)} \over {dE} })^{-1}
\label{T-of-rho}
\end{equation}
and
\begin{equation}
S = \mbox{ln}(\rho) .
\label{S-of-rho}
\end{equation}

In most practical cases, the stochastic variable $\Gamma$ that defines the 
fluctuating force is linked to the dissipation strength by the 
fluctuation-dissipation theorem \cite{Einstein05}.

Equations (\ref{Eq-Langevin-1}) and (\ref{Eq-Langevin-2}) are valid, if $m$ and
$\beta$ do not depend on the coordinate $q$ and the direction of motion.
In the general case, $m$ and $\beta$ are tensors, and the Langevin
equations must be adapted.
At low energies, pairing correlations and shell effects should be included
not only in the potential-energy surface, but also in the
level density \cite{Ward17}, in the friction tensor \cite{Mirea14} and in the mass tensor \cite{Mirea10}. However, this is rarely done.
For more detailed information on the application of the Langevin equations 
to fission and to other nuclear reactions we refer to the review article
of Fr\"obrich and Gontchar in Ref.~\cite{Froebrich98}.
 
Since a nucleus is an isolated system with fixed total energy and fixed
particle number, equations (\ref{Eq-Langevin-1}) and
(\ref{Eq-Langevin-2}) must be formulated in the fully micro-canonical version
(specified as option 1 in Table \ref{FISSION-Tab2}).
This entails that temperature $T$ and entropy $S$ are given by the level density with
equations (\ref{T-of-rho}) and (\ref{S-of-rho}), respectively,
at an energy equal to the 
total energy of the system minus the local potential and the actual collective energy.
This way, the potential energy does not explicitly appear in equation (\ref{Eq-Langevin-2}),
but its influence enters via temperature $T$ and entropy $S$.

This is important in applications to low-energy fission, where 
approximations that are often applied (specified as options 2, 3 and 4 in Table \ref{FISSION-Tab2}) 
only badly represent the statistical
properties of the nucleus. Also the influence of pairing correlations and shell effects
on the binding energy $and$ the level density should be properly considered, in particular
at low excitation energies.
This is not so critical at higher excitation
energies, where for example the use of 
%a heat bath of constant temperature and 
%the Fermi-gas level density or even 
Boltzmann statistics may be a suitable approximation. 
These aspects have been stressed in several 
places, for example by Fr\"obrich in Ref.~\cite{Froebrich07}. 
He also stresses that the driving force is not given by the
derivative of the potential, but by the derivative of the entropy
times the temperature that expresses the influence of the environment
on the selected degrees of freedom according to the laws of
statistical mechanics, see equation (\ref{Eq-Langevin-2}). Due to the complexity of the nuclear level density,
this can lead to very different results. 

If very strong friction is assumed, the motion becomes over-damped, and
the influence of the mass tends to vanish. This case
%, which is probably not realized in fission, 
is represented by the Smoluchowski equation \cite{Froebrich98} that requires 
less computational expense. As argued in Ref.~\cite{Randrup11a}, even less demanding
in computing resources is the replacement of the kinematic
equations (\ref{Eq-Langevin-1}) and (\ref{Eq-Langevin-2}) by a 
random-walk approach using Metropolis sampling \cite{Metropolis53}.
All these different approaches are presently in use.

\begin{table}[t] 
\begin{center}
\caption{Stochastic approaches to nuclear fission}
\label{FISSION-Tab2}
%\begin{tabular}{|p{3.8cm}|p{4.5cm}|p{3.5cm}|c|} \hline \hline 
%   Name     &  approximations  & associated\newline observables   \\ \hline \hline
% Langevin equations,\newline microcanonical  &  classical   &  selected collective\newline variables     \\  \hline
% Langevin equations,\newline not fully\newline microcanonical $^*$  &  classical +\newline simplified driving force\newline or state density &  selected collective\newline variables     \\  \hline
%   Smoluchowski\newline equation  & classical +\newline overdamped motion     & selected collective\newline variables  \\ \hline
%  Random walk & classical +\newline overdamped motion +\newline Metropolis sampling  &  selected\newline collective variables \\ \hline
%\end{tabular}
\begin{tabular}{|p{3.2cm}|p{3.6cm}|c|} \hline \hline 
   Name     &  Approximations     \\ \hline \hline
 Langevin equations,\newline micro-canonical  &  Classical dynamics $^*$   \\  \hline
 Langevin equations,\newline not fully\newline micro-canonical $^{**}$  &  Classical dynamics $^*$ +\newline simplified driving force\newline or state density \\  \hline
   Smoluchowski\newline equation  & Classical dynamics $^*$ +\newline over-damped motion    \\ \hline
  Random walk & Classical dynamics $^*$ +\newline over-damped motion +\newline Metropolis sampling  \\ \hline
\end{tabular}
\end{center}
\small{
$^*$) Certain quantum-mechanical features can effectively be considered in the classical 
Langevin equations, for example shell effects in the potential energy, 
contribution of the zero-point motion to fluctuation phenomena, etc.
\\
$^{**}$) Different kind of approximations, for example coupling to a heat bath of 
constant temperature, Boltzmann statistics etc.}
\end{table}

\subsubsection{Application to fission}

In practice, the application of stochastic classical approaches is performed in two
steps, like in the case of the first two self-consistent microscopic approaches listed in table \ref{FISSION-Tab1}, where a small number of collective degrees
of freedom is explicitly considered. In a first step,
the potential energy is computed on a grid in the space determined by the 
relevant degrees of freedom, usually by the macroscopic-microscopic model. In most
cases, the relevant degrees of freedom are the coordinates of a suitable shape
parametrization. Eventually, the potential energy is minimized individually on each
grid point with respect to additional shape parameters.
Also the dissipation tensor and the mass tensor must be defined,
for example on the
basis of one-body and two-body dissipation with phenomenological adjustments and the
hydrodynamical inertia with the Werner-Wheeler approximation for the velocity field
\cite{Davies76}, respectively. 
With these ingredients, Monte-Carlo sampling of the
fission trajectories with one of the stochastic approaches listed in Table 
\ref{FISSION-Tab2} is performed.  

\subsubsection{Selected results}

To our knowledge, stochastic approaches are being applied to low-energy fission with the
inclusion of shell effects since 2002. Ichikawa et al.~\cite{Ichikawa02} 
studied the fission of $^{270}$Sg with three-dimensional Langevin calculations
at an excitation energy of 10 MeV above the ground state. By considering the energy dependence of the shell
effect, the calculation was essentially microcanonical.
The shell effects were obtained with the two-center shell model \cite{Scharnweber71}.
The mass tensor was calculated using the hydrodynamical model with the
Werner-Wheeler approximation \cite{Krappe79} for the velocity field, and the 
wall-and-window one-body dissipation \cite{Nix77} was adopted for the dissipation
tensor. The distance of the fragment centers, the quadrupole deformation, assumed to be common 
to both fragments, and the mass asymmetry were chosen as shape parameters.
The measured mass distribution was well reproduced, while the total kinetic energy
(TKE) was overestimated.
The authors stressed the strong influence of the dynamics on the mass distribution.
This model has been applied in Ref.~\cite{Asano04} to study the multi-modal
fission of $^{256,258,264}$Fm. In Ref.~\cite{Asano06}, the influence of the
dissipation tensor on the fission trajectory was demonstrated.  

\begin{figure} [h]  % top
\begin{center}
\includegraphics[width=0.46\textwidth]{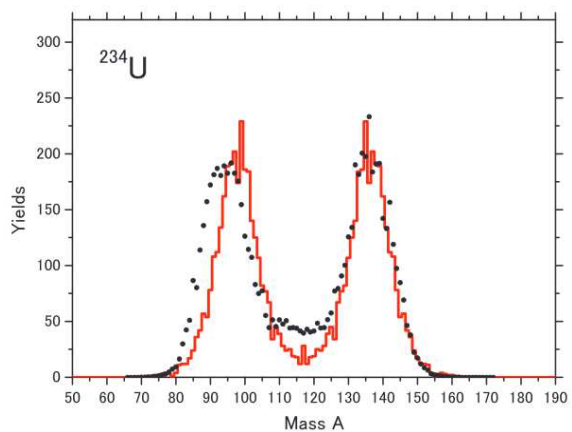}
\caption{(Color online) 
Mass distribution of fission fragments of $^{234}$U at
$E^\ast$ = 20 MeV. Calculation \protect\cite{Aritomo13} and evaluated 
data \protect\cite{Katakura12} are denoted
by histogram and circles, respectively.
In contrast to the evaluated data, the calculation does not
include the influence of multi-chance fission and 
prompt-neutron emission from the fragments.
The figure is taken from Ref.~\protect\cite{Aritomo13}.}
\label{FIG-Aritomo}
\end{center}
\end{figure}

Aritomo et al.~\cite{Aritomo13} succeeded to reproduce fairly well the measured fission-fragment
mass distributions and the TKE distributions of $^{234}$U, $^{236}$U, and $^{240}$Pu 
before prompt-neutron emission at an excitation energy of 20 MeV
with their stochastic approach, 
similar to the one applied before in refs.~\cite{Ichikawa02,Asano04,Asano06}
and using the same shape parametrization. Figure \ref{FIG-Aritomo} shows the comparison
of the calculated and evaluated mass distributions for $^{234}$U. Still some differences can be
observed:
The general shift to lower masses in the experiment can be attributed to the neglect of
prompt-neutron emission in the calculation. The larger peak-to valley ratio cannot be
explained by the influence of multi-chance fission, which is also disregarded in 
the calculation, because it would act in the opposite way. 
They failed to reproduce the transition
to single-humped mass distribution towards $^{226}$Th and $^{222}$Th and attributed this
to an insufficiently detailed shape parametrization. In particular, they concluded that 
the deformation parameters of the two nascent fragments should be chosen independently.
In Ref.~\cite{Aritomo14}, Aritomo et al. introduced a new shape parametrization, but sticked
to 3 dimensions. They tested the model against the mass-TKE distribution for the fission 
of $^{236}$U at an excitation energy of 20 MeV. The influence of pairing correlations
that may be assumed to be weak at this energy is neglected.
In Ref.~\cite{Aritomo15}, Aritomo et al. considered the $N$/$Z$ degree of freedom in their model,
which enabled them to calculate independent yields, that means fission-fragment yields specified in $A$ and $Z$.

The comprehensive data base of 5-dimensional potential-energy landscapes, 
calculated by M\"oller et al.~\cite{Moeller01} with the 
macroscopic-microscopic approach, was used by Randrup et al.
as a basis for wide-spread stochastic calculations of pre-neutron fission-fragment mass
distributions \cite{Randrup11,Randrup11a,Moeller12,Randrup13}.
$Z$ distributions were deduced with the unchanged-charge-density (UCD) assumption, which
means that the $N$/$Z$ of the fragments is equal to that of the fissioning nucleus.
Due to the relatively large number of 5 shape parameters
(overall elongation, constriction, reflection asymmetry, and deformations of the two
individual pre-fragments), some simplifications and approximations
in the dynamical treatment were applied in order to keep
the computational needs on an affordable level. 
A random-walk approach using the Metropolis sampling was applied, assuming over-damped 
motion, and the driving force was taken
as the derivative of the potential. These approximations prevent obtaining realistic results 
on the energetics of the fission process (for example kinetic and excitation energies 
of the fragments, prompt-neutron multiplicities, etc.).
The measured mass distributions of a large
number of systems, reaching from $^{180}$Hg to $^{240}$Pu are fairly well reproduced. 
These results are already depicted and were duly acknowledged in ref.~\cite{Andreyev18}.
The importance of a sufficiently detailed shape parametrization for the calculation
of fission-fragment mass distributions, in particular the freedom that the nascent 
fragments can take individual deformations, is demonstrated. 
Recently, a method to extend this approach to six dimensions by including the $N/Z$
(charge polarization) degree of freedom has been proposed \cite{Moeller15}.
In the most recent version of this approach \cite{ Ward17}, the Metropolis
walk is not driven by the potential energy any more, but by the number of available
levels. This means that the driving force was calculated microcanonically. 
%see Table \ref{FISSION-Tab2}. 
By using microscopic level densities,
it was possible to study the evolution of the fission-fragment mass distribution 
as a function of initial excitation energy.

\begin{figure} [h]  % top
\begin{center}
\includegraphics[width=0.42\textwidth]{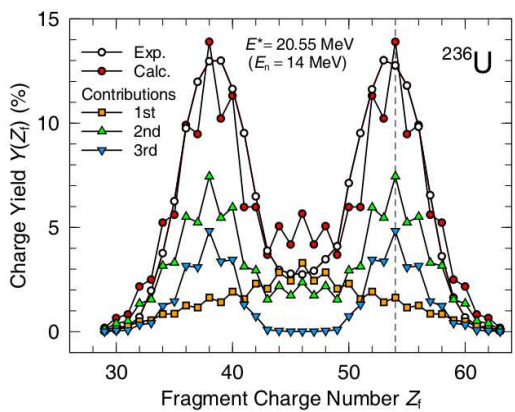}
\caption{(Color online) 
Calculated fission-fragment nuclear-charge distributions at
14 MeV incident-neutron energy, corresponding to 20.55 MeV
excitation after neutron absorption. The contributions from
first-, second-, and third-chance fission to the charge distribution are shown and added
up to a total yield. The fractions of the fission events from the different chances 
were taken from Ref.~\protect\cite{Madland82}.
An experimental evaluation \protect\cite{Katakura12} is also shown.
The figure is taken from Ref.~\protect\cite{Moeller17}.}
\label{Moeller-Multi-Chance}
\end{center}
\end{figure}

In a recent study \cite{Moeller17} on the basis of the Brownian shape-motion model with its recent extensions,
the evolution of fission-fragment charge distributions
with neutron number for the compound-system sequence $^{234}$U, $^{236}$U, $^{238}$U, and $^{240}$U was studied and compared with experimental data.
The evolution of the location of the
peak charge yield from $Z$ = 54 in $^{234}$U towards $Z$ = 52 in heavier isotopes, seen in the experimental data was reproduced fairly well.
Moreover, it was shown that it is necessary to take multi-chance fission into account to describe the yields already at an excitation energy of 20 MeV. The contributions of the different chances to the fission-fragment $Z$ distribution of the system $^{235}$U(n,f) at E$_n$ = 14 MeV and a comparison of the total distribution with empirical data is shown in figure \ref{Moeller-Multi-Chance}. The absence of an odd-even effect in the
empirical data can be understood by the limited experimental $Z$ resolution.

Ishizuka et al.~\cite{Ishizuka17} developed a four-dimensional Langevin model, which can treat the deformation of each fragment independently, and applied it to low-energy fission of $^{236}$U. 
The transport coefficients were calculated by macroscopic prescriptions.
Note that the deformation parameters of the two complementary fragments are set equal in most versions of the stochastic fission approach, except in the Brownian shape-motion model \cite{Randrup11}.
The choice of the collective variables, in particular the independent variation of the deformation of the two complementary fragments allowed them to perform a multiparametric correlation analysis among the three key fission observables, mass, TKE, and prompt-neutron multiplicity.
A  strong correlation is found between the mass-dependent deformation of fragments at the scission point and the sawtooth structure of prompt neutron multiplicity including their dependence on excitation energy.
In detail, the mass-dependent shapes of the complementary nascent fragments at scission develop in opposite directions as a function of the mass asymmetry.
The underlying feature, the saw-tooth shape of the mass dependent prompt-neutron multiplicities, is known from experiment since long (e.g. Ref.~\cite{Terrell62}) and has been investigated by different theoretical approaches, for example by the scission-point model of Wilkins et al.~~\cite{Wilkins76} or by the random-neck-rupture model of Brosa et al. \cite{Brosa90}. 
This demonstrates the importance of considering the shape parameters of the two fragments independently. 
 
The mass distributions and the total
kinetic energy of fission fragments for a series of actinides and Fm isotopes at various excitation energies 
were studied within the three-dimensional Langevin approach by Usang et al.~\cite{Usang17}.
The experimental results are fairly well reproduced. In particular, the obtained variation of the TKE in neutron-induced fission of $^{231}$Pa, $^{235}$U and $^{239}$Pu follows the experimental trend up to incident-neutron energies of 45 MeV, when microscopic transport coefficients are used, although the calculation was restricted to first-chance fission.
This agreement seems to be in conflict with the results from other models (for example the Brownian shape-motion model \cite{Moeller17} or the semi-empirical GEF model \cite{Schmidt16}), where a strong influence of multi-chance fission was observed for fission-fragment distributions \cite{Schmidt16,Moeller17} and kinetic energies \cite{Schmidt16} at excitation energies above the threshold for second-chance fission.  

Very recently, Sierk performed calculations with the Langevin approach on a five-dimensional
potential-energy surface \cite{Sierk17}. The approach is not fully microcanonical, since
the driving force is determined by the derivative of the potential.
Because dissipation is
explicitly considered by the surface-plus-window dissipation model, not only 
fission-fragment mass distributions, but also total
kinetic energies are obtained for neutron-induced fission of $^{233,235}$U and $^{239}$Pu
from thermal energy to a few MeV, as well as for spontaneous fission of $^{240}$Pu and $^{252}$Cf.
Measured pre-neutron mass distributions are considerably well reproduced, while the 
kinetic energies of the fragments close to symmetry are overestimated. 
The post-neutron fission-fragment mass distributions acquire the measured widths by 
assuming a random neck rupture and after a few corrections of the model. 
Ref.~\cite{Sierk17} offers a detailed discussion of the model assumptions,
approximations and uncertainties, and of possible reasons for the observed 
deviations from the experimental data.

At present, systematic calculations of fission-fragment mass distributions for
a large number of different fissioning systems have only been 
performed with the simplified dynamics of the Metropolis sampling \cite{Moeller15a}. Systematic 
calculations of fission-fragment mass distributions, total kinetic energies, spectra and
mass-dependent multiplicities of prompt neutrons, and other fission observables
for many systems as a function of initial excitation energy
with a fully micro-canonical Langevin approach appear to be possible, but they have
not yet been reported. 
Although this stochastic approach misses
the full inclusion of quantum-mechanical features, 
such results would be very interesting, because those calculations are still out of reach for microscopic, self-consistent models.

\subsection{General Fission Model GEF} \label{4-3}

\subsubsection{Basic considerations} \label{4-3-1}

Very recently, an approach to fission that exploits some mostly qualitative 
general concepts of mathematics and
physics\footnote{The term 'general' is used here to denote a concept or law of general validity
in contrast to ab-initio modeling, see for example Ref.~\cite{Reuter04}. Models that are based on general concepts often need empirical information for performing quantitative calculations, while ab-initio modeling strives for quantitative results without relying on empirical or fitted parameters.
As a critical remark about a sharp separation between these models, we would like to cite Niels Bohr, who stated that 'it appeared difficult to define what
one should understand by “first principles” in a world of
knowledge where the starting point is empirical evidence' \cite{Bortignon16}.}, combined with empirical information, has been successfully applied to develop
a model, named GEF (GEneral description of Fission observables) \cite{GEF}. 
This model covers the majority of the fission quantities and reproduces the
measured observables, especially in the range $A \ge 230$, where most of the experimental
data were obtained, 
with a remarkable accuracy that makes it suitable for technical
applications. To illustrate this with a few examples, we mention that
fission yields from GEF are being used to fill in unmeasured fission-yield
data in the JEFF-3.3 evaluation \cite{JEFF33}. 
Moreover, the GEF model was able to reveal that the ENDF
evaluation for the mass yields of the system $^{237}$Np(n$_{th}$,f) was based on erroneous data,
because the target was contaminated with a contribution of another fissile nucleus,
probably $^{239}$Pu \cite{Schmidt16}. 
The accuracy of GEF calculations can be inferred from the systematic comparison of 
fission data of different kind with GEF calculations shown in Ref.~\cite{Schmidt16} and
some recent publications,  
where they are compared to measured independent \cite{Gupta17,Pellereau17} and cumulative \cite{Privas16}
fission yields, and prompt-gamma spectra \cite{Rose17}.
The GEF model was also applied to study fragment-mass and excitation-energy-dependent
prompt-neutron yields \cite{Tudora16}, and for establishing
uncertainty propagation of fission product yields and decay heat \cite{Leray17}.
It was also used to estimate corrected fission-fragment yields, strongly diverging from the ENDF evaluation, in the interest to calculate
realistic anti-neutrino spectra for $^{235}$U(n,f) \cite{Sonzogni16}.
%As a consequence, a re-measurement of this system was
%set on the Nuclear Data High Priority Request List of the Nuclear-Energy Agency of the OECD.
Large deviations for fission-fragment yields between experimental data and GEF calculations have been reported in Refs.~\cite{Khan16,Wilson17}. In the first case, the authors admitted an error in the input options of their GEF calculation \cite{Khan18}; in the second case, the discrepancies may need clarification, if possible with an independent experimental approach.

In the domain of fundamental physics, GEF was used to interpret measured mass distributions after heavy-ion induced reactions $^{12}$C+$^{235}$U, $^{34}$S+$^{208}$Pb, $^{36}$S+$^{206}$Pb, and $^{36}$S+$^{208}$Pb
in terms of multi-chance fusion-fission and quasi-fission \cite{Khuyagbaatar15}.

GEF also proved to be a useful tool for the detailed planning of experimental methods
\cite{Boutoux13,Adili15}.

The calculation of many fission quantities that can be obtained from GEF still pose severe problems
to stochastic and microscopic models. A detailed documentation of 
the GEF model code, its underlying ideas and a 
presentation of a large variety of results can be found in refs.~\cite{JEFF24,JEFF24-2,Schmidt16}. 
GEF makes use of several long-standing mostly
qualitative ideas that were already able to explain many systematic trends and regularities 
in several fission observables and combines them with some innovative conceptions.
It owes its accuracy and a considerable predictive 
power\footnote{The predictive power of GEF is understood in the following sense, which is common to most, if not all, theories in nuclear physics: Once, the set of
model parameters is determined by a fit to a comprehensive set of measured observables for
a large variety of systems, the GEF model with this set of parameters is able to predict the
fission quantities of other systems with an accuracy comparable with the uncertainties of 
the experimental data used for the parameter fit, if these systems are not too far from the nuclei that were used
for determining the model parameters. This is corroborated by the good agreement, which was found in numerous cases,
between new data and GEF predictions that were made before the data became available, see
section \ref{4-4}.}
to the development of additional powerful ideas and the consideration of important 
experimental findings that were not fully understood before
or obtained only recently. In the following, we will describe the most important ideas 
and their successful application in the GEF model. 

\subsubsection{Topographic theorem: Accurate fission barriers} \label{4-3-2}

The modeling of the nuclear-fission process in stochastic and 
self-consistent models starts with the calculation of the
potential energy of the fissioning system in the space defined by the "relevant"
collective degrees of freedom. Besides the ground state of the nucleus, the saddle
point that defines the fission threshold (the lowest energy where the system can 
proceed to fission without tunneling) is a prominent point in the potential-energy
landscape. 
However, in contrast to the nuclear binding energy in the ground state, the 
binding energy at the fission saddle is not directly measurable. 
Empirical information on the fission threshold has been derived from measured
energy-dependent fission cross sections and/or fission probabilities assuming
penetrability through a multi-humped barrier, approximated by one-dimensional 
inverted parabolas. One of the most comprehensive studies of this type was made in \cite{Bjornholm80}.
The resulting values of the barrier parameters depend on the details of the model analysis, 
for example on 
the level-density description and, in particular, on the properties of the first 
excited states above the barriers.  
Therefore, the empirical fission-barrier heights are considered to be subject to an
appreciable uncertainty, usually presumed to be in the order of 1 MeV,
see for example Ref.~\cite{Rodriguez14a}. 
Although the barrier height deduced from this method can not be identified with the saddle-point energy of multi-dimensional
potential-energy surfaces from microscopic or macroscopic-microscopic theories, a link can
be established by matching the energy-dependent fission probabilities in the different scenarios.
Fission barriers derived with the one-dimensional penetrability approach have a 
considerable practical importance, because they are very much used to estimate 
the energy-dependent fission probabilities and neutron-induced-fission cross sections.
To our knowledge, the penetrability through a multi-dimensional barrier has not yet been calculated.

In this section, we will derive a well-founded estimation of the uncertainty of
fission barriers in the one-dimensional penetrability approach 
and propose a procedure for predicting accurate fission-barrier values.
This is an important information, because it allows to better assess the 
quality of a theoretical model by its ability to reproduce the empirical values
of the fission barrier. As we will see, one may conclude on the ability of the 
model for realistic estimations of the full potential-energy surface of the fissioning
system from this result.  

\begin{figure} [h]  % top
\begin{center}
\includegraphics[width=0.42\textwidth]{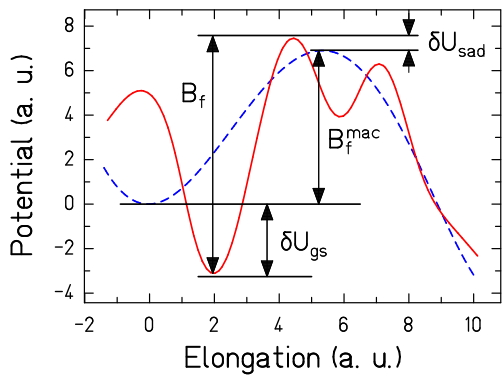}
\caption{(Color online) Schematic drawing of the potential energy on the fission path relative to
the macroscopic ground-state energy $E_{gs}^{mac}$ for a nucleus that is deformed
in its ground state. Spherical shape corresponds to zero elongation.
 Blue dashed line: macroscopic potential.
Red full line: full potential including the shell effects.
The figure is taken from Ref.~\protect\cite{Schmidt15EPJA}
with kind permission of The European Physical Journal (EPJ).}
\label{BF-FIG01}
\end{center}
\end{figure}

Myers and Swiatecki introduced the idea that the nuclear 
binding energy (or mass) at the fission threshold is essentially a macroscopic 
quantity \cite{Myers96}. That means that the mass at the highest one 
of the consecutive barriers between the ground-state shape and the scission configuration
can be estimated to a good approximation by the saddle mass of a macroscopic model.
They deduced this ``topographic theorem" from 
considerations on the topological properties of a surface in multidimensional
space with the specific properties of the 
potential-energy surface around the fission threshold.
The basic idea is illustrated in 
figure \ref{BF-FIG01}, where 
%the influence of the pairing condensation energy is neglected.
pairing effects are neglected.
%, because at this stage it is assumed that the pairing correction energy is independent of deformation and, thus, has no impact on the binding-energy difference.
The height of the fission barrier $B_f$ is given to a good approximation by the
difference of the macroscopic barrier $B_f^{mac}$ and the 
shell-correction energy in the ground state $\delta U_{gs}$. In practice, 
the ground-state shell correction $\delta U_{gs}$ is determined as the difference of the 
ground-state energy $E_{gs-nopair}$ that is averaged over odd-even staggering in $N$ and $Z$
and the macroscopic binding energy:
 $\delta U_{gs} = E_{gs-nopair} - E_{gs}^{mac}$.
If the ground-state mass is experimentally known, the fission barrier can be estimated on the
basis of a macroscopic model that provides the macroscopic ground-state mass and the 
fission barrier:

\begin{equation} 
  B_{f} \approx  B_f^{mac} - E_{gs-nopair}^{exp} + E_{gs}^{mac}.
\label{BF-EQ1}
\end{equation}

The condition
for this topological property of a surface in multi-dimensional space is that 
the wavelength of the fluctuations induced by the shell effects is smaller than
the wavelength of the variations induced by the macroscopic potential.   
This behavior can be understood, because
a local modification of the potential by a bump or a dip, for example by shell
effects, does not have a big effect on the height of the saddle: The 
fissioning nucleus will go around the bump, and it cannot profit from the depth
of the dip, because the potential at its border has changed only little.
This phenomenon is related to the observation that the potential 
at the fission saddle in calculations with a shape parametrization tends to take 
lower values by allowing for more complex shapes. The inclusion of additional
degrees of freedom gives access to a path that is energetically more favorable
and avoids the bump mentioned above.

A detailed investigation of the applicability of the topographic theorem 
was performed in Ref.~\cite{Karpov08}. 
The validity of the topographic theorem has been demonstrated before in a more qualitative way,
for example in Ref.~\cite{Swiatecki07}, and possible explanations for the observed 
deviations in the range of a few MeV are discussed. 
The topographic theorem has also been used before as a test of the 
ability of different theoretical models to describe the long-range behavior of the
fission threshold along isotopic chains \cite{Dahlinger82,Kelic06}. 

\begin{figure*}[h] 
\begin{center}
\includegraphics[width=0.8\textwidth]{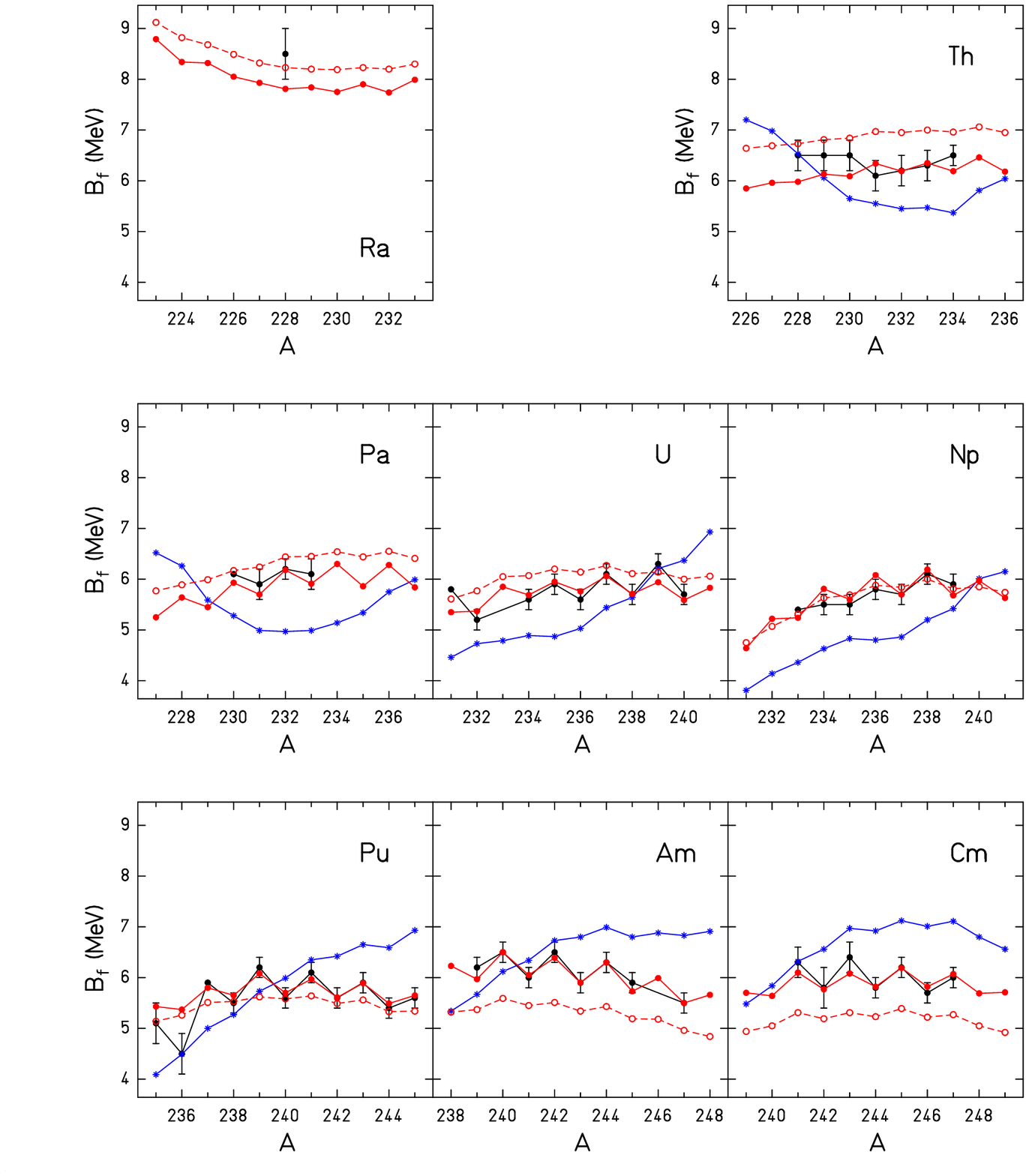}
\caption{(Color online) The empirical fission threshold of Ref.~\protect\cite{Bjornholm80} 
(full black circles) is compared to the value 
(open red circles) estimated from the topographic theorem according to equation 
(\ref{BF-topo}) for isotopic sequences
of different elements. In addition, a modified estimation (full red circles)  
with a $Z$-dependent shift and an assumed increased pairing parameter 
$\Delta_f = 14 / \sqrt A$ MeV at the barrier
(see text and Ref.~\cite{Schmidt15EPJA}) as well as 
the theoretical prediction of the microscopic-macroscopic approach of Ref.~\protect\cite{Moeller09} 
(blue asterisks) are shown. 
Empirical values without error bars are given without an uncertainty range in Ref.~\protect\cite{Bjornholm80}.
The figure is taken from Ref.~\protect\cite{Schmidt15EPJA}
with kind permission of The European Physical Journal (EPJ).
}
\label{BF-FIG02}
\end{center}
\end{figure*}

According to a previous analysis in \cite{Kelic06},
the average trend of the saddle mass along
isotopic chains is very well reproduced by the Thomas-Fermi model 
of Myers and Swiatecki \cite{Myers96,Myers99}.
(As will be shown below, this is not the case in the $Z$ dependence of the saddle mass.)
Therefore, the comprehensive set of empirical fission thresholds, that means
the maxima of the first and the second barrier heights, 
from Ref.~\cite{Bjornholm80} that are extracted from
experimental fission probabilities and cross sections are compared in figure \ref{BF-FIG02} 
with the quantity 
\begin{equation} 
  B_f^{topo} =  B_f^{TF} - E_{gs-nopair}^{exp} + E_{gs}^{TF}
\label{BF-topo}
\end{equation}

where $B_f^{TF}$ denotes the macroscopic fission barrier of
Ref.~\cite{Myers99}, represented by $B_f^{mac}$ in figure \ref{BF-FIG01},
and $E_{gs}^{TF}$ is
the macroscopic ground-state energy from the Thomas-Fermi model of Ref.~\cite{Myers96}. 
Both quantities do not contain neither shell nor pairing effects.
Note that there is no fit parameter in equation (\ref{BF-topo})! 
$E_{gs-nopair}^{exp}$ was taken from the 2012 Atomic Mass Evaluation \cite{Audi12,Wang12}, averaged 
over odd-even staggering in $Z$ and $N$. 
The quantity $E_{gs-nopair}^{exp} - E_{gs}^{TF}$ defines the empirical 
ground-state shell correction, represented by $\delta U_{gs}$ in figure~\ref{BF-FIG01}.

In accordance with Ref.~\cite{Kelic06}, figure \ref{BF-FIG02} shows that
the overall isotopic trend of the empirical barriers 
is very well reproduced by $B_f^{topo}$. 
However, there are some systematic deviations in the absolute values for the different isotopic chains:
Firstly, the barriers of thorium, protactinium and uranium
isotopes are overestimated, while the barriers of the heaviest elements plutonium, americium and curium
are underestimated. The deviation shows a continuous smooth trend as a function of $Z$
with a kink at $Z = 90$.
Secondly, a systematic odd-even staggering that is evident in the empirical barriers 
from protactinium to curium is not reproduced by the  $B_f^{topo}$ values 
estimated with equation (\ref{BF-topo}). 

\begin{figure}[!t]  % bottom
%\begin{figure}[!h]  % here
\begin{center}
\includegraphics[width=0.33\textwidth]{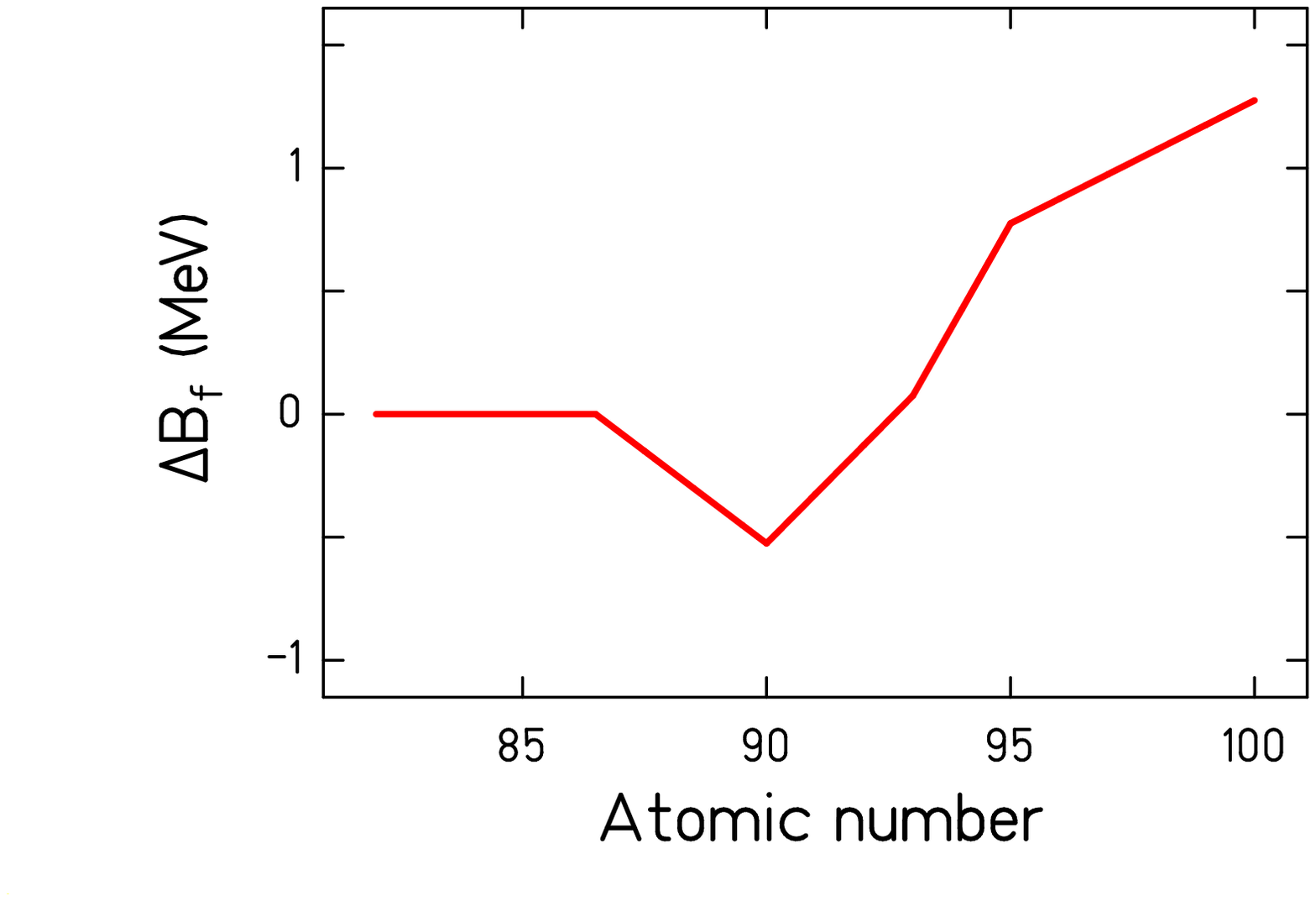}
\caption{
Empirical correction applied to the fission-barrier height obtained with the topographic
theorem as a function of the atomic number of the fissioning nucleus.
The figure is taken from Ref. \protect\cite{JEFF24}.}
\label{BF-FIG01a}
\end{center}
\end{figure}

In Ref.~\cite{Schmidt15EPJA}, an even better reproduction of the empirical barriers was obtained
by applying a simple $Z$-dependent correction $\Delta B_f$ to the values obtained by equation (\ref{BF-topo})
and by increasing the pairing-gap parameter $\Delta$
at the barrier in proton and neutron number to $14 / \sqrt{A}$ MeV, which is appreciably
larger than the average value of about $11 / \sqrt{A}$ MeV found in the nuclear ground state
\cite{Myers66}.
This was technically done by increasing the binding energy at saddle for nuclei with even $N$ by $3 / \sqrt{A}$ MeV and for even $Z$ by the same amount.
The correction term $\Delta B_f$ was parametrized in a way to account for the 
observed Z-dependent deviations as shown in figure \ref{BF-FIG01a}.

The modified barriers are also shown in figure~\ref{BF-FIG02}.
Possible explanations for the origin of this correction are given in Ref.~\cite{Schmidt16}.
Note that the correction $\Delta B_f$ depends only on $Z$. Thus, it shifts the sequence of barriers
of one element only by the same amount and does not change the structures in these curves.
Indications for an 
increased pairing gap at the barrier were already discussed by Bj{\o}rnholm and Lynn \cite{Bjornholm80} to explain the enhanced odd-even staggering at the saddle. 
They interpreted this finding as a possible evidence for surface pairing.

%\clearpage

In addition, figure \ref{BF-FIG02} shows the predictions from $Z$ = 90 to $Z$ = 96 of an elaborate 
theoretical model \cite{Moeller09} with the macroscopic-microscopic approach based on
a meticulous mapping of the
potential in five-dimensional deformation space \cite{Moeller01}. 
Although, as already mentioned, the fission barriers from this multi-dimensional model \cite{Moeller09} cannot directly be identified
with the one-dimensional fission barriers determined in Ref.~\cite{Bjornholm80}, the comparison
is not without interest.
The model values \cite{Moeller09} deviate appreciably from the empirical one-dimensional values \cite{Bjornholm80} with an rms deviation of 1.42 MeV (see Table \ref{BF-Tab1}).
In particular, the isotopic trend is not well reproduced. 
Moreover, the model does not 
show the observed odd-even staggering of the barrier height. 
Other models, 
macroscopic-microscopic or microscopic ones, show similar deviations.
In this context, it is interesting that nuclear ground-state masses can be obtained with 
an uncertainty of about 500 keV by the most reliable macroscopic-microscopic or microscopic models \cite{Sobiczewski14}. 
With this uncertainty and an additional uncertainty of the potential energy at the saddle in mind, uncertainties of the fission-barrier height from the macroscopic-microscopic approach 
in the order of 1 MeV are not unexpected.
Thus, it seems to be doubtful, whether the deviations between the empirical barriers \cite{Bjornholm80} and those of Ref.~\cite{Moeller09} are related to the multi-dimensional character of the calculation. 
They may rather be an indication for the accuracy that the macroscopic-microscopic approach can obtain on the multi-dimensional potential-energy surface, see also the related discussion in Ref.~\cite{Sierk17}. 

\clearpage

%\begin{figure}[!t]  % top

The most interesting feature, however, is that
the saddle-point masses (the nuclear binding energies at the
saddle) extracted by Bj{\o}rnholm and Lynn \cite{Bjornholm80}, which are derived from energy-dependent fission cross sections and fission probabilities, show no structural effects beyond a regular odd-even staggering. 
This is very remarkable, because the underlying measured energy-dependent fission probabilities include naturally the influence of all structure effects (for example shell effects and pairing) on the flux over the barrier.  
The saddle masses estimated by the topographic theorem show this characteristics, because they are macroscopic quantities by construction.

From our study, we may draw the following conclusions:

Considering that (i) the procedure used by Bj{\o}rnholm and Lynn for extracting the empirical
fission barriers from the measured energy-dependent fission cross sections and probabilities, and
(ii) the application of the topographic theorem for obtaining estimated values of the fission
barriers are completely independent, the good agreement of these sets of values
is a strong indication that, firstly, the empirical barriers deduced by Bj{\o}rnholm and Lynn
represent the $true$ fission thresholds in a one-dimensional picture with a remarkable accuracy
and that, secondly, the topographic theorem is a rather good approximation. 
From the rms deviation between these two sets of barriers, given in Table \ref{BF-Tab1},
it may be concluded that the uncertainties of the empirical barriers, determined 
by Bj{\o}rnholm and Lynn, are not larger than 500 keV, which is appreciably 
smaller than the presumed value of 1 MeV mentioned above.

%\clearpage

\begin{table} [h]
\begin{center}
\caption{Rms deviation between the empirical barriers and the values from different approaches.} 
\label{BF-Tab1}
~
\newline
\begin{tabular}{c|c|c|c} \hline \hline
  topographic    &  adjusted  &   M\"oller   &     RIPL 3       \\ \hline %\hline
  0.52     &   0.24    &   1.42     &    0.36        \\ \hline
\end{tabular}
\end{center}
\small{Note:
The table lists the rms deviations in MeV of the different sets of fission barriers shown in 
figure~\ref{BF-FIG02} and the values of RIPL 3 from the empirical values. References: empirical \cite{Bjornholm80},
topographic (this work, equation (\ref{BF-topo})), adjusted (this work 
and Ref.~\cite{Schmidt15EPJA}),
M\"oller et al.~\cite{Moeller09}, RIPL 3 \cite{Capote09}. 
Note that refs.~\cite{Capote09} and \cite{Moeller09}
do not cover all nuclei, for which empirical values are available. 
The typical uncertainty of the empirical values, given in Ref.~\cite{Bjornholm80},
is 0.2 to 0.3 MeV.} 
\end{table}

The fact that the deviations can even substantially be reduced by a simple $Z$-dependent
shift indicates that these deviations are caused by systematic shortcomings in the 
$Z$ dependence of the macroscopic model used for the estimations or by a slight violation 
of the topographic theorem, see Ref.~\cite{Schmidt16} for a detailed discussion.
By applying the deduced $Z$-dependent shift and an increased odd-even staggering at the saddle, 
the empirical barriers are reproduced 
within their uncertainties of 200 to 300 keV, as given in Ref.~\cite{Bjornholm80}. 
Regarding the absence of any systematic deviations of the fission barriers estimated with the topographic theorem from the barriers determined by Bj{\o}rnholm and Lynn \cite{Bjornholm80} along isotopic
chains, it seems that it is well justified to assume that reliable predictions of 
fission thresholds in an extended region of the 
chart of the nuclides can be made with this description.
The agreement of this parametrization with the empirical values proposed in 
RIPL 3 is less good (see Figs.~7 and 8 in Ref.~\cite{Schmidt16}), in particular in the structures 
along isotopic chains, which are not affected by the applied $Z$-dependent shift.
This gives more confidence to the empirical values of Bj{\o}rnholm and Lynn \cite{Bjornholm80}.

In a more fundamental sense, any noticeable structural effects on the fission-barrier height can be attributed
to the microscopic contributions to the ground-state mass and to a systematically
stronger odd-even staggering at the barrier, only. Any fluctuations of the saddle masses beyond the even-odd staggering stay within the given uncertainties 
of the empirical fission thresholds \cite{Bjornholm80} that amount to typically 200 keV. 
An influence of shell-correction energies on the saddle mass cannot 
strictly be excluded, but if there is any, it must 
show a gradual and smooth variation with $Z$ and $A$ that behaves like a macroscopic quantity. 
At present, theoretical estimates of the fission barriers do not yet attain this accuracy. 
They show deviations in the order of 1 MeV.
The best reproduction of empirical fission barriers has been reported in
Ref.~\cite{Kowal10}, where an rms deviation of 0.5 MeV has been obtained
within the framework of the macroscopic-microscopic approach.
This is also the accuracy to be expected for
theoretical calculations of the whole potential-energy landscape of the fissioning system.

\subsubsection{Fission probabilities} \label{4-3-3}

The value of the highest barrier, deduced in the preceding section, is not sufficient
to calculate the energy-dependent fission probability, which is required to estimate the probabilities of the different fission chances at higher excitation energies.
In the scenario of a one-dimensional double-humped barrier with parabolic shapes around the saddles and around the second well, among others the heights of both saddles and the curvature parameters are required.
In Ref.~\cite{Schmidt16}, it was found from a fit to the empirical barriers of Ref.~\cite{Bjornholm80} that the quantity
$B_A - B_B$ in the actinides varies smoothly as a function of $Z^3/A$ like
\begin{eqnarray}
(B_A - B_B) / \mbox{MeV} = 5.401 - 0.00666175 \cdot Z^3 / A \nonumber \\ 
 + 1.52531 \cdot 10^{-6} \cdot (Z^3/A)^2     
\label{DEV-EQ11}.
\end{eqnarray}
Together with the estimation of the highest barrier described above and the systematics of other fission-barrier parameters, for example from Ref.~\cite{Bjornholm80}, a rather accurate and complete set of fission-barrier parameters can be established for a region of nuclei, extending appreciably beyond the region covered by experimental information. 
These parameters are used in GEF to provide estimated fission probabilities in the framework of the transmission through, respectively the passage over, a one-dimensional, double-humped fission barrier. 
Additional ingredients are the nuclear level density, the gamma strength function and neutron transmission coefficients; for details we refer to \cite{Schmidt16}.
\clearpage
Using the formalism of Ref.~\cite{Bjornholm80} assures a good reproduction of the measured fission probabilities and is expected to give rather accurate estimates for neighboring fissioning nuclei, too. 
Further experimental information for specific fissioning systems is not required. However, this formalism does not provide any information on resonance structures that appear at energies below and slightly above the fission barrier, because the appropriate parameters can only be deduced from dedicated experimental data, see section \ref{4-4}. 
Fortunately, these resonances
have little influence on the characteristics of multi-chance fission due to the higher excitation energies that are mostly involved.

\subsubsection{Multi-chance fission} \label{4-3-4}

\begin{figure}[h] 
\begin{center}
\includegraphics[width=0.43\textwidth]{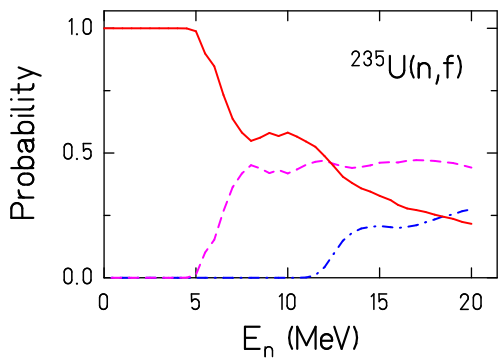}
\caption{(Color online)
Relative contributions of the different fission chances
in the system $^{235}$U(n,f) as a function of the
incident-neutron energy. Full line: first-chance fission, 
dashed line: second-chance fission, dot-dashed line: third-chance fission.
The figure is modified from Ref.~\protect\cite{Schmidt16}.
}
\label{Fig-MC-1}
\end{center}
\end{figure}

\begin{figure}[h]  
\begin{center}
\includegraphics[width=0.46\textwidth]{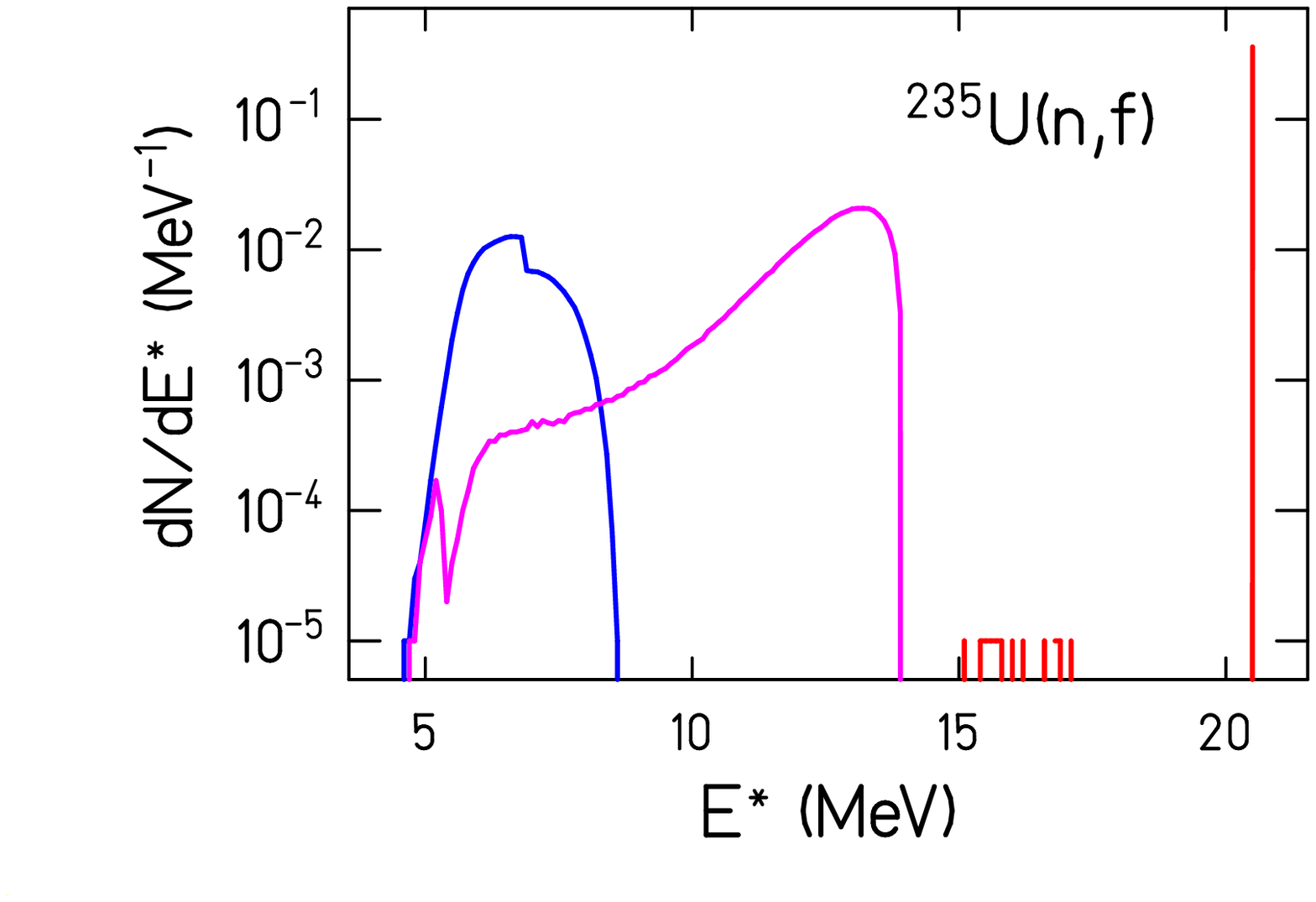}
\caption{(Color online)
Distribution of excitation energies $E^*$ at fission for the 
system $^{235}$U(n,f) at $E_n$ = 14 MeV. $E^*$ is the excitation energy of the
compound nucleus above its ground state, before it passes the fission barrier
towards scission. The right-most peak shows events from 
first-chance fission (fission of $^{236}$U), the middle curve corresponds to
second-chance fission (fission of $^{235}$U), and the left curve corresponds to
third-chance fission (fission of $^{234}$U). The broad distribution around 16 MeV 
corresponds to gamma emission before fission.
The figure is modified from Ref.~\protect\cite{Schmidt16}.
}
\label{Fig-MC-2}
\end{center}
\end{figure}

When fission after particle emission is possible, the measured fission observables originate from the fission of several nuclides with different excitation energies. 
In this case, the interpretation of the data and the modeling of the fission observables requires a good estimation of the characteristics of multi-chance fission. 
In GEF, the application of the topographic theorem and the scenario of the transmission, respectively passage, of a one-dimensional, double-humped barrier is used for this purpose.
Thus, quantitative estimations of the contributions from different fissioning systems and the corresponding excitation-energy distributions at fission are provided. Figures \ref{Fig-MC-1} and  \ref{Fig-MC-2} show the results for the system $^{235}$U(n,f).

\subsubsection{Hidden regularities of fission channels} \label{4-3-5}

\begin{figure*}[h] 
\begin{center}
\includegraphics[width=0.9\textwidth]{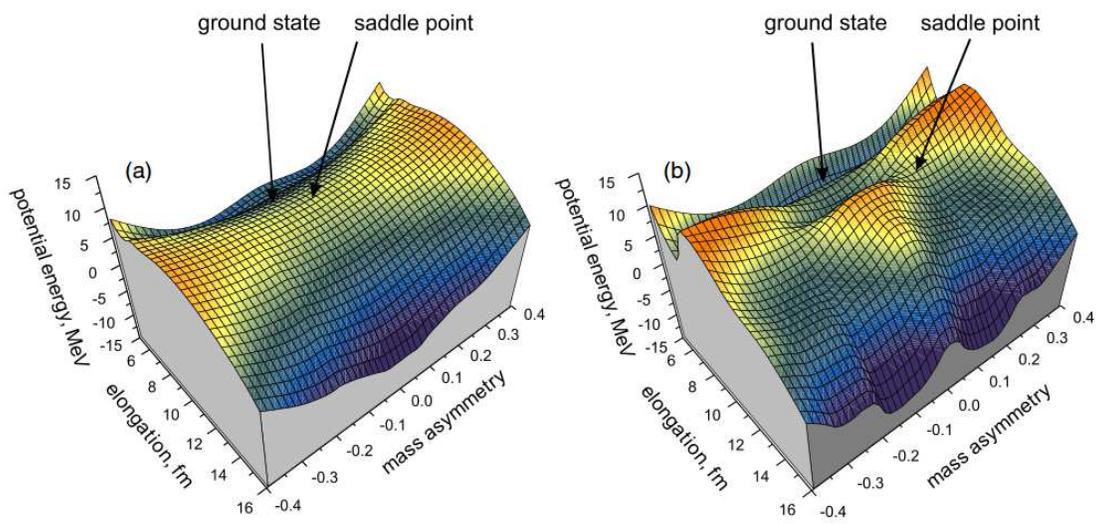}
\caption{(Color online)
Macroscopic (a) and  macro-microscopic (b) potential energy surface for the
$^{238}$U nucleus as a function of elongation and mass asymmetry.  
The macroscopic part is normalized to zero for the spherical shape of the compound nucleus. 
See Ref.~\protect\cite{Karpov08} for details. 
The figure is taken from Ref.~\protect\cite{Karpov08}.
}
\label{FIG-Karpov}
\end{center}
\end{figure*}

Although the good description of the fission barriers by the topographic theorem 
that is demonstrated in section \ref{4-3-2}
means that the saddle mass is essentially a macroscopic quantity, many other fission quantities
show very strong signatures of nuclear structure, for example, the evolution of the
shape and the potential on the fission path, in particular the existence of fission isomers,
triaxiality at the first barrier and mass asymmetry at the second barrier due to
shape-dependent shell effects in the actinides. Also the fission-fragment
yields are characterized by several components in the mass distributions from different 
fission channels that are
attributed to shell effects in the potential energy 
and by an odd-even staggering in
proton and neutron number due to the influence of pairing correlations. 
The potential-energy surface of the $^{238}$U nucleus, calculated with the two-center shell
model in Ref.~\cite{Karpov08}, demonstrates the structures created by the shell effects,
see figure \ref{FIG-Karpov}.

The mass-asymmetric deformation belongs to the relevant degrees of freedom of most
dynamical fission models, and the manifestation of shell effects in the fission-fragment 
mass distributions plays a prominent role in benchmarking these models.  
Macroscopic-microscopic models and, to some extent, also microscopic self-consistent models were rather
successful in reproducing the appearance of mass-asymmetric fission in the actinides \cite{Goutte05,Randrup11},
the features of multi-modal fission around $^{258}$Fm \cite{Ichikawa09}, the gradual transition from 
single-humped to double-humped distributions around $^{226}$Th \cite{Randrup13} and, most recently,
the appearance of complex mass distributions in a region around $Z \approx 80$ to 
$Z \approx 86$ from beta stability to the proton drip line \cite{Ghys14}. 
However, the deviations from the measured data remain important,
see the examples in figures \ref{FIG-Goutte} to \ref{FIG-Aritomo}.
A much higher accuracy
has been obtained with the semi-empirical description used in the GEF code \cite{Schmidt16} 
by exploiting regularities in the 
characteristics of the fission channels that are not obvious from microscopic models,
because these models treat each fissioning nucleus independently. In the following, we
will describe the theoretical considerations that are behind this semi-empirical approach.
They are not only important for high-accuracy estimates of fission-fragment yields,
kinetic energies, spectra and multiplicities of prompt neutrons and gammas, and other
fission quantities, but also
for a better understanding of the fission process by revealing an astonishingly high 
degree of regularity in the properties of fission channels. 

\clearpage

\paragraph{Early manifestation of fragment shells:}

When the two-center shell model \cite{Scharnweber71} became available, it
was possible to study the single-particle structure in a 
di-nuclear potential with a necked-in shape. Investigations
of Mosel and Schmitt \cite{Mosel71,Mosel71a} revealed that the single-particle
structure in the vicinity of the outer fission barrier indicates 
the existence of nearly independent single-particle states localized in the two fragments.
This leads to a coherent superposition of the shell-correction energies
from the two fragments.  
%Figure ? demonstrates that the single-particle
%energies remain almost constant little behind the outer
%saddle up to scission. 
The authors explained this result by
the general quantum-mechanical feature that wave 
functions in a slightly necked-in potential are already 
essentially localized in the two parts of the system. Also recent
self-consistent calculations show this feature (e.g. Ref.~\cite{Simenel14,Tanimura15}),
which is a direct consequence of the necking, independent
from a specific shape parametrization. Therefore, one may expect
that the complex structure of the fission modes can essentially be
explained by the influence of the shells in the proton and neutron
subsystems of the fragments. 
Potential-energy surfaces of fissioning systems 
calculated with the macroscopic-microscopic approach, for example
Ref.~\cite{Moeller09}, support this assumption.
The fact that in the actinides, for which a double-humped
fission-fragment mass distribution is observed, theoretical
models predict a mass-asymmetric shape at the outer saddle,
suggests also that fragment shells are already established to
a great extent at the outer saddle.

As a consequence, the shell effects on the fission path
from the vicinity of the second barrier to scission
can be approximately considered as the sum of the shell
effects in the proton- and neutron-subsystems of the light
and the heavy fission fragment. 
Thus, these shells do not
primarily depend on the fissioning system but on the 
number of neutrons and protons in the two fission fragments.
However, these shells may be substantially different from
the shell effects of the fragments in their ground state,
because the nascent fragments in the fissioning di-nuclear
system might be strongly deformed due to their mutual interaction \cite{Wilkins76}.

\begin{figure} [h]
\begin{center}
\includegraphics[width=0.4\textwidth]{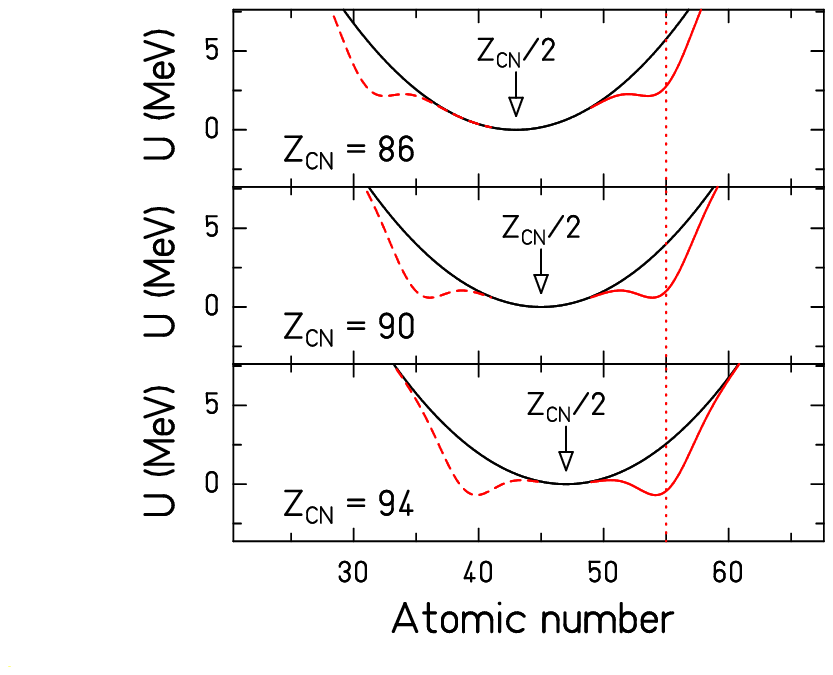}
\caption{(Color online) Schematic illustration of the potential
energy for mass-asymmetric shape distortions on the fission
path, after a qualitative idea of Itkis et al. \protect\cite{Itkis85}. The black curve
shows the macroscopic potential that is minimum at symmetry,
while the red curve includes the extra binding due to an
assumed additional potential well appearing at $Z$=55 in the heavy fragment. 
The assumed well position at $Z$=55 reflects the measured position of the 
asymmetric fission-fragment component (see figure \ref{Fig-Mode-Positions}). It is unclear, whether this is an indication for a proton shell around $Z$=55, see section \ref{5-4}.
The figure is taken from Ref.~\protect\cite{Schmidt16}.
}
\label{BF-mic-mac-pot}
\end{center}
\end{figure}

Therefore, the potential energy can be understood as the
sum of a macroscopic contribution, depending on the fissioning nucleus,
that changes gradually on the fission path and from one system to another one, and
a microscopic contribution that depends essentially only on the
number of protons and neutrons in the nascent fragments.
Thus, in nuclear fission, the macroscopic-microscopic approach turns
out to be particularly powerful. The distinction of the two
contributions to the potential is accompanied with an assignment of
these contributions to different systems: The macroscopic potential
is a property of the total system, while the shell effects are
attributed to the two nascent fragments \cite{Schmidt08}. 
Figure \ref{BF-mic-mac-pot} illustrates, how the interplay
of these two contributions can explain why symmetric fission is
energetically favored for fissioning nuclei below thorium, while
asymmetric fission is favored for nuclei above thorium.

%\clearpage

The shell effects used 
for the three nuclei on figure \ref{BF-mic-mac-pot} are the same. The changes in the total 
potential are caused by the shift to higher $Z$ values of the minimum of 
the macroscopic potential, which is located at symmetry.

In the following sub-section,
additional considerations on the dynamic evolution of the system are performed
in order to estimate, how the levels of the nascent fragments are populated.

%\clearpage

\paragraph{Quantum oscillators of normal modes:}

The early manifestation of fragment shells provides the explanation
for the appearance of fission valleys in theoretical calculations
of the potential-energy landscape of fissioning nuclei, in particular 
in the actinides. As demonstrated in figure \ref{FIG-Karpov}, 
these are valleys in direction of elongation, starting
in the vicinity of the second barrier until scission, with
an almost constant position in mass asymmetry. For the dynamic
evolution of the fissioning system, this means that each valley can be
considered as a quantum oscillator in the mass-asymmetry degree of freedom.
The initial flux in each valley, corresponding to a specific fission channel, 
is decided at the outer barrier. 
Depending on the height of the ridge between neighboring fission valleys, 
there might be some exchange of flux further down on the way to scission. 
The positions of the asymmetric minima that
are created by shell effects and the shape 
of the corresponding oscillator potentials stay approximately constant until scission, 
but the excitation energy of each quantum oscillator
tends to increase on the way towards scission by the feeding from the
potential-energy gain along the fission path.
It is assumed that the statistical ensemble of a large number of fission 
events\footnote{As introduced by Gibbs \cite{Gibbs02}, a statistical ensemble is 
an idealization consisting of a large number of virtual copies of a system, 
considered all at once, each of which represents a possible state that the 
real system might be in.
%a probability distribution for the state of the system.
}
establishes an excitation-energy distribution that can formally be replaced by
the distribution of a quantum oscillator in
instantaneous equilibrium with a heat bath of temperature $T$, whereby the 
temperature increases on the way to scission.  
With these ideas in mind, one can formulate the evolution of the mass-asymmetry
degree of freedom of the fissioning system on the way to scission.
Deviations from instantaneous equilibrium by a dynamical freeze-out will be 
discussed in the next section.

Stochastic calculations show that the excited nucleus stays in the 
ground-state minimum and later in the second minimum for quite a time, see figure 4 in 
\cite{Aritomo14}. One expects that this is still true if tunneling is considered, 
because the transmission probability decreases by about 5 orders of magnitude per 1 MeV barrier height \cite{Patyk89}. 
Therefore, an excited nucleus has enough time to re-arrange its available energy before passing the fission barrier. 
The probability for the passage
of the fission barrier increases considerably, if the nucleus
concentrates enough of its energy on the relevant shape
degrees of freedom for avoiding tunneling as much as the
available energy allows. If the available energy exceeds the
barrier, this excess can be randomly distributed between the different states
above the barrier without any further restriction, such
that the barrier is passed with maximum possible entropy
on the average \cite{Wigner38}, replacing again event averaging by
instantaneous equilibrium. 
For this reason, the fissioning system
has no memory on the configurations before the barrier,
except the quantities that are preserved due to general
conservation laws: total energy, angular momentum and
parity.\footnote{This assumption is supported by microscopic calculations
reported in Ref. \cite{Regnier16}, which showed that the 
dynamic evolution of the fissioning system beyond the fission barrier 
depends only little on the initial configuration inside the first minimum chosen in the calculation.} 
Thus, the starting point for calculating the 
properties of the fission fragments is the configuration above
the outer fission barrier.

Considering again the statistical ensemble of fissioning systems, 
the evolution of the entropy plays a decisive role in the fission process. 
The concentration of a large amount of energy into the elongation degree of 
freedom at the barrier leads to a decrease of the thermal energy and
induces a reduction of the entropy\footnote{We would like to 
stress that this reduction of entropy
is not in contradiction to the Second Law of Thermodynamics, because
the laws of thermodynamics have a statistical nature. Thermodynamical
quantities such as the entropy are subject to fluctuations that become sizable in
mesoscopic or microscopic systems like nuclei. A proper way to treat
such systems is the explicit application of statistical mechanics \cite{Gross06}.}.
%to
%defined as the logarithm of the density of occupied levels, averaged over 
%a large number of fissioning systems.
%Moreover, the levels
%at the barrier are populated according to their statistical
%weights \cite{Wigner38}. 
After passing the barrier, the entropy may increase again due to dissipation.
If dissipation is low, this increase may be small, and the passage from the barrier
to scission may be adiabatic, leaving the entropy essentially unchanged.
However, we think that the arguments for a reduction of the entropy on the way from the first minimum to the barrier, formulated above, are very strong.
Therefore, we think that the
approximation of treating fission as an isentropic process \cite{Diebel81,Pei09}
is not a generally valid assumption.

\begin{figure} [th]  % top
\begin{center}
\includegraphics[width=0.4\textwidth]{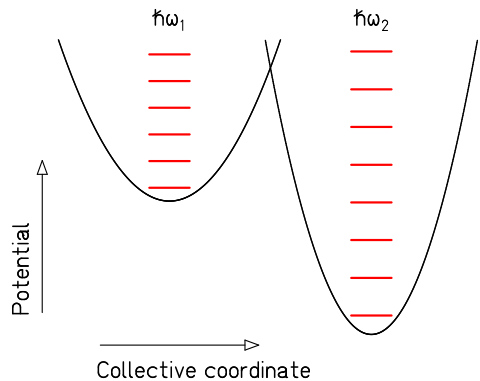}
\caption{(Color online)
Schematic drawing of the potential energy as a function of a
collective coordinate that is orthogonal to the fission direction
at a fixed elongation. In the present context, 
the two harmonic oscillator potentials with 
different depths and $\hbar \omega$ values represent the potential in 
two fission valleys for mass-asymmetric distortions that 
are related to different fission channels. The energies of the
stationary states are indicated by the red horizontal lines.
The overlapping of the two potential-energy
curves illustrates the possibility that the fission valleys are 
divided by a higher ridge that becomes perceptible when a continuous 
transition from one valley to the other is established in a deformation 
space with a sufficiently high dimension, see for example Ref.~\protect\cite{Pashkevich08}.
The figure is taken from Ref.~\protect\cite{Schmidt16}
with kind permission of The European Physical Journal (EPJ).}
\label{Fig-Osci}
\end{center}
\end{figure}

Beyond the outer barrier, the distribution of the mass-asymmetry 
coordinate is given by the occupation probability of the states of
the quantum oscillators in the respective fission valleys.
The situation is schematically illustrated in figure \ref{Fig-Osci} for
the mass-asymmetry coordinate in two fission valleys that
are well separated, assuming that the potential pockets have
parabolic shape. The fission-fragment mass distribution 
is given by the evolution of the respective collective
variable, until the system reaches the scission configuration.
It is defined by the number and the energy distribution of occupied states in
the different valleys.

In the case of weak coupling
and in thermal equilibrium with a heat bath of temperature $T$,
the ratio of the yields $Y_i$ of the two fission
channels corresponding to the population of the two harmonic quantum
oscillators depicted in figure \ref{Fig-Osci} is given by
\begin{equation}
Y_2 / Y_1 = e^{- \Delta E / T} \cdot {{\hbar \omega_1} \over {\hbar \omega_2}} \approx 
 e^{- \Delta E / T}.
\label{EQ-Yields-Osci}
\end{equation} 
$\Delta E = E_2 - E_1$ is the potential-energy difference between the lowest states in the two
quantum oscillators, and $\hbar \omega_i$ is the level spacing in the respective oscillator
potential.\footnote{
Equation (\ref{EQ-Yields-Osci}) may be derived as follows: It is assumed that $T > \hbar \omega_i$ 
and that the coupling between the oscillators and between each of the oscillators and the heat
bath is strong enough to allow for thermal equilibration, but it is so weak that the states of the oscillators are essentially undisturbed. In this situation, the population probability of a state $n_i$ in oscillator 1 and the population probability of a state $n_j$ in oscillator 2 are given by 
$P_{1,ni} = exp[(-n_i \cdot \hbar \omega_1)/T]$ and $P_{2,nj} = exp[(-\Delta E - n_j \cdot \hbar \omega_2)/T]$, respectively. 
The total population probabilities $Y_1$ and $Y_2$ in the two oscillators are obtained by summing up the individual population probabilities over $n_i$ and $n_j$, respectively:
$Y_1 = \sum_{n_i = 0}^{\infty} P_{1,ni}$ and $Y_2 = \sum_{n_j = 0}^{\infty} P_{2,nj}$.
By replacing the discrete states by a continuous state density, the sums become integrals. This leads to the solution in equation (\ref{EQ-Yields-Osci}).
}
(Because the $\hbar \omega$ values of the oscillators in the different fission values are normally rather similar, the factor $\hbar \omega_1 / \hbar \omega_2$ in equation (\ref{EQ-Yields-Osci}) is set to one as an approximation.)  
%As indicated, the relation is strongly dominated by the exponential term.
The distribution of the collective coordinate of the quantum oscillator
for asymmetric distortions in one fission channel 
is a Gaussian function with a variance $\sigma^2$ that is given by 
the well known equation \cite{Landau58}
\begin{equation} 
\sigma^2 = { {\hbar \omega} \over {2 C} } \mbox{coth}( { {\hbar \omega} \over {2 T} } ) .
\label{EQ-Sigma-Osci}
\end{equation} 
$C$ is the second derivative of the potential near its minimum in the direction of
mass asymmetry.

If the exchange of flux between different fission channels beyond the second barrier 
is negligible, the temperature parameter in equation (\ref{EQ-Yields-Osci}) is the value at the second barrier. As argued above, it is assumed to be given by the nuclear temperature
at the second barrier as derived from the nuclear level density. 
However, the width of mass asymmetric distortions, described by the
temperature parameter in equation (\ref{EQ-Sigma-Osci}), evolves on the way to 
scission. The width of the fission-fragment mass distribution is given by the temperature
at the dynamical freeze-out that is described in the following. 

\paragraph{Dynamical freeze-out:}

It is well known \cite{Oganessian85} that the statistical model, applied
to the scission-point configuration, is unable of explaining
the variances of the mass and energy distributions and
their dependence on the compound-nucleus fissility. 
Also stochastic \cite{Aritomo13,Aritomo14} and self-consistent models \cite{Goutte05} 
show the importance
of dynamic effects on the width of the fission-fragment mass distributions,
especially in low-energy fission.
Studies of Adeev and Pashkevich \cite{Adeev89} suggest that dynamical
effects due to the influence of inertia and dissipation 
on mass-asymmetric distortions can
be approximated by considering the properties of the 
system at an earlier time. Nifenecker \cite{Nifenecker80} and Asghar
\cite{Asghar80} explained the fluctuations in the charge polarization 
at scission by a freeze-out\footnote{The term "freeze-out" in connection with fission-fragment
yields was used by Asghar \cite{Asghar80} for denoting the inability of the wave function of
the fissioning system to adjust adiabatically when approaching scission.} of the giant-dipole resonance before neck rupture.
For all processes, which are connected with a transport
of nucleons from one nascent fragment to the other one (e.g. evolution of mass asymmetry \cite{Nix65} or charge polarization \cite{Nifenecker80,Asghar80}), the effective mass increases 
dramatically during the necking-in of the fissioning system.
This makes it more difficult for the system to adjust to the
evolution of the potential-energy surface when approaching scission. 
That means that the statistical
model may give reasonable results for the distribution of any
observable related to a certain normal mode, if it is applied to a
configuration that depends on the typical time constant
of the collective coordinate considered. The relaxation time
is specific to the collective degree of freedom considered.
It is relatively long for the mass-asymmetric distortions
\cite{Karpov01} and rather short for the charge-polarization degree of
freedom \cite{Nifenecker80,Asghar80,Myers81,Karpov02}, due to the
difference in the associated inertia. 
Thus, the shape of the potential and the
value of the corresponding collective temperature that are
decisive for the distribution of the respective observable
are those that the system takes at the respective relaxation
time before scission.

From these considerations, it may be concluded that the 
observed fission-fragment mass distribution of a specific fission channel can approximately
be understood as the equilibrium distribution of the quantum oscillator in
the mass-asymmetry degree of freedom in the corresponding fission valley at the time 
of freeze-out on the fission path with the local second derivative of the mass-asymmetric
potential and temperature. 

\paragraph{Empirical extraction of universal fragment shells:}

\begin{figure} [h]  
\begin{center}
\includegraphics[width=0.5\textwidth]{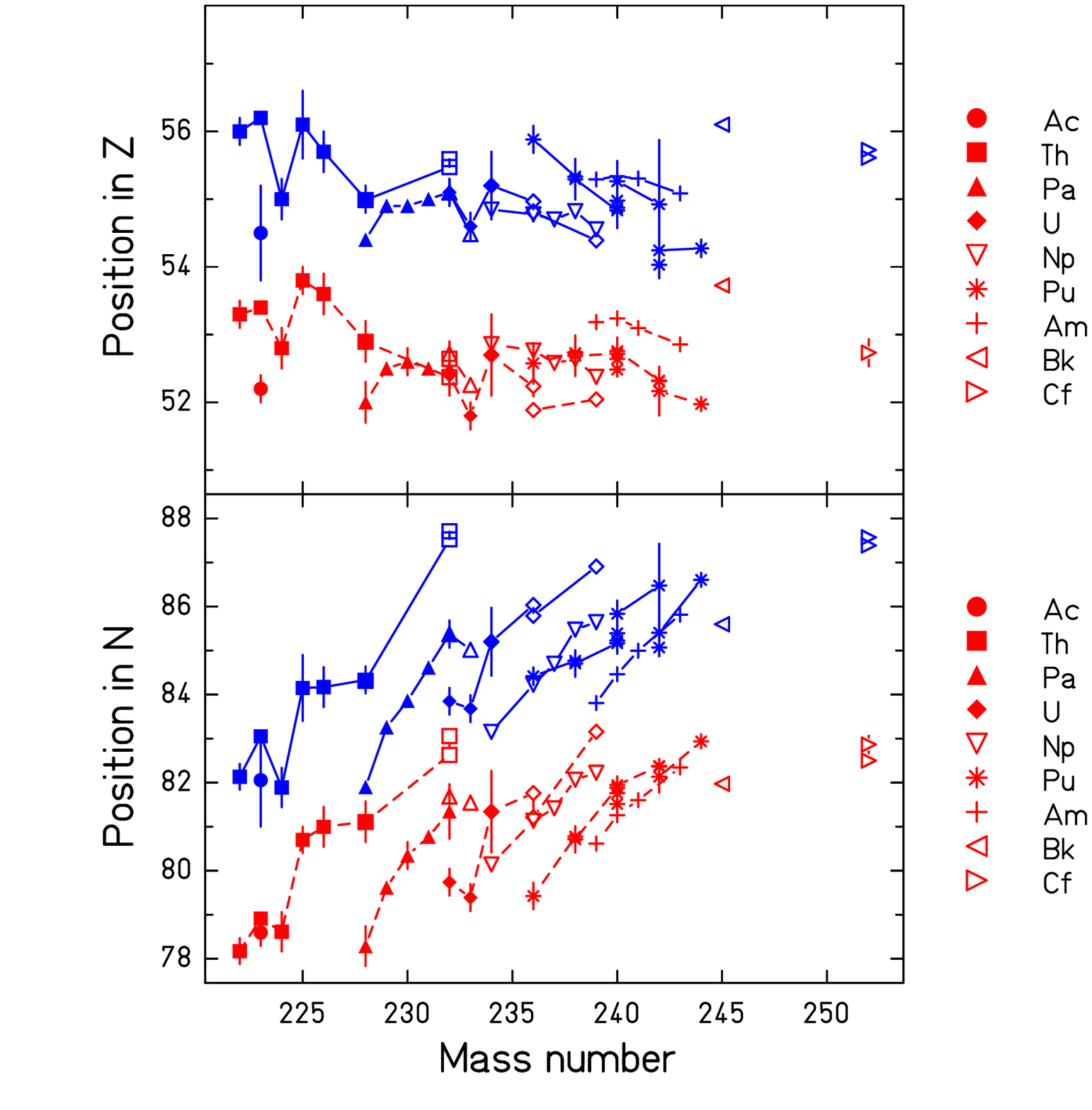}
\caption{(Color online)
Mean positions of the standard fission channels in atomic number (upper part) and neutron
number (lower part) deduced from measured fission-fragment mass and element distributions.
Values were converted from measured atomic numbers or mass
numbers using the unchanged-charge-density assumption and neglecting neutron evaporation. 
The shape of the symbol
denotes the element as given in the legend of the figure. 
Data from Ref.~\protect\cite{Schmidt00} are marked by solid symbols.
The values of standard 1 (standard 2) for the isotopes of a given element are connected by dashed (full) lines and marked
by red (blue) symbols.
The figure is taken from Ref.~\protect\cite{Boeckstiegel08}.}
\label{Fig-Mode-Positions}
\end{center}
\end{figure}

In the macroscopic-microscopic approach, 
the potential energies at the bottom of the different
fission valleys that determine the relative yields of the
fission valleys according to equation (\ref{EQ-Yields-Osci}) are the sum of 5 terms: the
macroscopic potential and the shell energies of the 
proton and neutron subsystems of the two nascent fragments.
The stiffness of the macroscopic potential against mass-asymmetric
distortions evolves gradually as a function of the
fissility \cite{Adeev88}. An empirical function has been deduced
with a statistical approach \cite{Rusanov99}
from the widths of measured mass distributions of
the symmetric fission channel at higher excitation 
energies, where shell effects are essentially washed out.
For describing the yields and the variances of the
contribution of each fission channel to the fission-fragment
mass distributions with equations (\ref{EQ-Yields-Osci}) and (\ref{EQ-Sigma-Osci}), 
the following 3 parameters are 
required in addition to the curvature of the macroscopic potential: 
the position, the magnitude and the second derivative of each shell in each
of the four subsystems (protons and neutrons in the light
and the heavy pre-fragment).
These parameters are expected not to depend on the fissioning system, and 
to stay constant on the way to scission, 
once the nascent fragments have acquired their individual properties.

%\clearpage

\subparagraph{\textit{Fragment shells deduced from the fission of actinides:}}

It is known since long that the mean mass of the $heavy$ component in
asymmetric fission of the actinides is approximately constant at
$A \approx 140$ \cite{Unik74}. This is an indication that shells in the heavy fragment are
dominant. B\"ockstiegel et al.~\cite{Boeckstiegel08} compiled a systematics of fission
channels, deduced from measured fission-fragment mass and element
distributions, partly from two-dimensional mass-TKE distributions.
The systematics of the mean proton and neutron numbers of the 
standard 1 and the standard 2 fission channels, following the nomenclature
of Brosa et al. \cite{Brosa90}, is shown in figure~\ref{Fig-Mode-Positions}.
Obviously, the standard 1 and the standard 2 channels are located at 
the proton numbers $Z \approx 52$ and $Z \approx 55$, respectively.
This feature is most clearly evidenced by the 
data from the long isotopic chains measured in electromagnetic fission
of relativistic secondary beams \cite{Schmidt00}, but it had already
been observed for proton-induced fission of isotopic chains of heavier 
elements by Gorodisskiy et al.~\cite{Gorodisskiy03}.
In contrast, the neutron number varies systematically as a function of
the mass of the fissioning nucleus.

%\clearpage

This means that the most prominent asymmetric fission channels, standard 1 and
standard 2, are caused each by a fragment shell that fixes the number of protons
in the heavy fragment at $Z \approx 52$ and $Z \approx 55$, respectively. 
%These numbers cannot directly be interpreted as fragment shells,
%because the nucleons in the neck will add up to the
%nucleons of the pre-formed fragments at scission. 
However, while the value of $Z$ = 52 is compatible with the $Z$ = 50 proton shell, if some
contribution from the neck is taken into account, 
one should be cautious to conclude that this observation is an indication for
a proton shell near $Z$ = 55 or a bit lower due to the neck contribution.
A discussion of this problem
in view of theoretical studies of nuclear shell structure
is found in section \ref{5-4-1}.

The shell strength behind the S2 fission channels was kept constant for all 
systems. The shell strength behind the S1 fission channel was assumed to vary 
as a unique function of the $N/Z$ value of the fissioning system, starting from a maximum
value for $N/Z = 82/50$, except for the plutonium isotopes, where this shell was
enhanced by 0.4 MeV in order to fit the experimental mass distributions. 
This enhancement will come into play in the next section again.

%The ideas outlined above with a few refinements were applied for describing
%the mass-dependent fission yields in 
%the semi-empirical GEF model \cite{Schmidt16}.

\begin{figure*} [h]  
\begin{center}
\includegraphics[width=1.0\textwidth]{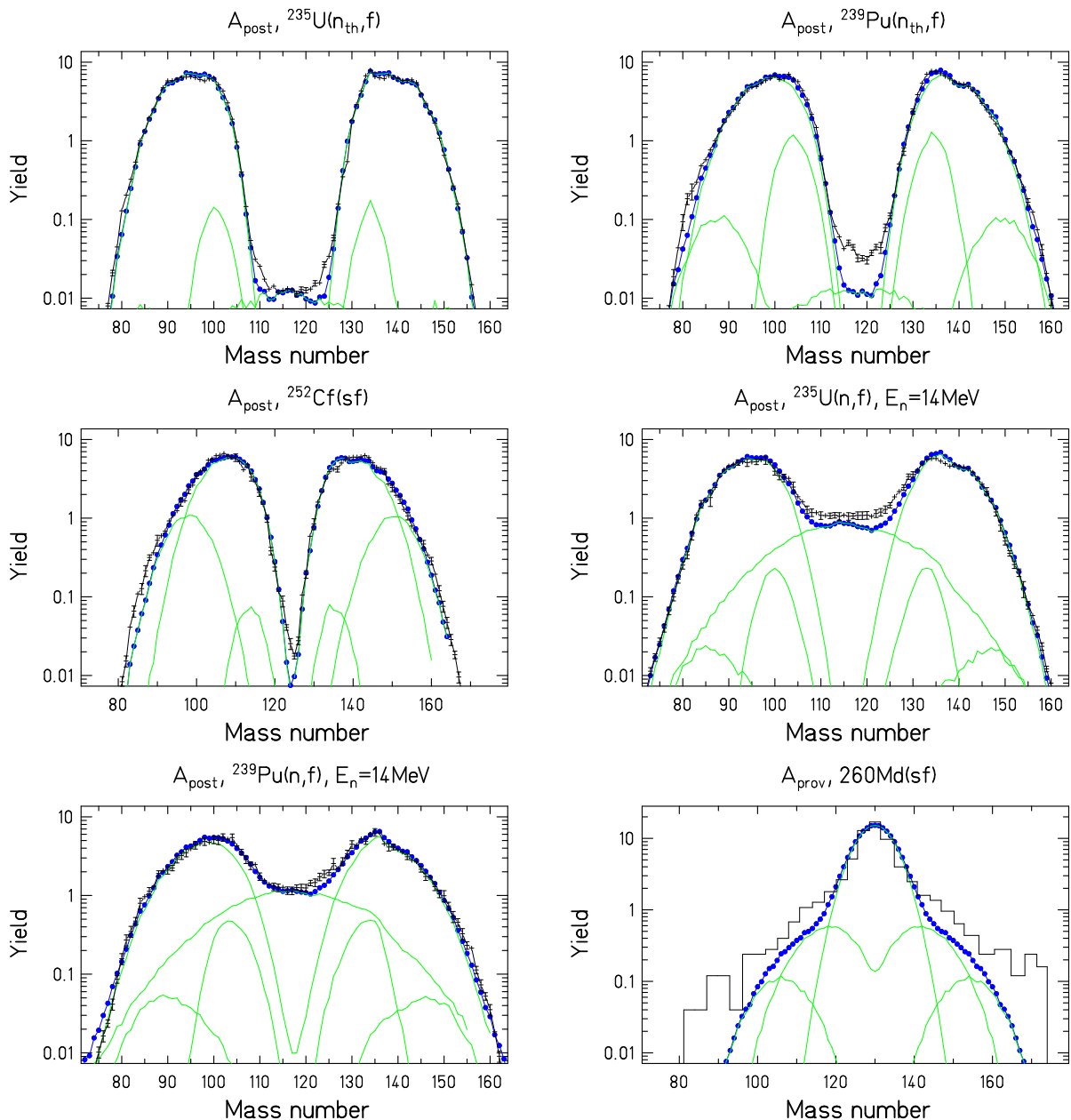}
\caption{(Color online)
Evaluated and measured mass distributions (black symbols) of fission fragments in comparison with the result of
the GEF model (blue symbols). The mass distributions after prompt-neutron emission are taken from the evaluation 
of Ref.~\protect\cite{England94}.
The provisional masses from spontaneous fission of $^{260}$Md (black histogram) 
were directly deduced from the ratio of the fragment energies with a finite resolution
of about 4 mass units
without applying a correction for prompt-neutron emission. 
The data of the corresponding mass distribution are taken from Ref.~\protect\cite{Hulet89}. 
The green lines show the calculated
contributions from the different fission channels.
The figure is taken from Ref.~\protect\cite{Schmidt15EPJA}
with kind permission of The European Physical Journal (EPJ).}
\label{Fig-GEF}
\end{center}
\end{figure*}

The ideas outlined above with a few refinements were applied 
for describing with a remarkable precision the fission 
yields of a great number of fissioning systems ranging from $Z=90$ to 
$Z=112$ in the semi-empirical GEF model using a unique set of parameters 
\cite{Schmidt16}. Figure \ref{Fig-GEF} shows 
calculated fission-fragment mass distributions 
compared to evaluated data for some selected nuclei. 
Note that the masses in the underlying data 
are unambiguously identified by radiochemical methods \cite{Laurec10}. 
Thus, the GEF results can directly be compared with the evaluated data. 
This is not the case for mass distributions determined in kinematical measurements (e.g. double-energy measurements), which are distorted by a limited mass resolution \cite{Gaudefroy17} and ambiguities in the calibration \cite{Adili12}, as already mentioned in section \ref{3-2}. 
In addition, the provisional masses of $^{260}$Md(sf) are shown in comparison with the GEF result.
A much more detailed comparison covering 
%a large variety of fission observables for
many fissioning systems and different energies can be found in \cite{Schmidt16}.

\subparagraph{\textit{Fragment shells deduced from fissioning systems in the lead region:}}

Let us remind that,
when moving from the actinides to lighter fissioning systems, asymmetric fission gradually
disappears, and symmetric fission takes over \cite{Schmidt00}. 
However, in the lead region, complex structures appear \cite{Itkis88,Andreyev13}, showing up
as double-humped or triple-humped mass distributions \cite{Ghys14}.
It is tempting to search for regularities in this phenomenon. This would also allow to improve the 
description of fissioning systems in this region with the GEF model.
This appears to be a difficult task, because the data are very scarce. 

\begin{figure*} [h]  
\begin{center}
\includegraphics[width=0.9\textwidth]{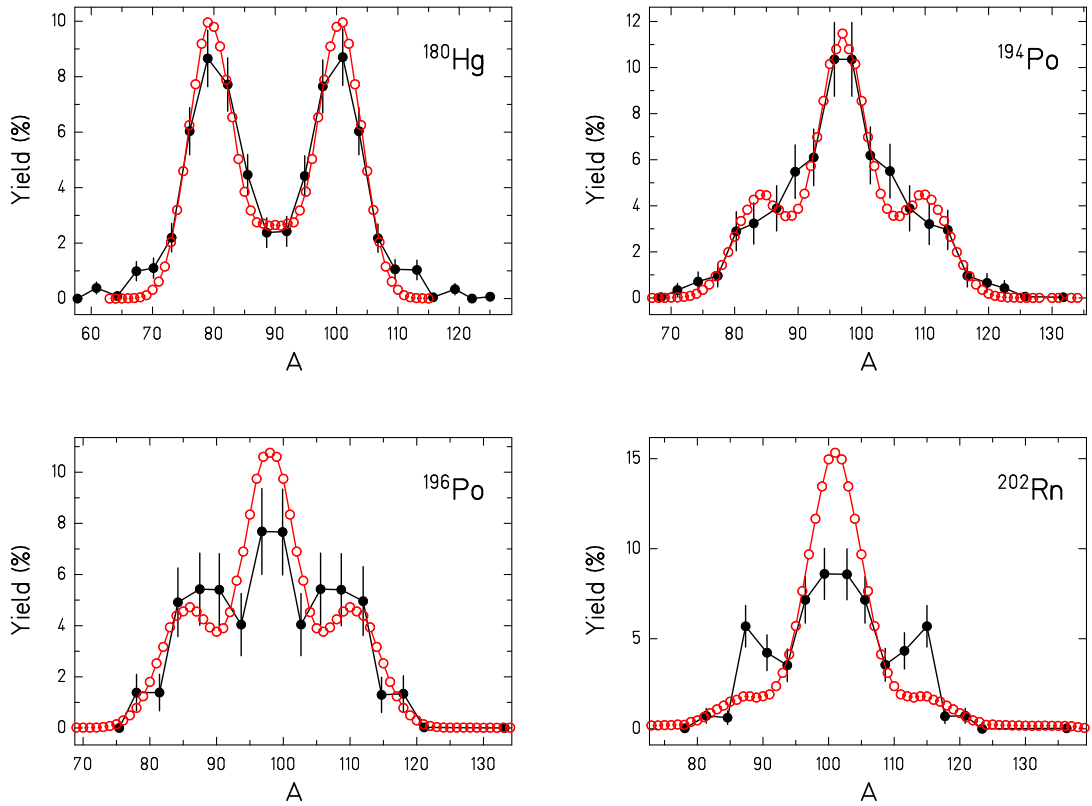}
\caption{(Color online)
Measured pre-neutron fission-fragment mass distributions from beta-delayed fission of the compound nuclei
$^{180}$Hg, $^{194}$Po, $^{196}$Po, and $^{202}$Rn \protect\cite{Elseviers13, Ghys14}
(full black symbols)
in comparison with the result of the GEF code (GEF 2017/V1.2) (open red symbols).}
\label{BDF} 
\end{center}
\end{figure*}

\begin{figure*} [h]  
\begin{center}
\includegraphics[width=1\textwidth]{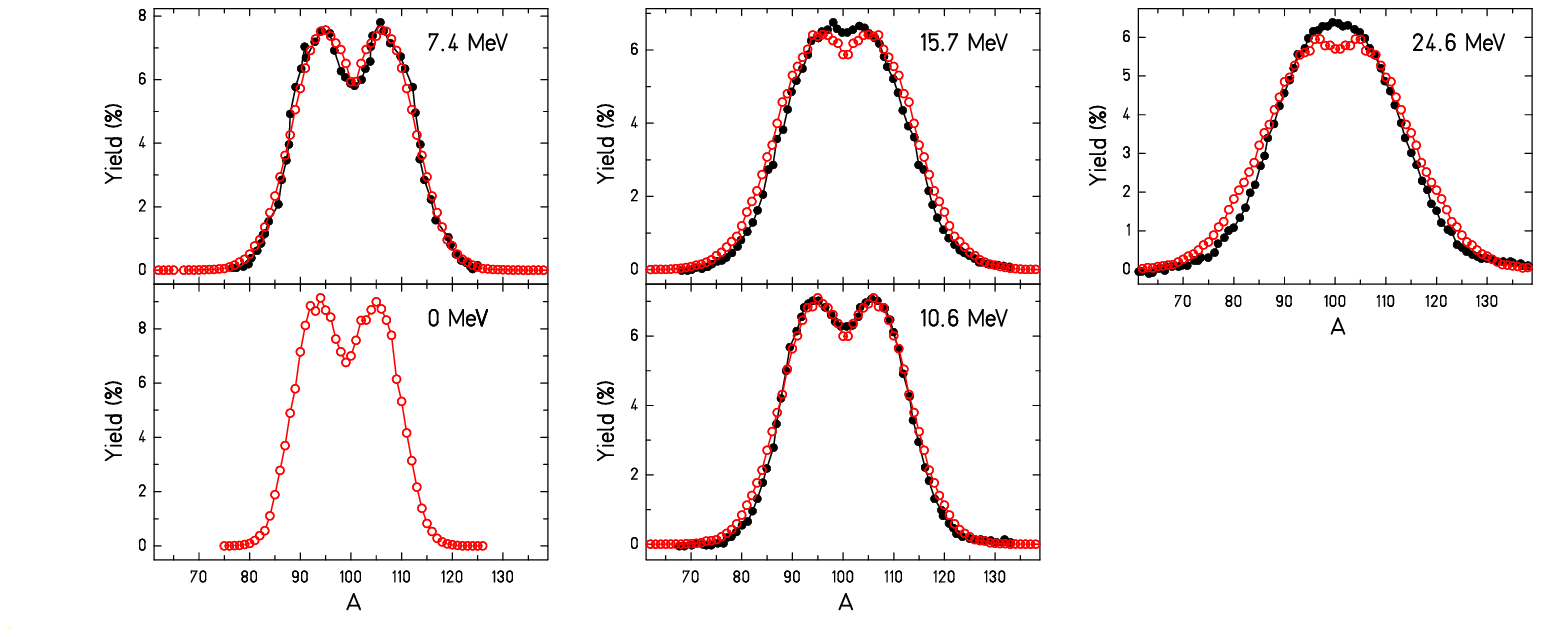}
\caption{(Color online)
Measured pre-neutron fission-fragment mass distributions from proton-induced fission of the compound nucleus
$^{201}$Tl \protect\cite{Itkis88} (full black symbols) in comparison with the result of the GEF code (GEF 2017/V1.2) (open red symbols).
The excitation energies above the fission barrier are indicated.}
\label{Tl201-all} 
\end{center}
\end{figure*}

\begin{figure} [h]  
\begin{center}
\includegraphics[width=0.5\textwidth]{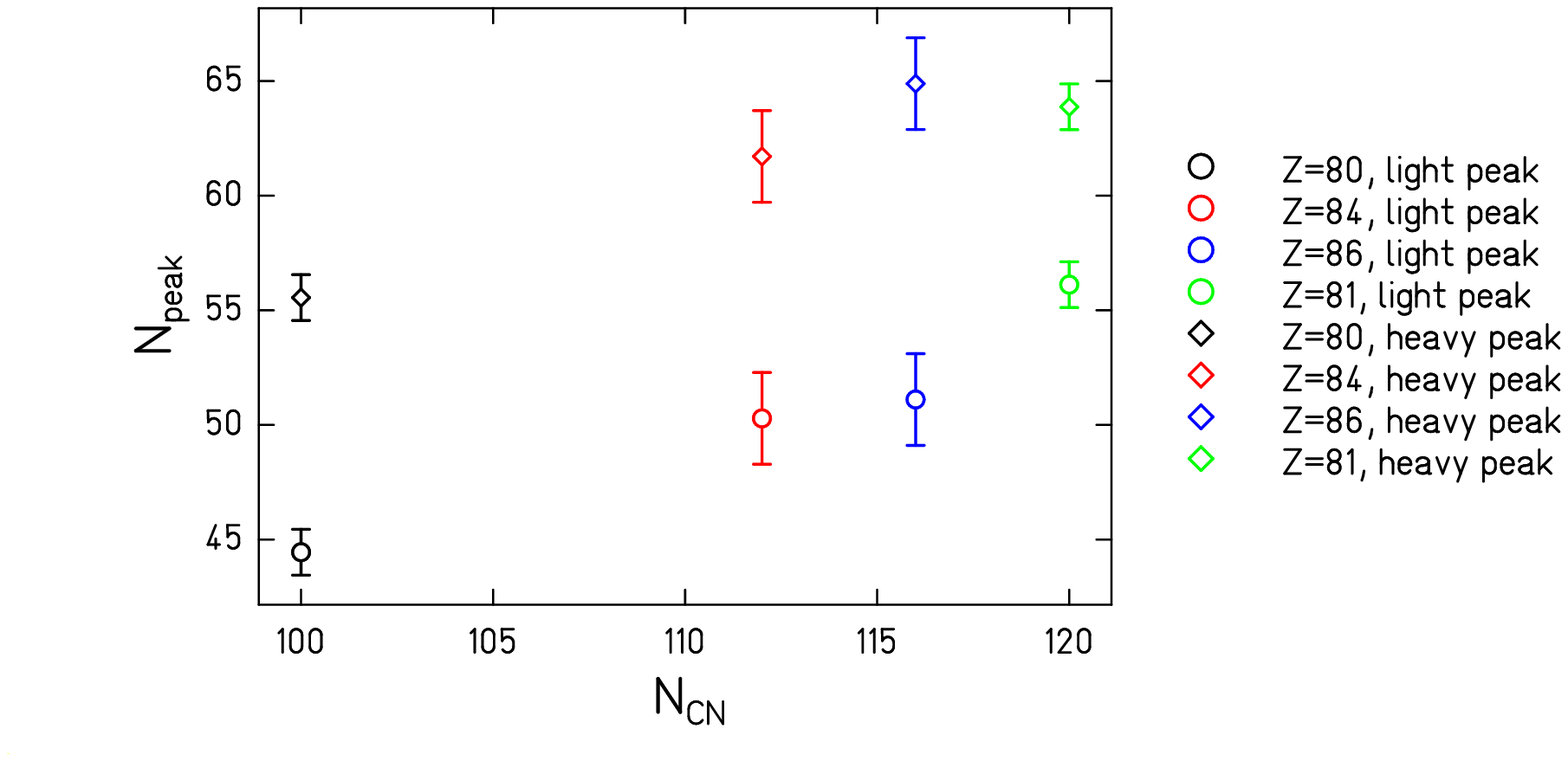}
\caption{(Color online)
Mean positions of the asymmetric fission components of the mass distributions shown in figure 
\ref{BDF} (beta-delayed fission of the compound nuclei $^{180}_{~80}$Hg, $^{196}_{~84}$Po, and $^{202}_{~86}$Rn),
and figure \ref{Tl201-all} ($^{201}_{~81}$Tl at 7.4 MeV above the fission barrier). The mass values were transformed into neutron numbers by the UCD assumption.}
\label{Peakpos-N} 
\end{center}
\end{figure}

\begin{figure} [h]  
\begin{center}
\includegraphics[width=0.5\textwidth]{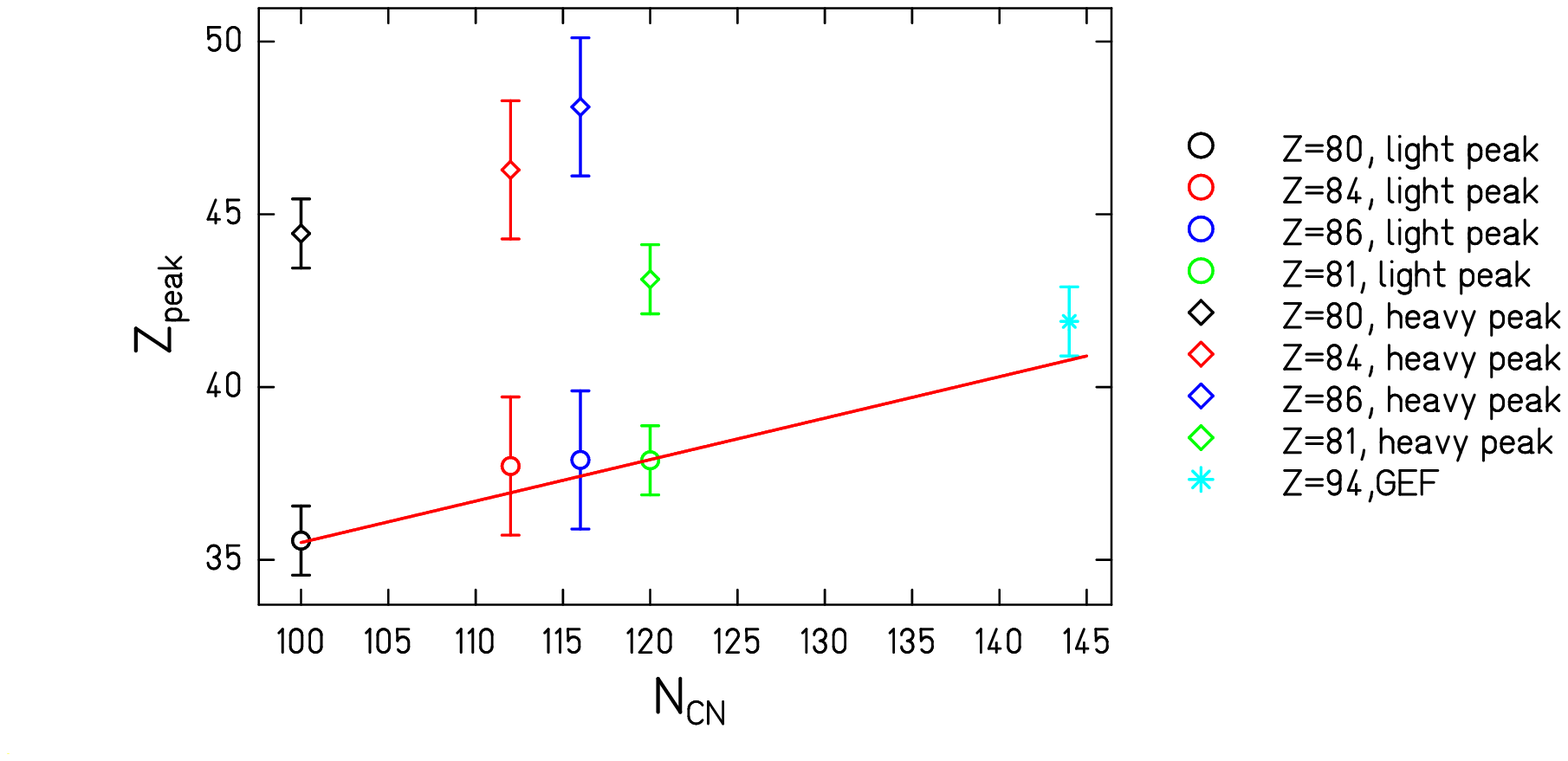}
\caption{(Color online)
Mean positions of the asymmetric fission components of the mass distributions shown in figure \ref{BDF} (beta-delayed fission of the compound nuclei $^{180}$Hg, $^{196}$Po, and $^{202}$Rn)
and figure \ref{Tl201-all} ($^{201}$Tl at 7.4 MeV above the fission barrier). The mass values were transformed into proton numbers by the UCD assumption. In addition, the mean number of protons in the light fragment of the fission of Pu isotopes, belonging to the S1 fission channel, is shown.}
\label{Peakpos-Z} 
\end{center}
\end{figure}

Figures \ref{BDF} and \ref{Tl201-all} show the
measured mass distributions with the most pronounced structures in this region.
For most of these distributions, the mean masses of the asymmetric components and their
uncertainties were
determined in \cite{Ghys14}.
Figures \ref{Peakpos-N} and \ref{Peakpos-Z} show the deduced mean positions of these peaks in
neutron and proton number, respectively, both in the light and in the heavy fragment.
The UCD assumption has been used to infer the number of protons and neutrons in the fragments
from the measured mass distributions.
Obviously, the positions in neutron number move in an irregular way. 
The same is true when the position in proton number of the heavy fragment is considered.  
The position of the light peak in proton number, however, shifts much less and moves in
a rather smooth way. 
%Another argument for attributing the origin of theses structures
%to the light fragment is the observation that it is not observed in systems 
%approaching Z = 90.  
Figure \ref{Peakpos-Z} also shows the position of the light fragment
produced in the fission of plutonium isotopes, which corresponds to the S1 fission channel.
As reported above, the yield of the S1 fission channel shows a clear enhancement if compared
to the regular trend. One is tempted to attribute this enhancement to the simultaneous formation
of fragments near the doubly magic $^{132}$Sn in the heavy fragment and the formation of fragments in the fission valley that is responsible for 
the double-humped peaks of fissioning systems in the lead region. 
The straight line shows a linear fit to the positions of the light peak in the fission of
$^{180}$Hg and $^{201}$Tl. Figures \ref{BDF} and \ref{Tl201-all} 
show, in addition to the data, the description with the GEF\footnote{
In this case and in some others, where newly implemented features are involved, the calculations were made with the GEF version Y2017/V1.2, although this version is still under development, and the parameter values are still preliminary.
This is indicated in the corresponding figure captions.
In all other cases, the version Y2016/V1.2 was used.} code that was achieved after the
implementation of a fission valley with the position defined by the straight line in 
figure \ref{Peakpos-Z}. 

In this context, it is interesting to note that also the mean $Z$ values of the S1 and the S2
fission channels show a modest linear variation as a function of mass $A_{CN }$ or neutron number
$N_{CN}$, see figure
\ref{Fig-Mode-Positions}. The magnitude of this variation is similar to the one observed
in figure \ref{Peakpos-Z} for the shell around $Z = 36$ (around $\Delta Z = 1$ for $\Delta N_{CN} = 10$, but the sign is opposite. May be, the different sign is related to the fact that 
the shells behind the asymmetric fission channels in the actinides are located in the heavy fragment,
while the structures in the measured mass distributions in the lead region are caused by a shell in the light fragment.

\begin{figure} [h]  
\begin{center}
\includegraphics[width=0.4\textwidth]{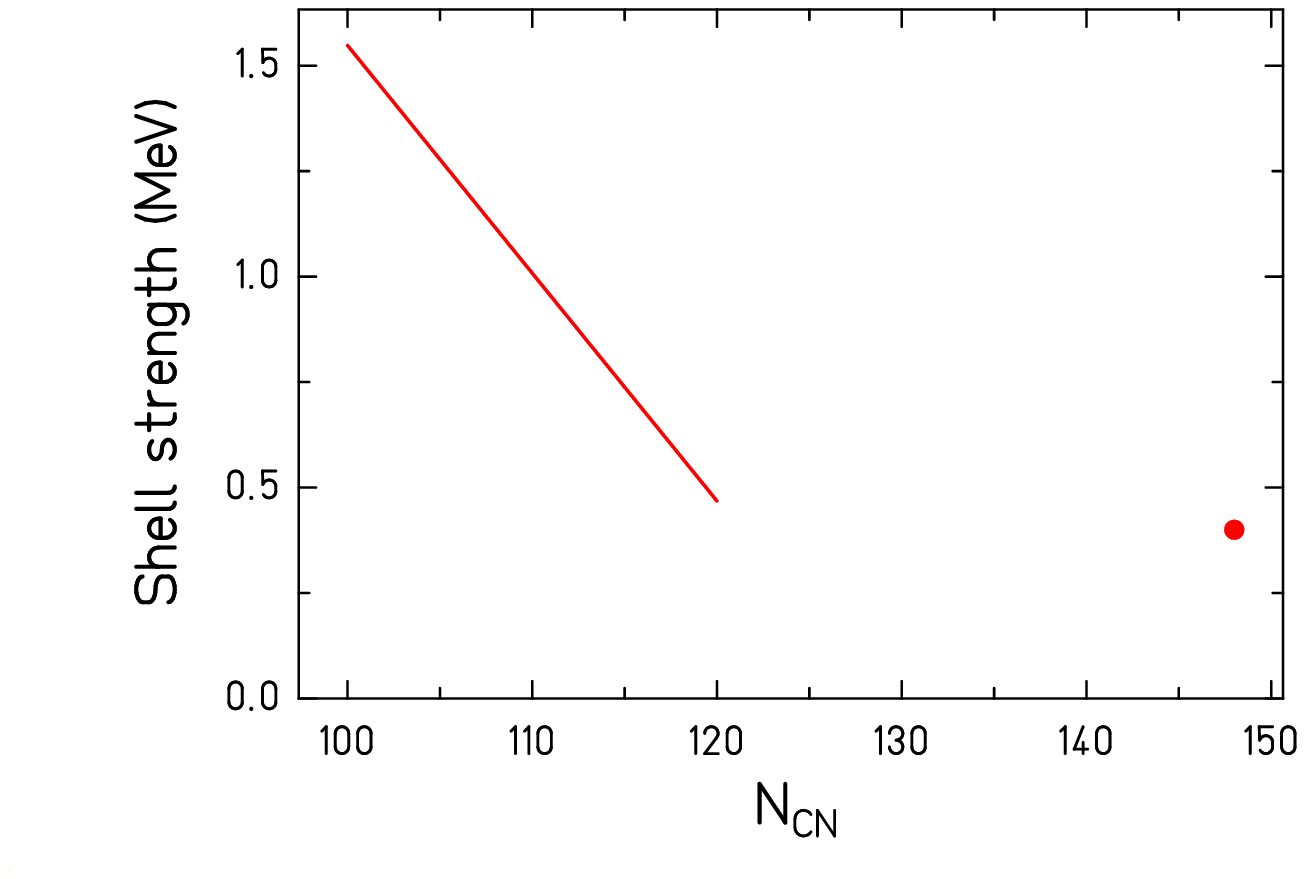}
\caption{(Color online) Line: Strength of the shell that is responsible in GEF for 
the asymmetric fission valley of the systems shown in figures \ref{BDF} and
\ref{Tl201-all}. The full symbol denotes the strength of the shell in the light fragment
that enhances the S1 fission channel in the Pu isotopes. See text for details.
}
\label{Shell} 
\end{center}
\end{figure}

In order to reproduce the data, the strength of this shell was
supposed to vary as a function of the number of neutrons in the fissioning nucleus
$N_{CN}$ as shown in figure \ref{Shell}. In addition, figure
\ref{Shell} shows the strength of the shell in the light fragment that is assumed to enhance
the S1 yield in the plutonium isotopes. 
The value of 0.4 MeV is only little below the value found for $^{201}$Tl, suggesting that the linear variation of the shell strength, represented by the straight line in figure \ref{Shell} 
cannot simply be extrapolated to larger values of $N_{CN}$. 

The mass distribution after proton-induced fission of $^{201}$Tl has been carefully measured as
a function of excitation energy \cite{Itkis88}. Figure \ref{Tl201-all} shows that this variation is reasonably well
reproduced by the GEF code with the available formalism without any specific adjustments.

We conclude that the characteristics of the mass distributions in the lead region,
as far as measurements exist, are compatible with
the appearance of a fission valley (in proton number of the light fragment) that varies smoothly in strength and position with the neutron number of the fissioning system, if one disregards the
substantial deviations found between GEF results and the measured mass distribution
of $^{202}$Rn, see figure \ref{BDF}. Unfortunately, the statistical uncertainties for this
system are rather large. 

More data from future experiments will allow to verify the tentative description of the
structural effects in the fission-fragment distributions of nuclei in the lead region 
deduced in this section. In any case, this analysis revealed some prominent systematic
trends in the available data.  
%The new description for fissioning systems in the lead region is rather similar to the one applied in the
%GEF model for the modeling of the fission-channels in the actinide region.
%More experimental data are required in order to verify this idea.
%If confirmed, the physics behind this finding remains a puzzle at this moment. 

\clearpage

\subsubsection{Heat transport between nascent fragments} \label{4-3-6}

The transformation of energy between potential 
energy, intrinsic and collective excitations as well as kinetic
energy is a very important aspect of the 
nuclear-fission process. It determines the partition of the fission Q
value (plus eventually the initial excitation energy of the
fissioning system) between kinetic and excitation energy
of the final fragments. Moreover, the division of the total
excitation energy between the fragments is of considerable
interest, because it strongly influences the number of
prompt neutrons emitted from the fragments. Thus, it also induces 
a shift towards less neutron-rich isotopes.

\paragraph{Dissipation on the fission path:}

As mentioned above, 
the description of dissipation in the fission process,
in particular in low-energy fission, where pairing correlations
play an important role, still poses severe problems to theory.
In the range of pairing correlations, 
Bernard et al.~\cite{Bernard11} developed the 
Schr\"odinger Collective Intrinsic Model for describing the 
coupling between collective and intrinsic two-quasi-particle excitations 
on the fission path in an extended version of the generator coordinate method.
Tanimura et al.~\cite{Tanimura15} recently observed 
deviations from the adiabatic limit of the microscopic
transport theory by single-particle levels crossing
in the vicinity of the fission barrier.
Once the nascent fragments
acquire their individual properties, the single-particle levels 
stay approximately parallel, and the process is essentially adiabatic. 
Shortly before scission, one-body dissipation becomes stronger due 
to the fast shape changes connected to the neck rupture. 

It is expected that the effects on the fission observables from these two
processes are rather different. Because the relaxation time of intrinsic
excitations is short compared to the estimated saddle-to scission 
time\footnote{The characteristic time for the thermalization of the
intrinsic excitation energy of a nucleus is a few times the time a 
nucleon needs to travel over
the diameter of the nucleus with the Fermi velocity. This is in the order 
of a few times 10$^{-22}$ s.
The estimated saddle-to-scission time is appreciably longer, about 10$^{-20}$ s
\cite{Davies76} or even longer \cite{Bulgac16}. 
}, one may 
assume that the induced nucleonic excitations are, on the average, equally distributed 
over all intrinsic degrees of freedom of the fissioning system when it
reaches the scission configuration.
%\footnote{The coupling between collective
%and intrinsic degrees of freedom might not be strong enough to establish 
%statistical equilibrium of intrinsic and collective excitations \cite{Noerenberg74}.} 
The dissipation near scission, however, occurs so shortly before neck rupture
that the equilibration of the energy may be expected to happen
to a great part after scission, where exchange between the fragments is
inhibited.  

The dissipated energy is fed by the potential-energy difference between
outer barrier and scission for which Asghar and Hasse derived a general
estimation \cite{Asghar84}. This energy difference gives an upper limit
of the dissipated energy, because it is shared by
intrinsic excitations, collective excitations and pre-scission 
kinetic energy.

\paragraph{Statistical properties of the nascent fragments:}

As already mentioned, we assume that the intrinsic excitation energy,
consisting of the intrinsic excitation energy above the outer barrier 
plus the energy dissipated in the region of strong level crossing
behind the barrier is, averaged over many fission events, equally 
distributed over all intrinsic degrees of freedom of the
fissioning system when it reaches scission. The division of this
excitation energy $E_{tot}$ among the nascent fragments in statistical
equilibrium is determined by the number $N$ of states available in the
two nuclei. 
Thus, the distribution of excitation energy $E_1$ of one fragment before
neck rupture is calculated by the statistical weight
of the number $N$ of states with a certain division of excitation energy between the fragments
\begin{equation}
   \frac{dN}{dE_1} \propto \rho_1(E_1) \cdot \rho_2(E_{tot}-E_1).
\label{DEV-EQ28}
\end{equation}
Note that $\rho_1$ and $\rho_2$ are the level densities of the fragments in their shape
just before scission, not in their ground-state shape! 
The remaining energy $E_{tot} - E_1$ is taken by the other fragment. 

There exist several analytical level-density descriptions, 
e.g. refs.~\cite{Gilbert65,Ignatyuk75,Svirin06,Capote09,Egidy09}, and, recently, also a few
microscopic calculations \cite{Houke09,Hilaire12,Bonett13}.
In the present context, we are not interested in describing the peculiarities 
of specific systems, but to understand the main thermodynamical properties 
of a nucleus that determine the average behavior of the energy
division between the nascent fragments. For this purpose, using a global
level-density description is better suited and more transparent than using
individual results of microscopic models for specific nuclei. 
General investigations of the validity of recommended parametrizations can be found in 
refs.~\cite{Svirin06,Koning08,Capote09}. These were rather oriented in benchmarking
the level-density descriptions against empirical data derived from level counting,
neutron resonances and evaporation spectra. However, in Ref.~\cite{Schmidt12} it was
pointed out that many level-density descriptions violate basic theoretical 
requirements, in particular in the low-energy range where pairing correlations play
an important role. 
These violations are often not easily recognized or checked by a comparison
with experimental data due to the incompleteness and uncertainty of the data, but they can
be important when considering the evolution of nuclear properties in terms of 
statistical mechanics.

\begin{figure} [h]  % top
\begin{center}
\includegraphics[width=0.5\textwidth]{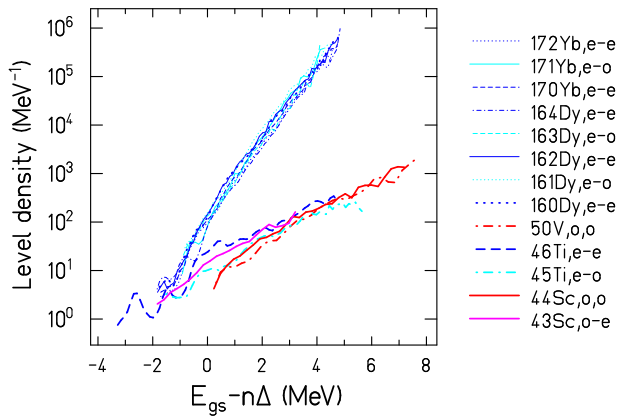}
\caption{(Color online)
Experimental level densities of various nuclei in a reduced
excitation-energy scale $U = E_{gs} - n \cdot \Delta (\Delta = 12 / \sqrt{A})$. 
The excitation energy above the ground state $E_{gs}$
is reduced by $2 \cdot \Delta$ $(n = 2)$ for even-even (e-e) nuclei, by $\Delta$ $(n = 1)$ for even-odd (e-o) or
odd-even (o-e) nuclei and left unchanged $(n = 0)$ for odd-odd (o-o) nuclei.
The line colors characterize the different categories (ee,eo,oe,oo), and the line width
is specific to the lighter or the heavier group of nuclei.
The figure is modified from Ref.~\protect\cite{Jurado15}, where also the references of the data can be found.}
\label{Fig-levdens-cT}
\end{center}
\end{figure}

The requirements, formulated in \cite{Schmidt12}, are:

\begin{itemize}
\item
The level density below the critical pairing energy \cite{Decowski68}, the excitation energy
where pairing correlations disappear\footnote{Strictly speaking, this transition
is not sharp due to the small size of the nucleus \cite{Moretto72}.},
is characterized by an approximately
exponential function, corresponding to a constant temperature. This is qualitatively
explained by the phase transition from super-fluidity to a Fermi gas that stores 
any additional energy in creating additional degrees of freedom by quasi-particle 
excitations instead of an increasing temperature. In addition to the empirical evidence,
for example 
from experiments performed at the Oslo Cyclotron Laboratory \cite{Guttermson15}, 
a theoretical justification 
on the basis of the BCS approximation was given recently by Moretto et al.~\cite{Moretto15},
where the thermodynamical nuclear properties are considered as a function of excitation
energy instead of the temperature as usually done before, e.g.~\cite{Moretto72}. 
There are even experimental indications \cite{Voinov09} and theoretical considerations
\cite{Dinh17}
that the constant-temperature behavior extends up to
an excitation energy of 20 MeV.  
Empirical constant-temperature parametrizations, e.g. Ref.~\cite{Egidy09} or Ref.~\cite{Ignatyuk01}, 
represent the level density in this energy range rather well.

\item
The level densities of neighboring even-even, odd-mass and odd-odd nuclei are essentially 
identical, except the gradual systematic dependence on the mass and the variation of the 
shell effect,
in a reduced energy scale $U$ where the excitation energy above the ground state $E_{gs}$ is
shifted to exactly eliminate the odd-even staggering of the binding energies.
%\footnote{An absolute energy scale that includes the ground-state mass would 
%lead to the same result.}.
The main differences are the additional levels below the pairing energy
$\Delta$ or $2 \cdot \Delta$ in odd-mass and even-even nuclei, respectively, 
if compared to odd-odd nuclei.
This feature has already been described by Strutinsky \cite{Strutinskii58} with an
analytical solution of the pairing problem in a Boltzmann gas. (See also figure 9 
of Ref.~\cite{Steinhaeuser98}.) It is stressed again
in Ref.~\cite{Moretto15}. 
As can be seen in figure \ref{Fig-levdens-cT}, the experimental level densities obtained
with the Oslo method fulfill fairly well this requirement: The level densities of
neighboring nuclei tend to converge when the excitation energy is shifted accordingly,
and the systematic even-odd staggering of the level density is removed.
The remaining deviations are not very important in view of the strong systematic
mass dependence of the inverse logarithmic slope. They are discussed in more detail below.
 
\item 
The level density above the critical pairing energy is well represented by the 
Bethe formula of independent fermions \cite{Bethe36}, which is also known as the
Fermi-gas level density, with an energy shift by the pairing condensation energy 
with respect to the ground state\footnote{The pairing condensation energy is
defined as the energy difference between the (hypothetical) ground state without the pairing
effect and that with the pairing effect \cite{Decowski68}.}. 
This energy shift includes an odd-even staggering 
that eliminates the odd-even staggering of the nuclear binding energy
(see previous point). In addition, the collective enhancement \cite{Bjornholm74}
 is considered
by the application of an appropriate factor. Shell effects can additionally be taken into account, for
example by the analytical formula of Ignatyuk et al.~\cite{Ignatyuk75}.
\end{itemize}

The resulting level-density description, proposed in Ref.~\cite{Schmidt12} resembles the 
composite level-density formula of Gilbert and Cameron \cite{Gilbert65}, however with
an increased matching energy in the order of 10 MeV, see figure 1 of Ref.~\cite{Schmidt12}. 
This value of the matching energy, which can also be interpreted as the critical pairing energy, is in good agreement with results of an analysis of measured angular distributions of fission fragments \cite{Itkis73} and energy-dependent fission probabilities \cite{Ignatyuk75}.

\paragraph{Energy sorting:}

From the previous discussion, we conclude that a fissioning nucleus on the way to scission
develops from a mono-nucleus to a di-nuclear system, where two nascent fragments acquire 
their individual thermodynamical properties well before scission. Because they are still
connected by a neck, they can exchange nucleons and excitation energy \cite{Schmidt11}.
At moderate excitation energies, the two nascent fragments form a rather peculiar system:
They act like microscopic thermostats. Each fragment can be considered as a heat bath of constant temperature
for the other fragment, although the system has a rather small fixed number of particles and
a rather low fixed amount of total energy. Disregarding shell effects, the nuclear temperature
in the constant-temperature regime
decreases systematically with the fragment mass: $T \propto A^{-2/3}$ \cite{Egidy09}.

Figure \ref{Fig-levdens-cT} illustrates the variation of the logarithmic
slope of the level density of the fission fragments as a function of mass in the reduced
energy scale. The nuclei are situated in the extreme light, respectively heavy, wings of the fission-fragment distributions in order to clearly show the effect.
This indicates 
that the light fragment has a systematically higher temperature than the heavy one.
Accordimg to an estimate on the basis of the empirical level-density description of Egidy et al. \cite{Egidy09} (which considers the systematic variation of the temperature with the nuclear mass and the influence of shell effects)
and using shell effects at scission from the fit parameters of GEF, 
the influence of shell effects may inverse this tendency only in nearly symmetric splits.
This may eventually be the case for the S1 fission channel in the actinides. 
The few measured mass-dependent neutron multiplicities (e.g. \cite{Mueller84,Naqvi86}) do not show any indication for such a reversed energy sorting, but, in any case, it would be difficult to be observed due to the small yield of the S1 fission channel in these systems.

As a consequence, the intrinsic energy (heat) tends
to flow to the heavier nascent fragment that has the lower temperature. This process
of energy sorting has first been described in Ref.~\cite{Schmidt10}.
The process of heat exchange proceeds in rather large and fluctuating steps
that lead to an averaging of the level densities and make possible the 
application of thermodynamical concepts \cite{Schmidt11}. 
This averaging significantly smoothes out the fluctuations in the level density present
at the lowest excitation energies, which are caused by the first quasi-particle excitations.
With increasing initial excitation energy, the fragments enter the Fermi-gas domain, and
the energy sorting gradually disappears \cite{Schmidt11a}. Asymptotically, at high
excitation energy, the heat is shared by the fragments in proportion to their masses.

\paragraph{Prompt-neutron yields:}

There are several observables that provide information on the energetics of the
fission process. The total kinetic energy of the fragments, and the
energy spectra and multiplicities of prompt neutrons and prompt gammas are the most
prominent ones.
Among those, the prompt-neutron
multiplicity gives the most direct and the most detailed information, because the
emitted neutrons
can individually be attributed to a specific fragment by their kinematical properties
in a moving-source fit \cite{Svirin08}. 
Moreover, neutron 
evaporation is by far the most probable decay channel, when the excitation energy
exceeds the neutron separation energy. Thus,  
the excitation energy of a specific
fragment is given to a good approximation by the sum of the neutron separation
energies and the mean neutron kinetic energies, which can be estimated rather
reliably, plus an offset, i.~e.~an amount of energy that ends up in prompt-gamma emission.
This offset is rather independent of
the initial excitation energy of the fragment. It amounts
to about half the neutron separation energy of the final fission product plus
the rotational energy, the fragment still has after prompt-neutron emission.

\begin{figure} [h]
\begin{center}
\includegraphics[width=0.4\textwidth]{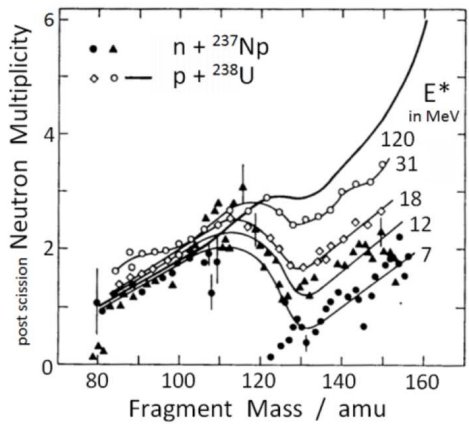}
\caption{Post-scission mean prompt-neutron multiplicities as a function
of fragment mass for different systems. 
The lines are drawn to guide the eye.
The initial excitation energy of the fissioning system is listed.
The figure is taken from \protect\cite{Hilscher92}.
}
\label{Fig-Hilscher}
\end{center}
\end{figure}

More than 40 years ago, the measurement of prompt-neutron multiplicities was already a 
subject of great interest, see for example \cite{Britt64,Cheifetz68,Cheifetz70,Burnett70,Bishop70}. 
Several experiments were performed to determine
the mass-dependent average neutron multiplicity as a function of the initial
excitation energy of the fissioning system. Figure \ref{Fig-Hilscher} shows  
this kind of data for neutron- and proton-induced fission of $^{237}$Np and $^{238}$U, respectively, for different
excitation energies. It should be noted that the observed events from proton-induced fission 
sum up from different fission chances. That means that the excitation-energy
distribution at the saddle deformation reaches from the initial excitation energy down to energies
in the vicinity of the fission barrier.

%\begin{figure} [th]  % top
%\begin{center}
%\includegraphics[width=0.5\textwidth]{Wahl-nu-bar.eps}
%\caption{Evaluation of measured prompt-neutron multiplicities as a function 
%of fragment mass for proton- and neutron-induced fission of $^{238}$U \cite{Wahl08}.
%Explanation of symbols:  $^{238}$U + p, $E_p$ = 30 MeV (o), $E_p$ = 50 MeV ( 
%$\bigtriangleup$), $E_p$ = 85 MeV ($\times$); $^{238}$U(n$_{fast}$,f) (line).
%The figure is reproduced from Ref.~\cite{Wahl08} with permission by the IAEA.}
%\label{Fig-Wahl}
%\end{center}
%\end{figure}

The curve for the fast-neutron-induced fission of $^{238}$U in figure \ref{Fig-Hilscher} 
shows a saw-tooth behavior that is typical for the 
prompt-neutron multiplicities in the actinides. An explanation in terms of
fragment shells that determine the deformation of the fragments at scission
was given by Wilkins et al.~in their scission-point model \cite{Wilkins76}:
As a result of their Strutinsky-type calculations, 
the energetically favorable deformation of the light and the heavy fragments 
increases with the mass of the fragment. This deformation energy is thermalized
after scission and feeds the evaporation of prompt neutrons. The minimum around $A = 130$
is attributed to fragments near the doubly magic spherical $^{132}$Sn.
Another salient feature of these data is that almost all additional energy
induced by an increasing incoming-particle energy ends up in the heavy fragment. 
This feature remained unexplained, in spite of many attempts. The discovery of
energy sorting provides a convincing explanation for the 
transport of essentially all additional excitation energy that is brought into 
the system to the heavy fragment. 
Similar features are found in mass-dependent prompt-neutron multiplicities of
proton-induced fission of $^{242}$Pu with $E_p$ between 13 and 55 MeV \cite{Kniajeva04}.
Previous model calculations could not
reproduce these data, because the division of excitation energy
at scission was estimated on the basis of the Fermi-gas level density. 
In these calculations, the
particularities at energies below the critical pairing energy due to pairing
correlations where not considered.
%\footnote{

\begin{figure} [h]  % top
\begin{center}
\includegraphics[width=0.45\textwidth]{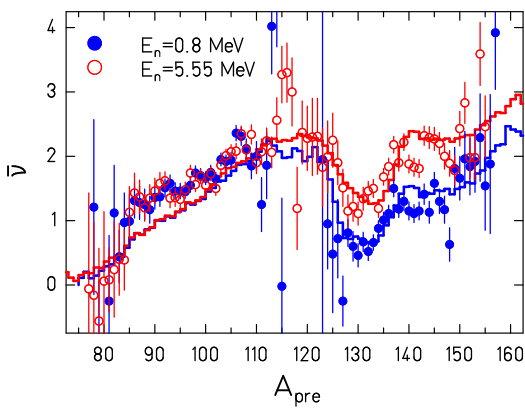}
\caption{(Color online)
Measured prompt-neutron multiplicity in $^{237}$Np(n,f) as a function of 
pre-neutron fragment mass at two different incident-neutron energies \protect\cite{Naqvi86} (data points) in
comparison with the result of the GEF model \protect\cite{Schmidt16} (histograms).
The figure is taken from Ref.~\protect\cite{JEFF24}.}
\label{Fig-Naqvi}
\end{center}
\end{figure}

Tudora and collaborators published a detailed
investigation on the mass-dependent prompt-neutron multiplicities
as a function of excitation energy like those shown in figures \ref{Fig-Hilscher} and 
\ref{Fig-Naqvi} and succeeded to reproduce the observed features by the Point-by-Point method, without invoking 
the energy-sorting mechanism \cite{Tudora15,Tudora16}. We are not convinced about the reliability
of these investigations, because they contain several inconsistencies. We mention only three:
(i) The ground-state shell effects were used for describing
the level density in the nascent fragments at scission, although the shell effects
are very sensitive to the shape of the nucleus \cite{Nilsson55,Brack72}. 
(ii) The application of the Fermi-gas
level-density formula at scission implies that pairing correlations are absent \cite{Schmidt12}. 
This contradicts the observation of a pronounced odd-even effect in 
fission-fragment $Z$ distributions \cite{Caamano11,Jurado15}.
(iii) In the cases considered in \cite{Tudora15}, the assumed intrinsic excitation energy 
at scission exceeds the available energy, which is the sum of the excitation energy at the fission 
barrier plus the estimated potential-energy lowering from saddle to scission \cite{Asghar84}.
%}  

The quantitative estimation of the mass-dependent prompt-neutron multiplicities
in proton-induced fission is complicated by the contributions from multi-chance fission. 
Rather accurate experiments on mass-dependent prompt-neutron multiplicities below 
the threshold for second-chance fission were performed by M\"uller et al. 
\cite{Mueller84} and Naqvi et al.~\cite{Naqvi86} with incident neutrons of different 
energies. The data of Ref.~\cite{Naqvi86} are compared in figure~\ref{Fig-Naqvi} with
a calculation performed with the GEF code that considers the constant-temperature
behavior of the level density in the range of pairing correlations 
\cite{Schmidt12}. Due to energy sorting, the prompt-neutron multiplicity in the light
fission-fragment group remains the same, in spite of an increase of the initial
energy by almost 5 MeV.

%Recently, several attempts to estimate the division of excitation energy between
%the fragments have also been performed with microscopic self-consistent approaches ... 

We conclude that the application of statistical mechanics is an efficient way 
to handle the division of excitation energy between the nascent
fragments on the fission path. The results reproduce the experimental data on
energy sorting with a good accuracy. Unfortunately, high-quality data of this 
type are still scarce.

\paragraph{Odd-even effect in fission-fragment yields:}

The odd-even effect in fission-fragment distributions, i. e. the enhanced production
of fission fragments with an even number of protons and/or neutrons, 
is one of the most prominent manifestations of nuclear structure. 
This phenomenon can be studied in analogy to the energy sorting, described above:
Also with respect to pairing correlations, the individual fragment properties
are assumed to be established well before scission \cite{Krappe01}, and
statistical equilibrium before scission may be assumed.

As already mentioned,
the nuclear level densities considered on an absolute energy scale evolve 
smoothly without any noticeable odd-even 
staggering as a function of the number of protons
and neutrons, except the appearance of additional levels compared to 
odd-odd nuclei, with a fully
paired configuration in even-even nuclei and with fully paired configurations
in the proton respectively neutron subsystem in odd-$A$ nuclei, see figure \ref{Fig-levdens-cT}. 
Therefore, in a statistical consideration the appearance of odd-even staggering in fission yields must be connected
in some way with these fully paired configurations. 
Indeed, at reduced energies above the ground-state level of odd-odd nuclei,
the statistical weight of excited states (see equation \ref{DEV-EQ28})  is equal in all classes of nuclei (even-even, even-odd,
odd-even and odd-odd), if the smooth mass dependence is taken out. Also the number of available 
states in even-odd and odd-even nuclei above the ground-state level of odd-A nuclei is the same. That means that
the overproduction of fragments with even number of protons can be traced back to even-even
light fragments that are formed fully paired at scission when statistical equilibrium is assumed.

The odd-even effect in fission-fragment proton or neutron number before neck rupture
can quantitatively be calculated by the statistical weight of configurations with even and odd
numbers of protons, respectively neutrons, in the nascent fragments.

\begin{figure} [h]
\begin{center}
\includegraphics[width=0.5\textwidth]{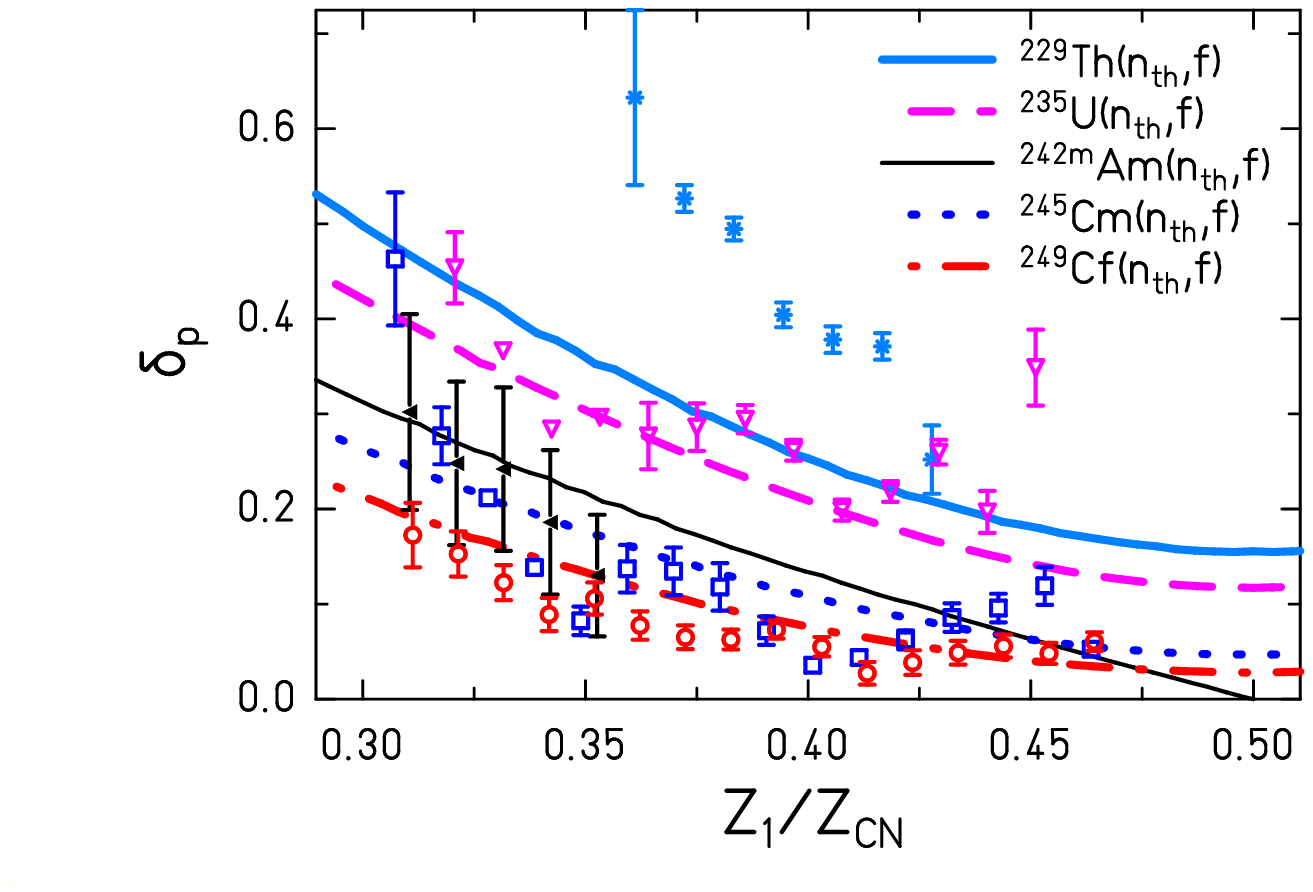}
\caption{
Local logarithmic four-point difference $\delta_p$ of the fission-fragment $Z$ 
distributions as a function of asymmetry, represented by the ratio of the
nuclear charge of the light fragment $Z_1$ and the nuclear charge of the fissioning nucleus $Z_{CN}$. 
The symbols show experimental data from the compilation of Ref.~\protect\cite{Caamano11} 
and denote the target nuclei: $^{229}$Th
(stars), $^{235}$U (open triangles), $^{242}$Am (full triangles), 
$^{245}$Cm (open squares), $^{249}$Cf (open circles).
The lines correspond to the results of the model of Ref.~\protect\cite{Jurado15} described in the text.
The figure is taken from Ref.~\protect\cite{Jurado15}.
}
\label{Fig-even-odd}
\end{center}
\end{figure}

A schematic model following these ideas has been developed in Ref.~\cite{Jurado15}. 
See also Ref.~\cite{Schmidt16} for the implementation in the GEF code. 
For an even-even fissioning nucleus, the number of configurations with fragment charge $Z_1$ even, which implies
that also the charge of the complementary fragment $Z_2$ is even,
at fixed total reduced energy $U_{tot}$ is given by:
\begin{equation}
 N_{Z_1=e}^{ee}(Z_1) = \int \limits _{-2 \Delta_1} ^{U_{tot}+2 \Delta_2} 
   \rho_1 (U_1)_{(ee)} \rho_2(U_{tot}-U_1)_{(ee)} dU_1 + 
\end{equation}
\begin{eqnarray*}  
   \int \limits _{- \Delta_1} ^{U_{tot}+\Delta_2} \rho_1(U_1)_{(eo)} \rho_2(U_{tot}-U_1)_{(eo)} dU_1
\end{eqnarray*}
where $\rho_i(U_i)_{(ee)}$ and $\rho_i(U_i)_{(eo)}$ are the level densities of 
even-even and even-odd fragments, respectively.
The mass numbers $A_1$ and $A_2$ of the two fragments (used to calculate the level density) are the closest integer numbers (even or odd for even-even or even-odd fragments, respectively) to the values calculated from $Z_1$ and $Z_2$ 
by the UCD assumption.
The reduced energy $U$ is shifted with respect to the excitation energy $E$ 
available in the two nascent fragments 
$U = E - n \Delta$, $n = 0,1,2$ for odd-odd, odd-mass, and even-even fragments, respectively.
This ensures the use of a common energy scale 
%without odd-even staggering 
in the frame of the fissioning system, 
which is a basic requirement for the application of statistical mechanics.
Long-range variations of the total available excitation energy as a function of mass asymmetry 
\cite{Moeller14}
are neglected in the schematic model presented here. They might be additionally considered in 
a more refined model.

The number of configurations with $Z_1$ odd, which implies that also $Z_2$ is odd, for an even-even fissioning nucleus is:
\begin{equation}
 N_{Z_1=o}^{ee} (Z_1) = \int \limits _{-\Delta_1} ^{U_{tot}+\Delta_2} 
   \rho_1(U_1)_{(oe)} \rho_2(U_{tot}-U_1)_{(oe)} dU_1 +
\end{equation}
\begin{eqnarray*}
  \int \limits _{0} ^{U_{tot}} \rho_1(U_1)_{(oo)} \rho_2(U_{tot}-U_1)_{(oo)} dU_1
\end{eqnarray*}
where $\rho_i(U_i)_{(oe)}$ and $\rho_i(U_i)_{(oo)}$ are the level densities of representative
odd-even and odd-odd nuclei, respectively, with mass close to $A_1$ or $A_2$. 
The mass numbers $A_1$ and $A_2$ of the two fragments are again related to $Z_1$ and $Z_2$ 
by the UCD assumption.
The yield
for even-$Z_1$ nuclei is $Y_{Z_1=e}^{ee}(Z_1) = N_{Z_1=e}^{ee}(Z_1)/N_{tot}^{ee}(Z_1)$ with
$N_{tot}^{ee}(Z_1) = N_{Z_1=e}^{ee}(Z_1) + N_{Z_1=o}^{ee} (Z_1)$. 
Similar equations hold for odd-even, even-odd and odd-odd fissioning systems.
The total available reduced intrinsic excitation energy $U_{tot}$ is assumed to be a fraction
of the potential-energy difference from saddle to scission plus the initial excitation energy 
above the barrier \cite{Asghar84}. Thus, it also increases with the
Coulomb parameter $Z^2/A^{1/3}$ of the fissioning nucleus. 

The result of these considerations is that the odd-even effect in 
fission-fragment $Z$ distributions is caused by the statistical weight of
configurations with a concentration of all
intrinsic excitation energy and unpaired nucleons in the heavy fragment and the formation
of a light fragment in a fully paired state.

This approach reproduces the observed salient features of the proton odd-even effect \cite{Caamano11}: (i) The global
odd-even effect 
($ \Sigma Y_{Z=e} -  \Sigma Y_{Z=o} $) / ($ \Sigma Y_{Z=e} +  \Sigma Y_{Z=o} $) 
decreases with the Coulomb parameter and with increasing initial
excitation energy. (ii) The local odd-even effect, represented by the 
logarithmic four-point differences,
\newline
$\delta_p(Z + 3/2) = 1/8 (-1)^{Z+1} ( \ln Y(Z+3) - \ln Y(Z) - 3 [\ln Y(Z+2) - \ln Y(Z+1) ] ) $,
\newline
 increases towards mass asymmetry.
(iii) The local odd-even effect for odd-$Z$ fissioning nuclei is zero at mass symmetry and
approaches the value of even-$Z$ nuclei for large mass asymmetry.
As shown in figure \ref{Fig-even-odd},
the quantitative reproduction is satisfactory, except for the system $^{229}$Th(n$_{th}$,f).
The disagreement found for this system may be caused by the neglect of fluctuations 
in the dissipated energy. In fact, for a great
part of the fission events the available energy of this system may be so low that they reach the scission point
in a completely paired configuration due to the threshold character of the first quasi-particle
excitation.

%According to these ideas, 
%the odd-even effect in fission-fragment $Z$ distribution 
%is a sign of an extreme energy-sorting process as discussed in Ref.~\cite{Jurado15}. 
%It documents the entropy-driven 
%enhanced production of a cold even-even fully paired light fragment in its ground state.  

It is expected that the same ideas are valid for the odd-even effect in fission-fragment 
$N$ distributions at scission, that means before the emission of prompt neutrons. However,
in the measured number of neutrons in post-neutron fission fragments this initial
odd-even effect is masked by the influence of the neutron-evaporation process, which imposes its own odd-even fluctuations \cite{Ricciardi04,Ricciardi11,Mei14}.
This idea explains, why the measured
values of $\delta_N$ for electromagnetic and neutron-induced fission are very similar, see
figure \ref{N-EO}.
It was successfully implemented in the SPACS code for calculating the 
nuclide yields of spallation residues \cite{Schmitt14}. 

We would like to stress that the model of Ref.~\cite{Jurado15} does not include the effect of the neck rupture. 
By many authors (see Ref.~\cite{Goennenwein14}), it is advocated that any
production of odd-$Z$ fragments starting from fully paired configurations at saddle   
is exclusively caused by pair breaking during the fast shape changes connected with the rupture
of the neck. 
However, no quantitative estimation has been proposed.
%while we reproduce the experimental data rather well with our quantitative calculation,.  
This would imply that the motion from saddle to the configuration before neck rupture is totally
adiabatic and that a sizable fraction of the unpaired nucleons, emerging from the quasi-particle 
excitations at neck rupture, end up in different fragments. These assumptions can only be 
verified by elaborate microscopic models. 
The result of Tanimura et al.~\cite{Tanimura15} from TD-EDF theory
seems to contradict the first assumption, because they obtained a sizable amount of dissipation
before scission in the region of many level crossings in the vicinity of the second barrier. 
The second assumption is not obvious neither to us, because we expect that the localization of the 
wave functions in the di-nuclear regime, discussed before, also leads to a localization of the 
pairing correlations in the two nascent fragments already before scission. 
The later transfer of single nucleons from one nascent fragment to the other one might be improbable during the short duration of the scission process.
Finding a valid answer to these questions is an important task for dynamical quantum-mechanical 
models. In any case, the complete sorting of the available intrinsic 
excitation energy, consisting of the initial excitation energy of the system
above the outer fission barrier and the energy, dissipated between outer barrier and scission, and its effect on the enhanced 
presence of an even number of protons and neutrons in the light pre-fragment is expected 
to well describe the situation of the system before neck rupture, when the excitation energies
of the fragments stay in the superfluid regime.

The odd-even effect in fission-fragment $Z$ distributions is one of the complex features of
nuclear fission that can only be fully understood by dynamical quantum-mechanical models.  
These models need to be further developed by simultaneously handling dissipation, 
statistical mechanics and quantum localization in a realistic way.
At present, the application of statistical models \cite{Jurado15} and considerations 
on the influence of dynamical processes at scission \cite{Goennenwein14} give an idea about 
the processes that are involved in the problem and that should be further studied.

\subsection{Other results of the GEF model} \label{4-4}

While the underlying theoretical ideas of the GEF model were presented in section \ref{4-3},
together with the rather directly related observables, the present section deals with
observables that are related to the basic ideas of GEF in a more complex way. 
The confrontation with experimental data tests, whether the interplay of the different
elementary processes implemented in GEF is able to describe the complex behavior
of the fission process in a realistic way in a more general sense. This section is not meant
to be exhaustive. For additional information we refer to the detailed description of the 
GEF model in \cite{Schmidt16} and other specific publications, e.g. \cite{Schmitt18}.

\subsubsection{Fission-fragment yields at higher excitation energies} \label{4-4-1}

\begin{figure} 
\begin{center}
\includegraphics[width=0.5\textwidth]{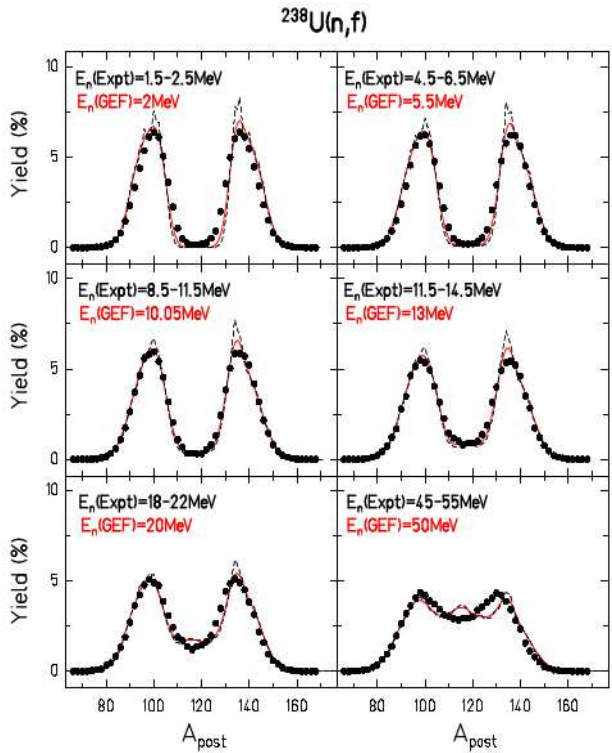}
\caption{(Color online)
Measured fission-fragment mass distributions from neutron-induced fission of $^{238}$U for
a series of incident-neutron energies \protect\cite{Zoller95} (black symbols) in comparison with
the GEF-model results without (black dashed line) and with (red full line) account of
experimental resolution. 
The neutron-beam-energy interval in the experiment as well as the energy used for GEF are indicated. The figure is taken from \protect\cite{Schmitt18}.
}
\label{Zoeller}
\end{center}
\end{figure}

Fission-fragment mass distributions from neutron-induced fission of $^{238}$U 
have been measured in the thesis of P. Z\"oller \cite{Zoller95} over a wide energy range. 
(Unfortunately, these data remained unpublished.) 
%The traditional double-energy method was applied, from which masses before and after prompt-neutron emission can be derived.
These data are shown in figure \ref{Zoeller} in comparison with results from GEF
calculations, which reproduce the data rather well. At the higher energies, 
multi-chance fission plays an important role.

\begin{figure} 
\begin{center}
\includegraphics[width=0.45\textwidth]{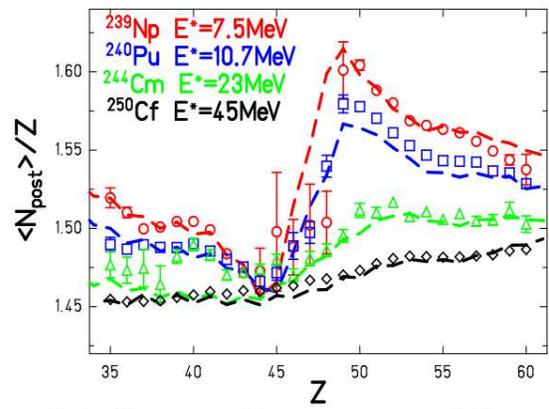}
\caption{(Color online)
Mean post-neutron fragment $N$/$Z$ ratio as function of fragment $Z$ for various fissioning systems as measured
at VAMOS (symbols) \protect\cite{Ramos17}, and compared with GEF calculations (dashed lines).
The figure is taken from \protect\cite{Schmitt18}.
}
\label{VAMOS-N-over-Z}
\end{center}
\end{figure}

However, for a full specification of the fission-fragment yields on the chart of the nuclides,
also the charge polarization (a different $N$/$Z$ ratio in the two fragments formed at scission)
and prompt-neutron emission, which induces a shift to less neutron-rich fragments, needs to be considered. 
(As already mentioned, fission-fragment yields before prompt-neutron emission, fully defined in $A$ and $Z$, are practically inaccessible to experiment.)
Data on the evolution of the $N$/$Z$ coordinate at higher excitation energies have emerged
only recently from the VAMOS experiment. Measured average numbers of neutrons after
prompt-neutron emission over $Z$ are shown in figure \ref{VAMOS-N-over-Z} and compared with
the result of GEF calculations for different fissioning systems at different excitation energies.
A pronounced structure above $Z$ = 45 is observed, which is gradually washed away with
increasing mass and increasing excitation energy of the system.
This structure is a combined effect of the charge polarization in the 
fission-fragment distribution before prompt-neutron emission, which favors the formation 
of neutron-rich heavy fragments in the S1 and the S2 fission channels, 
and the mass dependent saw-tooth shape of the prompt-neutron multiplicity.
For a detailed discussion see for example ref. \cite{Schmidt16}. 
The data are rather closely reproduced by the GEF model. Further investigations with the model
suggest that the washing out of the structure can be attributed almost exclusively to the
increasing excitation energy. 

\subsubsection{Total kinetic energy} \label{4-4-2}

\begin{figure} [h]
\begin{center}
\includegraphics[width=0.45\textwidth]{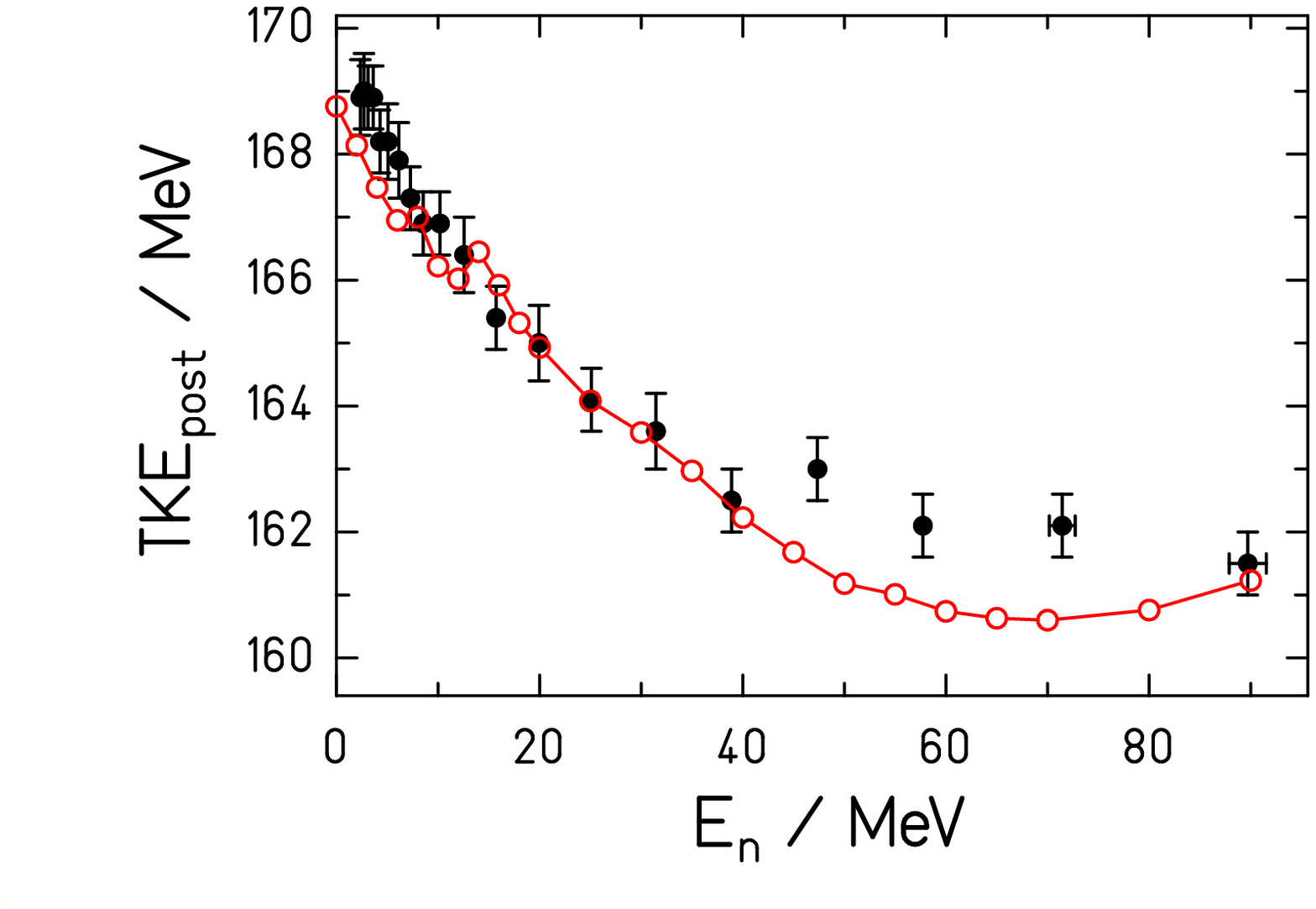}
\caption{(Color online)
Measured average total-kinetic energies after prompt-neutron emission from neutron-induced fission \protect\cite{Loveland17}
of $^{235}$U (black full symbols) in comparison with the result of the GEF code (red open symbols).
}
\label{Fig-Loveland-TKE}
\end{center}
\end{figure}

Although the total kinetic energy (TKE) is given by the available energy of the system,
that means by the sum of initial excitation energy and the Q value, minus the total
excitation energy of the fragments, when emission of particles and gamma radiation 
before scission is neglected, it contains additional information on the fission
process. In particular, the mean TKE, averaged over all partitions in $Z$ and $A$
also reflects the changes of the Q value due to variations of the fission-fragment
distribution. TKE values measured after prompt-neutron emission also reflect the total
corresponding mass loss of the system before and after scission. Thus, model predictions 
on the fragment yields can be tested also indirectly by other experiments than
direct measurements of the yields. 
This also shows again the interconnections and the correlations between the different 
fission quantities. 

Figure \ref{Fig-Loveland-TKE} shows recently measured average TKE values from neutron-induced fission
of $^{235}$U \cite{Loveland17} in comparison with the GEF code. These data are fairly well reproduced, although some deviations above $E_n$ = 40 MeV appear, whose origin is not clear.
Detailed investigations show that also the total prompt-neutron yields are well reproduced (they have been measured up to $E_n$ = 50 MeV \cite{Hyde64,Howe84}), 
which lead to a reduction of the fragment masses and, thus, explain part of the TKE decrease
at higher energies. Good agreement with the measured mass
distributions of $^{238}$U(n,f) \cite{Zoller95}, demonstrated in figure \ref{Zoeller}, indicates that also the fission-fragment mass distribution, 
which determines the average Q value, is well reproduced.

The kink of the calculated values in figure \ref{Fig-Loveland-TKE} at about $E_n$=14 MeV is related to multi-chance fission.
Here, 3rd-chance fission, that means fission of the compound nucleus $^{234}$U sets in. This leads
to a reduced yield of symmetric (with systematically lower TKE), and an enhanced yield 
of asymmetric fission (with systematically higher TKE). The nucleus $^{234}$U has
a large fission probability at an energy range somewhat above the fission barrier, because the fission barrier $B_f$ is lower than the neutron separation energy $S_n$.
This effect is weaker at the onset of 2nd-chance fission because in $^{235}$U we 
have $B_f > S_n$. This kind of structure disappears at higher energies, because 
it is washed out by the fluctuations of the kinetic energies of the pre-fission neutrons.

\subsubsection{Emission of prompt neutrons and gammas} \label{4-4-3}

\paragraph{Pre-scission neutrons:}
There are at least two sources of neutrons emitted before scission. The first 
pre-scission-neutron source is
caused by the competition of particle emission and fission of the initial excited
nucleus in the first well, eventually under the influence of transient effects \cite{Grange83}. Transient effects delay the onset of the quasistationary
probability flow over the fission barrier under the influence of dissipation after the formation of the fissioning system.
If the initial excitation energy is high enough, fission can occur from 
the initial nucleus and, after emission of mostly neutrons, from the respective daughter
nuclei. This phenomenon, called multi-chance fission, has influence on the fission
probabilities and the fission-fragment properties, as already discussed
in section \ref{4-3-4}, but it has also influence on the energy spectra, the angular distribution, and
the multiplicity of the prompt neutrons.
The second pre-scission-neutron source is fed by the evaporation of neutrons on the fission path, essentially beyond the outer fission saddle. 
This phenomenon has intensively been investigated for
determining the nuclear dissipation strength \cite{Hilscher92}. 
In an experiment, neutron emission before scission can be identified by a kinematical
multi-source fit \cite{Isaev08}, but the emission in the first well and the one beyond the fission
barrier cannot be distinguished by this method.

The dependency of the neutron emission between saddle and scission on the properties of the
fissioning systems is complex and not fully understood, see e.g. Ref. \cite{Shareef16}.
Therefore, in the GEF code an heuristic approach is used. Guided by the saturation of
the total multiplicity of the neutrons emitted from the fragments at about 4 neutrons found
in Ref.~\cite{Keutgen04}, saddle-to-scission
neutrons are assumed to be emitted as long as the excitation energy of the fissioning 
nucleus at scission exceeds 40 MeV. 
%Considering these neutrons 
%leads to a saturation of the multiplicity of the neutrons emitted from the fragments as
%a function of the initial excitation energy of the system.

\paragraph{Post-scission neutrons:}

After neck rupture, the fragments still interact by the Coulomb force,
which leads to an increase of the relative velocities, until
they reach their final kinetic energies. Only for a very short time, when
the gradient of the force across the fragments is still large, also
additional interactions may occur like Coulomb excitation and the generation 
of collective fragment angular momentum \cite{Hoffman64}. 
After this, the fragments can be considered separately as independent
nuclei, carrying intrinsic and collective excitations as well as a well
defined angular momentum. The fragment shape at scission that is influenced
by the interaction with the complementary fragment relaxes 
to the equilibrium shape, and the
gain in binding energy by the smaller deformation energy thermalizes and adds up to the initial 
intrinsic excitation energy. Also the excitation energy stored in collective
excitations transforms into intrinsic excitations, while preserving
the fragment angular momentum.

Thus, the fragments after scission can essentially be considered as
compound nuclei with well defined excitation energy and angular momentum.
They will undergo a statistical de-excitation process that can be 
described by a standard evaporation code. At sufficiently high excitation 
energies, some particle emission during the acceleration phase may
be needed to be considered, but most part of this process occurs from 
the fully accelerated fragments.

The characteristics of the emitted particles (mostly neutrons) and
of the gamma radiation are important by two reasons:
Firstly, they carry precious information on the evolution of the
fissioning system before scission. Secondly, they are important for
design and operation of nuclear-power reactors.

Therefore, a statistical-model code is implemented in the GEF code that
calculates the prompt neutrons and the prompt-gamma radiation emitted from
the fragments after scission. The kinematical properties of the neutrons
refer to the velocity of the emitting source. Isotropic emission in the
respective frame of the moving fragment is assumed. The angular momentum is assumed to be
conserved during the de-excitation process, because changes are expected
to be small due to the low angular-momentum values involved, where the
influences of the angular-momentum-dependent level density and of the
slope of the yrast line tend to compensate \cite{Huizenga60}. 
For details, see Ref.~\cite{Schmidt16}.

\subparagraph{\textit{Neutron multiplicities:}}

\begin{figure} [h]
\begin{center}
\includegraphics[width=0.5\textwidth]{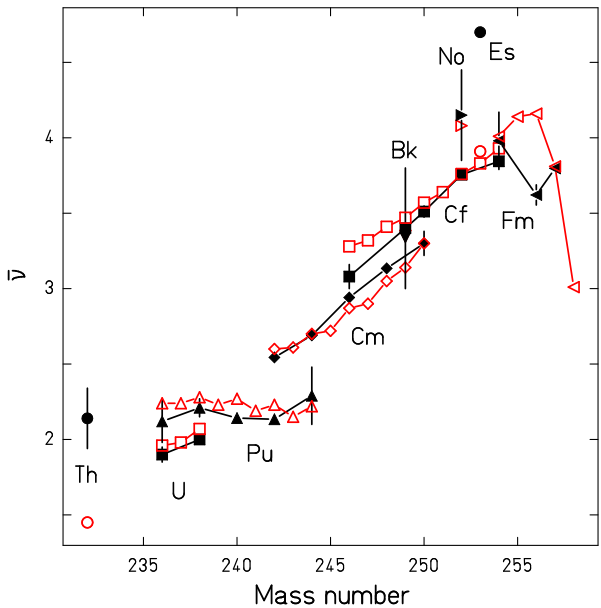}
\caption{(Color online)
Systematics of prompt-neutron multiplicities for spontaneous fission.
Measured mean prompt-neutron multiplicities for spontaneous fission (black full symbols)
as a function of the mass number of the fissioning nucleus \protect\cite{Malinovskij85}
in comparison with the result of the GEF model (red open symbols). 
Experimental error bars are not shown when they are smaller than the symbols.
The value for $^{253}$Es is reported without an experimental uncertainty.
The figure is taken from \protect\cite{Schmidt16}.}
\label{Fig-NMULT-1}
\end{center}
\end{figure}

%\begin{figure} [h]
%\begin{center}
%\includegraphics[width=0.6\textwidth]{NTH.EPS}
%\caption{(Color online)
%Systematics of prompt-neutron multiplicities for n$_{\mathrm{th}}$-induced fission.
%Measured mean prompt-neutron multiplicities for 
%thermal-neutron-induced fission
%as a function of the mass number of the target nucleus
%\cite{Mills95} (black full symbols), 
%\cite{Malinovskij85} (blue shaded symbols), 
%and \cite{Chadwick11} (green open symbols)
%in comparison with the result of the GEF model (red open symbols).
%We assumed that the value 3.132 for $^{232}$U given in 
%\cite{Malinovskij85} (blue open symbol) is wrong due to a misprint. The
%tentatively corrected value (2.132) is marked by a blue shaded symbol.  
%Experimental error bars are not shown when they are smaller than the symbols.
%}
%\label{Fig-NMULT-2}
%\end{center}
%\end{figure}

Direct measurements of post-scission prompt-neutron multiplicities are
difficult and subject to sizable uncertainties. Very precise values
have indirectly been deduced from integral and other kind of experiments.
Those, however, may deviate from the true values, because they are tuned
together with other not accurately known fission quantities to exactly 
reproduce the results of the indirect experiments. Therefore, 
results from direct experiments on prompt-neutron multiplicities in
spontaneous 
%and thermal-neutron-induced 
fission (sf) of several systems are
compared with the results of the GEF code in figure \ref{Fig-NMULT-1}.
% and \ref{Fig-NMULT-2}.
The data 
%for spontaneous fission 
are reproduced with a standard deviation of 0.1 units
%, while
%the deviations for the thermal-neutron-induced fission amount to about
%0.2 units. The difficulties due to the background of scattered neutrons 
%in experiments with thermal neutrons may contribute to the larger
%deviations in these cases. 
One can observe the sharp drop of the calculated
prompt-neutron multiplicities in the transition from $^{256}$Fm(sf) to $^{258}$Fm(sf).
Also the peculiar slope of the prompt-neutron multiplicities as a function of mass in the
Pu systems that differs from the average trend is a structural effect due to the strong and strongly varying
yield of the S1 fission channel in these systems. These features demonstrate
the importance of these data for the understanding of the fission process.

More results on the characteristics of prompt-neutron multiplicities are
shown in \cite{Schmidt16}.

\subparagraph{\textit{Neutron-energy spectra:}}

\begin{figure*}[h]
\centering
\includegraphics[width=1.0\textwidth]{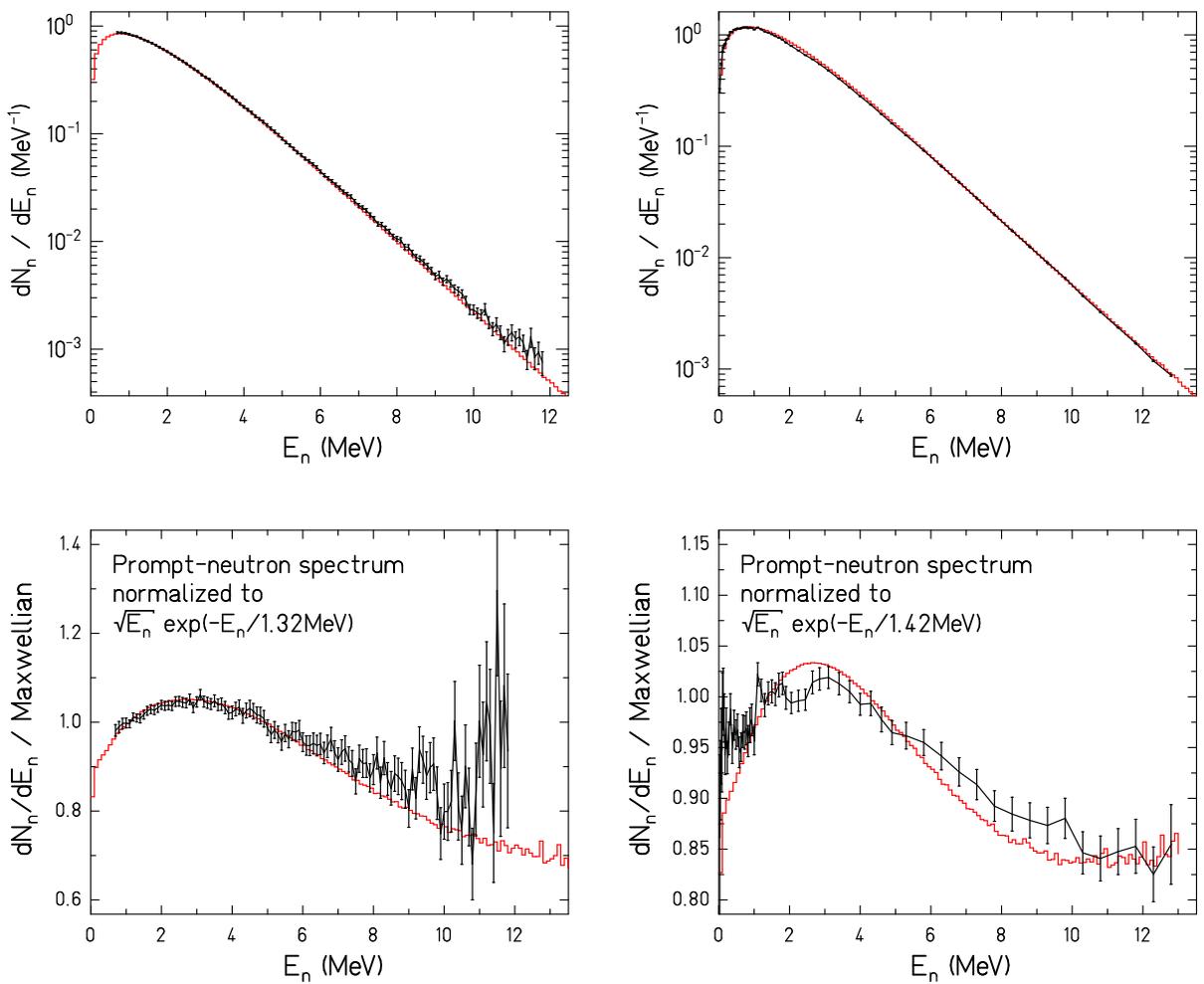}
\caption{Upper panels: Experimental prompt-fission-neutron energy spectra (black lines and error bars) for $^{235}$U(n$_{th}$,f) \protect\cite{Kornilov10}
(left part) and $^{252}$Cf(sf) \protect\cite{Mannhart89} (right part) in comparison with the result of the GEF model (red histograms) in logarithmic
scale. In the lower panels, the spectra have been normalized to a Maxwellian with $T$ = 1.32 MeV and $T$ = 1.42 MeV, respectively.
The figure is taken from ref.~\protect\cite{Schmidt17}.} 
\label{En}       % Give a unique label
\end{figure*}

In figure \ref{En}, the energy spectra of prompt post-scission neutrons are
compared with experimental data for $^{235}$U(n$_{th}$,f) and $^{252}$Cf(sf).
The GEF code establishes a strong correlation between the shape of the
spectrum (in particular its hardness) and the prompt-neutron
multiplicity. The  high-energy tail is sensitive to the yields
and the deformation of the heaviest fragments of the light group, because
they have the highest nuclear temperatures. The model parameters were slightly
tuned to simultaneously describe the measured mass yields, the mass-dependent prompt-neutron
multiplicities and the high-energy tails of the spectra for $^{235}$U(n$_{th}$,f) and $^{252}$Cf(sf) shown in figure \ref{En} without deteriorating the agreement with other fission observables. 
Due to the rather good agreement of the fragment properties at scission with
measured data provided by the GEF model, 
calculations with these parameters are expected to give also realistic
predictions for prompt-neutron spectra and multiplicities for other
systems in this region, also for those where no experimental data are 
available. 

\paragraph{Prompt-gamma emission:} 

As a weak decay channel, gamma radiation is emitted in competition with
neutron emission and fission from the excited nuclei from the beginning on, while it becomes the
dominant or finally the only decay channel when the excitation energy
falls below the neutron-separation energy and the fission barrier. 
The same is true for gamma radiation emitted from the fragments, although here
the fission branch is negligible.
Gamma radiation observed in coincidence with fission is largely dominated
by emission from the fragments after scission. Non-statistical gammas in coincidence
with fission are exclusively observed from the fragments.

\subparagraph{\textit{Statistical~gammas:}}

At energies sufficiently high above the yrast line, E1 radiation is 
dominant. It is essentially continuous and mostly governed by the
gamma strength of the giant dipole resonance.

\subparagraph{\textit{Non-statistical~gammas:}}

At energies close to the yrast line, distinct gamma lines appear, and,
finally, E2 radiation takes over that most efficiently removes the
fragment angular momentum.  
The transition energies are specific to the nucleus considered, and 
an accurate theoretical modeling is very demanding, e.g. \cite{Egido16}.
Therefore, if possible, tables of experimental spectroscopic data or, eventually,
empirical descriptions are often used, like the variable-momentum (VMI) model \cite{Mariscotti69}.
Compared to other advanced versions of the VMI model, e.g. \cite{Batra91},
the implementation of the VMI model in GEF \cite{Schmidt16} provides an improved modeling 
of the transition from rotational to vibrational behavior near closed shells, see \cite{Schmidt16}.

\begin{figure} [h]
\begin{center}
\includegraphics[width=0.5\textwidth]{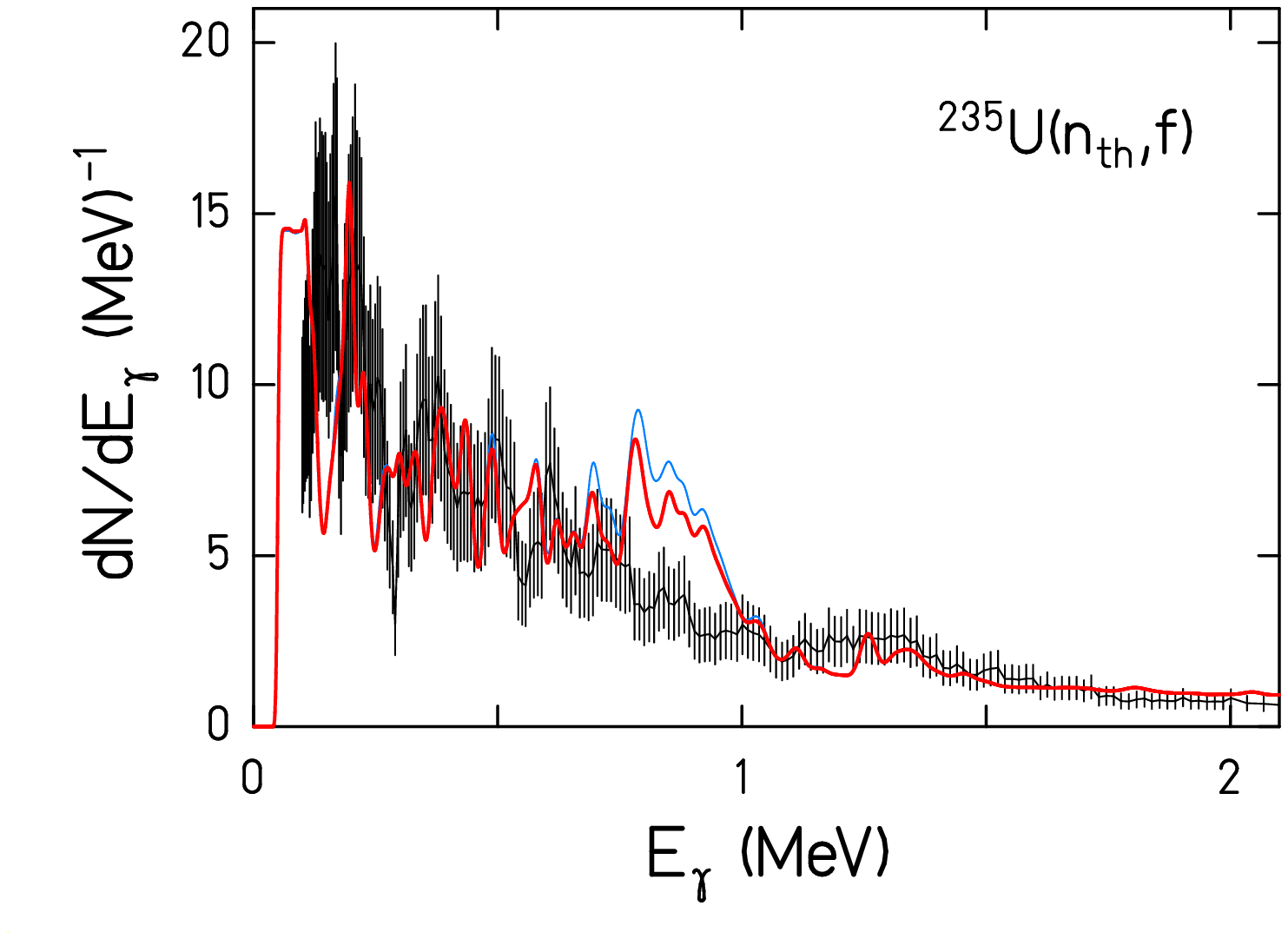}
\caption{(Color online)
Low-energy part of the experimental gamma-energy spectrum (black line with error bars) 
\protect\cite{Oberstedt14} for 
thermal-neutron-induced fission of $^{235}$U in comparison with the GEF prediction (red full line).
The GEF result is convoluted with the experimental energy resolution.
In addition, the calculated spectrum that includes the known delayed isomeric gamma transitions reported in JEF 3.1.1
is shown by the light blue line.} 
\label{GAMMAS-Fig4}
\end{center}
\end{figure}

The low-energy part of the
measured prompt-gamma spectrum of $^{235}$U(n$_{th}$,f) is compared in figure \ref{GAMMAS-Fig4} 
with the result of the GEF code. In this calculation, 
the E2 gamma cascade along the
yrast line is only included, until the fragment meets the angular momentum of a known
isomeric yrast state, which is listed in the JEFF 3.1.1 decay library.
The measured spectrum is fairly well reproduced with the exception
of a structure around 0.9 MeV, which is overestimated by GEF. 
Many gamma lines in this region stem from nuclei close to the N=82 or the Z=50 closed shell, 
where many isomeric transitions have been observed. This is illustrated by the blue line,
which includes the full E2 gamma cascades below the known isomeric states that are contained in the JEFF 3.1.1
decay tables. A possible explanation for the overestimation by the first-mentioned GEF calculation could be that it still contains
some isomeric transitions, which are not listed in JEFF 3.1.1, because they are too 
short-lived or not yet known and do not fall into the detection time window of the
experiment.  

\clearpage
 
\subsubsection{Delayed processes} \label{4-4-4}

Due to the curvature of the beta-stability line on the chart of the
nuclides, the fission fragments are in general unstable against
$\beta^-$ radioactive decay. $\beta$ decay is a slow process with
half lives in the ms region or longer. Thus, this process is well
separated in time from the emission of prompt neutrons and gammas.
The beta-decay chain ends, when a stable nucleus is reached.

\paragraph{Beta decay}

Beta decay proceeds to excited levels or to the ground state of the
daughter nucleus with different probabilities, given by the 
beta-strength function. The available energy (Q value minus rest energies
of electron and anti-neutrino) is shared between the emerging
electron and the accompanying anti-neutrino \cite{Balantekin16}.

\paragraph{Delayed gamma emission}

If the beta decay populates an excited level in the daughter nucleus,
this nucleus decays in most cases by one or more
gamma transitions to their respective ground state, eventually further delayed
by an isomeric state. 

\paragraph{Delayed-neutron emission}

\begin{figure*} [h]
\begin{center}
\includegraphics[width=1\textwidth]{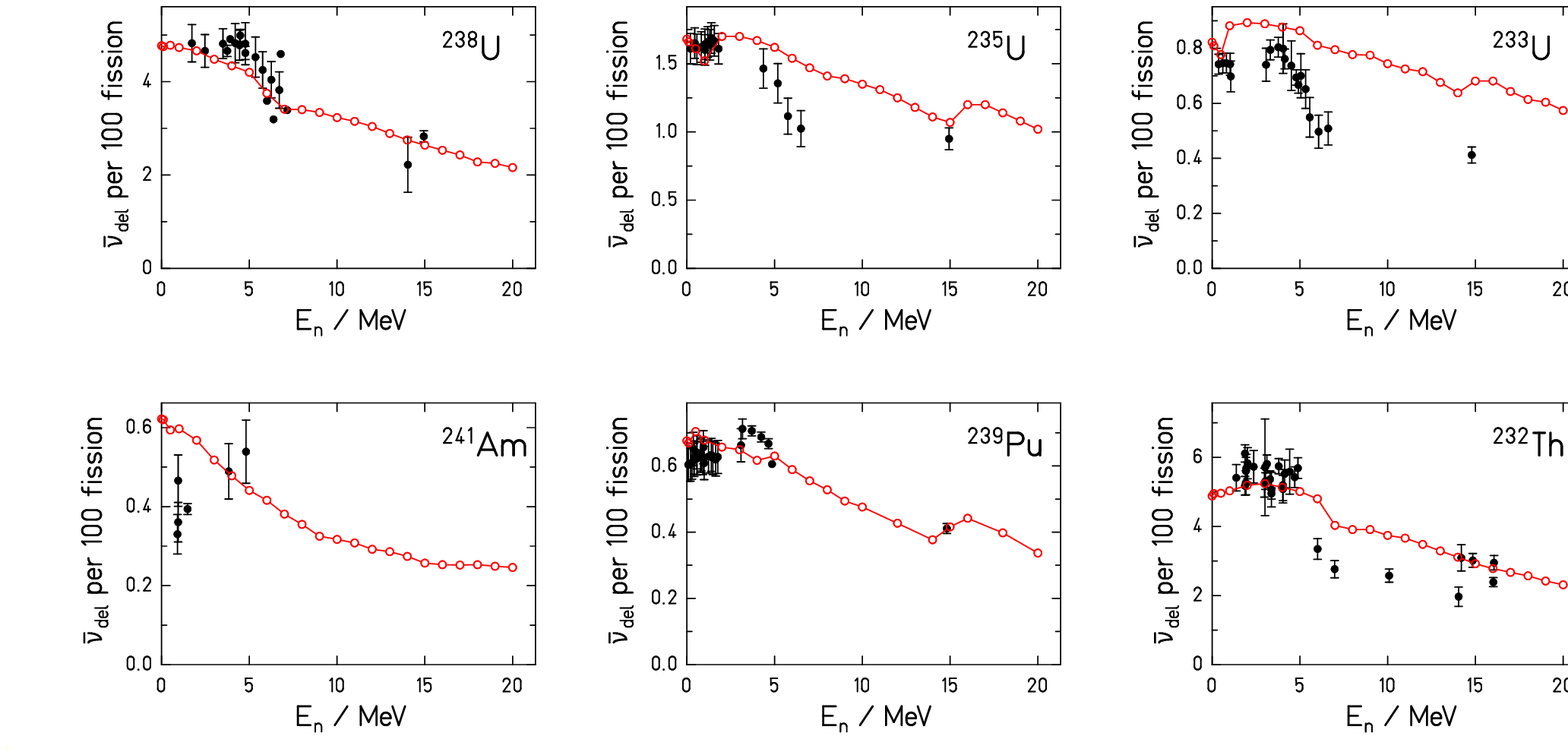}
\caption{(Color online)
Probability for the emission of delayed neutrons for neutron-induced fission
of different systems as a function of the incident-neutron energy. Experimental
data (full black symbols) from
Experimental Nuclear Reaction Data (EXFOR) \protect\cite{EXFOR} are compared with the result of the GEF code (version Y2017/V1.2) (open red symbols, connected by red lines.) 
} 
\label{Ndel}
\end{center}
\end{figure*}

When the $\beta$ Q value of very neutron-rich fission fragments exceeds the neutron separation energy, emission
of delayed neutrons opens up as a possible decay channel.  
Delayed neutrons play a major role in reactor control, and
many coordinated efforts have been undertaken to explain the complex characteristics, in particular in the excitation-energy dependence, of the delayed-neutron yields 
\cite{NEA90,NEA02,INDC11}, but this problem is not yet solved \cite{IAEA17}.

\begin{figure} [b]
\begin{center}
\includegraphics[width=0.4\textwidth]{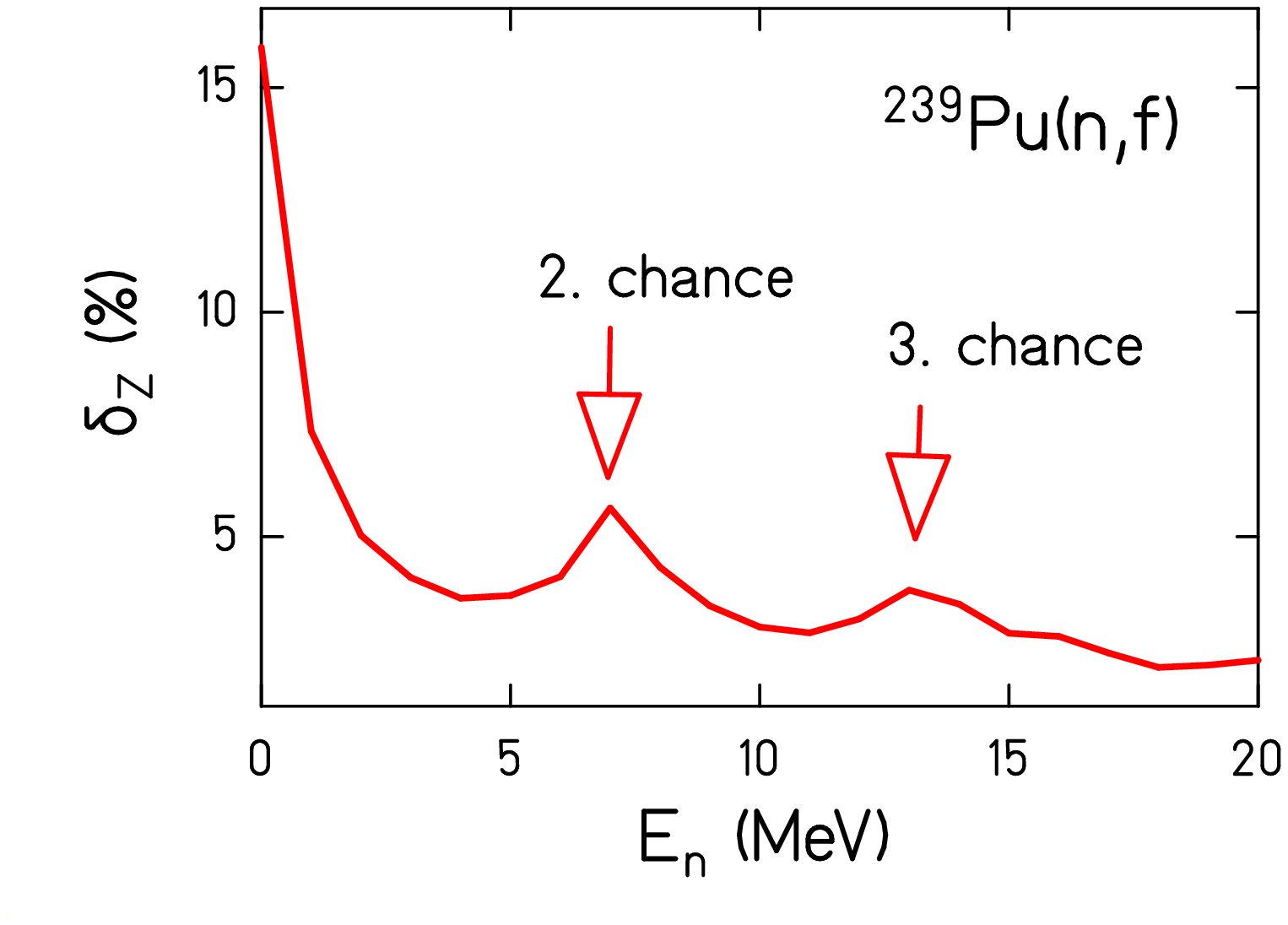}
\caption{(Color online)
Global odd-even effect ($\Sigma ( Y_{even} - Y_{odd})/ \Sigma ( Y_{even} + Y_{odd})$) in the fission-fragment $Z$ distribution of the system $^{239}$Pu(n,f)
as a function of the incident neutron energy from the GEF code.
} 
\label{even-odd}
\end{center}
\end{figure}

The GEF code follows the radioactive-decay chains, calculates the
cumulative yields\footnote{
The cumulative yield Y(A,Z) of nuclide (A,Z) is the total number of
atoms of that nuclide produced over all time after one fission. 
} 
and provides a list of delayed-neutron emitters with 
their contributions to the delayed-neutron yields. The decay tables of
JEFF 3.1.1 are used. 
Figure \ref{Ndel} shows calculated delayed-neutron yields
in comparison with experimental data for neutron-induced fission
of different systems. 
The calculated curves show some systematic features:

 (i) There is a more or 
less pronounced plateau at low energies up to about 4 MeV. This plateau is caused by
the combined effect of increased prompt-neutron emission, which tends to decrease the yields of the most neutron-rich fragments, and 
the gradual decrease of the odd-even staggering in the $Z$ yields, which tends to enhance the yields of odd-$Z$ fragments. Note that
delayed neutrons are predominantly emitted from isotopes of odd-$Z$ elements, which have
a systematically higher beta Q value. 
Both effects tend to compensate each-other.
The plateau disappears,
when the odd-even staggering in the $Z$ yields is switched off in the calculation.
The plateau is most pronounced for $^{232}$Th(n,f), which shows the largest
odd-even staggering in the fission-fragment $Z$ yields (see figure \ref{Fig-even-odd}). 
It is absent in odd-$Z$ fissioning systems, for example in $^{241}$Am, because the odd-even staggering in the $Z$ yields is much smaller, except at large mass asymmetry \cite{Caamano11}.

(ii) Above this plateau, there is a fall-off, which is most cleary seen in the case
of even-$N$ target nuclei.  
Often, this fall-off is attributed to the loss of one neutron at the onset of second-chance fission \cite{Mas69,Bey74}. 
%But we think that this has only a minor effect, because the total prompt-neutron yield does not show a pronounced structure at the onset of second-chance fission.
The mechanism becomes clear when the energy-dependence of the odd-even effect in the fission-fragment $Z$ distribution is considered, which is shown in figure \ref{even-odd}: When second-chance fission sets in, the odd-even staggering in the fission-fragment $Z$ distribution shows a peak. This effect is especially strong for even-$N$ target nuclei due to the
high fission probability slightly above the fission
barrier of the even-$N$ isotope that
is formed after the emission of one neutron in the case of an even-$N$ target nucleus.
This reduces the relative yields of the
odd elements, which are predominantly responsible for the delayed-neutron production.

(iii) Above about 6 to 7 MeV, there is a gradual decrease of the delayed-neutron
yield with increasing energy of the incident neutron in all systems. This gradual decrease is understood by the
shift of the isotopic distributions towards less neutron-rich isotopes by the increasing
emission of prompt neutrons. It is again partly compensated by the gradual reduction of the
odd-even staggering in the contribution from second-chance fission, in a similar way as in the first plateau, thus reducing
the slope of the energy-dependent delayed-neutron yield in this region.

\begin{figure*} [h]
\begin{center}
\includegraphics[width=0.75\textwidth]{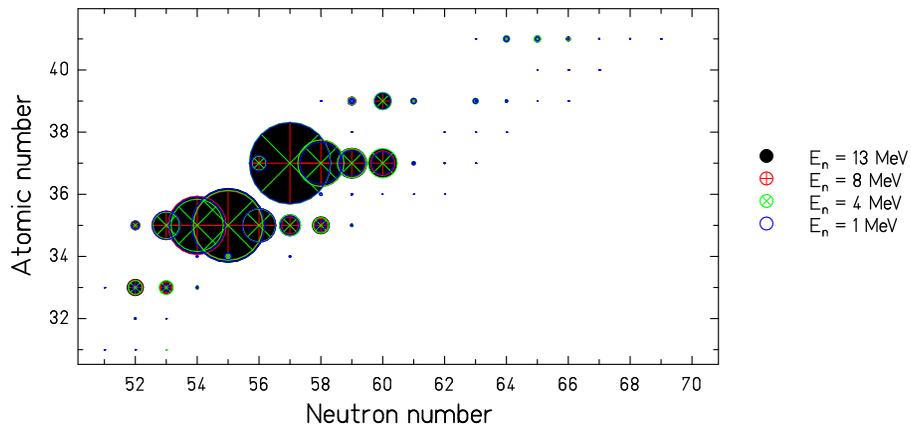}
\caption{(Color online)
Beta-delayed neutron yields in the light fission-fragment group from 
neutron-induced fission of $^{238}$U for different incident neutron energies on the chart of the nuclides, calculated with
the GEF code (version Y2017/V1.2). The radius of the symbol is proportional to the yield. 
} 
\label{beta-del-n-light}
\end{center}
\end{figure*}

\begin{figure*} [h]
\begin{center}
\includegraphics[width=0.75\textwidth]{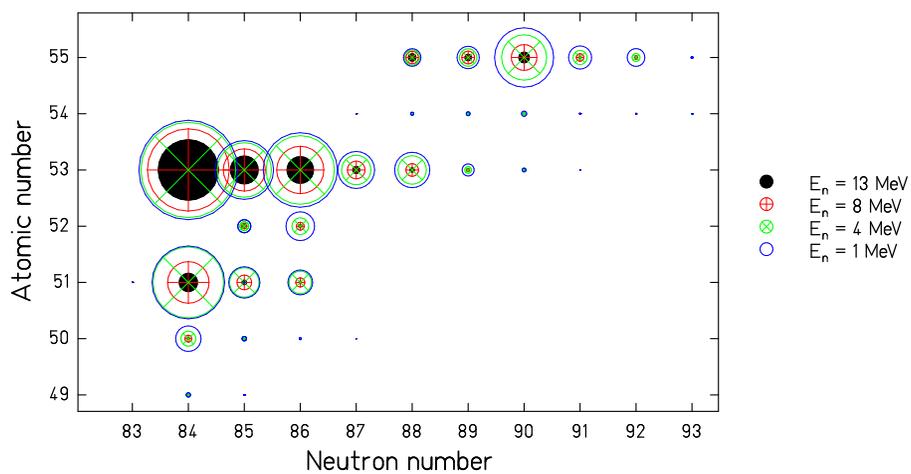}
\caption{(Color online)
Beta-delayed neutron yields in the heavy fission-fragment group from 
neutron-induced fission of $^{238}$U for different incident neutron energies on the chart of the nuclides, calculated with
the GEF code (version Y2017/V1.2). The radius of the symbol is proportional to the yield. 
} 
\label{beta-del-n-heavy}
\end{center}
\end{figure*}

\clearpage

(iv) A bump appears in some cases near $E_n$=15 MeV, slightly above the threshold of third-chance fission. 
This structure may be connected with the high fission probability slightly above the fission
barrier of the even-$N$ isotope that
is formed after the emission of two neutrons in the cases of an odd-$N$ target nucleus, but
the mechanism is not clear.

From the comparison with the measured data, one can deduce that the strong variation of the delayed-neutron yield from system to system is fairly well reproduced by the GEF code. 
The first plateau and the fall-off show up rather clearly in the measured delayed-neutron yields of all uranium isotopes and of  $^{232}$Th. 
For $^{238}$U and $^{232}$Th, the GEF results also show a clear fall-off, although it is not as pronounced as in the data. Some indications for a fall-off are also seen in the GEF results for $^{235}$U and $^{233}$U.
The data are too scarce, and they scatter too much, to pin down the slope of the gradual descent above the fall-off, but it seems that it is somewhat overestimated by GEF.
The measured data are also too scarce to verify the predicted bump at the onset of third-chance fission, may be with the exception of $^{232}$Th, where, however, this bump does not appear in the calculation. 

A much more detailed documentation of the GEF calculations is shown in figures \ref{beta-del-n-light} and \ref{beta-del-n-heavy}, where the contributions of the delayed-neutron emitters in the light and the heavy fission-fragment group to the delayed neutron yield are shown on a chart of the nuclides for the system $^{238}$U(n,f).
Evidently, the neutron yields in the light group stay almost constant, in contrast to the yields in the heavy group, which decrease strongly with increasing excitation energy.
This different behavior is certainly caused to a great part by the energy sorting, discussed in section \ref{4-3-6}, which assures that the isotopic distributions in the light fission-fragment group do not move when the excitation energy varies. 
Only at the onset of second- (or higher-chance) fission a slight shift is expected, but obviously it does not have a noticeable effect on the delayed-neutron yields.
Figure \ref{beta-del-n-heavy} also reveals a strong reduction of the contributions from
even-Z fragments to the delayed neutron yields with increasing incident energy, which is most clearly seen in the heavy fission-fragment group. This confirms the important role of the
odd-even effect in producing the plateau in the delayed-neutron yields below 4 MeV.

It seems that the GEF calculations come rather close to understand the origin of the observed structural features in the energy-dependent delayed-neutron yields, although some quantitative deviations show up.
Two effects can clearly be identified that strongly determine the evolution of the delayed-neutron yields with increasing excitation energy in the model:
The odd-even staggering in the fission-fragment yields and
the shift of the fission fragments towards less neutron-rich isotopes by prompt-neutron emission. 
But also the growing yield of the symmetric fission channel and other changes in the 
fission-fragment yields have certainly some influence. 
It is noteworthy that the cases, in which the sequence of the plateau - fall-off - gradual decrease is least pronounced in the GEF results, are those with the lower total
delayed-neutron yields. In these cases, the neutrons are provided by nuclei with low delayed-neutron branchings.  This gives rise to suspicion that the uncertainties in the decay data may also play a role.

%The GEF model is able to explain the prominent features of the delayed neutron yields as a function of the excitation energy. Still quantitative deviations remain.
In the moment, the reason for the observed deviations is not clear. It is also not so easy to adapt the model in some way, because there are constraints from other observables.
For example, the measured energy-dependent prompt-neutron multiplicities (see figure 77 in \cite{Schmidt16}), 
TKE values (see figure \ref{Fig-Loveland-TKE}) and fission-fragment yield distributions
(see figure \ref{VAMOS-N-over-Z}) are rather well reproduced by the model.
Still, there are no sufficiently accurate data available to pin down the shape of the function that describes the excitation-energy dependence of the odd-even staggering in the
fission-fragment $Z$ distribution \cite{Pomme93} due to the inability for varying the
excitation energy precisely in a well defined way.

Already in previous theoretical considerations, mostly the prompt-neutron emission \cite{Ohsawaa07,Nasrun15} and the odd-even fluctuations in the fission-fragment yields \cite{Alexander77} were connected with the observed features of the delayed-neutron yields.
%, but a conclusive explanation of the data was not found. 
In GEF, all the effects of the comprehensive description of the fission process are included in a consistent way, as much as they are known.
The GEF model is able to reproduce the prominent characteristics of the delayed-neutron probabilities to a certain extent and to connect them with certain features of the yields and the excitation energies of the fission fragments, although the
delayed-neutron data were not yet exploited during its development.
There is good agreement in the absolute values and their variation from system to system, while the structures are at least qualitatively reproduced.  
Thus, the complex features observed in the energy dependence of the delayed-neutron
yields seem to be essentially understood. The key is the general coverage of practically all fission oabservables by the GEF model. 
GEF is certainly also well suited to develop a more accurate quantitative description of the delayed-neutron yield, which is consistent with all other relevant observations, when the delayed-neutron data are included in the further development of the model. 

\paragraph{Decay heat}

\begin{figure} [h]
\begin{center}
\includegraphics[width=0.5\textwidth]{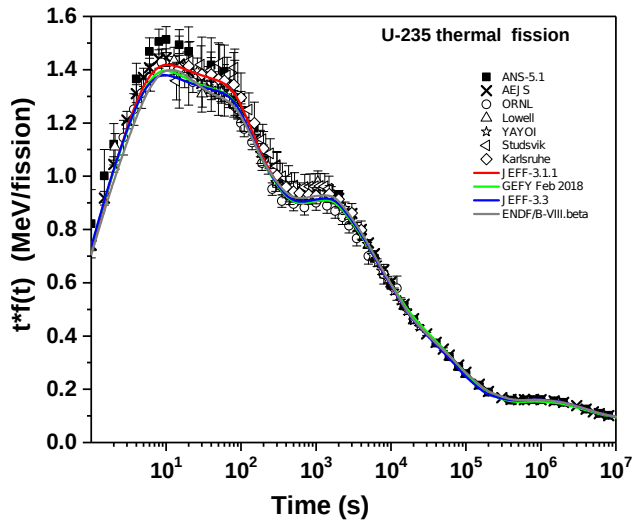}
\caption{(Color online)
Decay heat (thermal energy as a function of time, $t^{\ast} f(t)$) following a thermal fission burst of $^{235}$U. The calculated decay heat on the basis of fission-fragment yields from JEFF 3.3, JEFF 3.1.1, and ENDF/B-VII-beta are shown together with experimental data. In addition, GEFY Feb. 2018 denotes the calculated decay heat from GEF fission yields (version GEF Y2017/V1.2). 
In all cases, the JEFF-3.3 decay-data library was used.
The calculations were performed and the figure was produced by Alexey Stankovskiy.
} 
\label{decay-heat}
\end{center}
\end{figure}

Decay heat is the energy released by beta decay, and $\beta$-delayed neutrons and gammas
of the fission products.
It is the only source of heat in the nuclear fuel rods after reactor
shutdown \cite{Yoshida07}. 
Knowledge of the amount and form of energy emitted in the radioactive decays of fission products is critical for the determination of safety procedures for nuclear-power-plant operation and for the cooling of nuclear fuel after an accidental or planned reactor shutdown.

The calculated decay heat following a thermal fission burst of $^{235}$U is shown in figure \ref{decay-heat}, using different evaluated fission-fragment yields and compared
with experimental data. Apparently, the values based on the GEF yields are rather close to the values based on the evaluations. 
The parameters of GEF2017/1.2 were adjusted to the evaluated fission yields from ENDF/B-VII. No specific adjustment to the measured decay-heat data was performed.

\subsubsection{Correlations} \label{4-4-5}

The GEF model is unique in providing a global description of the complete fission process, in which all features are described in a consistent way and linked to each-other with all their 
interdependencies in full detail. 
This is an important asset for advancing the understanding of the fission process.
This also entails that the GEF model has the potential to revolutionize the application of nuclear
data for technology by replacing completely separate or insufficiently linked
descriptions of specific features, which are adjusted independently to
a small subset of observables and which are often incompatible with each-other.
In particular, the presently used mostly technically
defined covariances or correlations between specific fission observables
\cite{Talou11,Pigni15,Fiorito16} are supplemented by 
correlations, which are defined by the 
physics of the fission process, between any pair of fission quantities.
Correlations are also provided between fission observables of
different fissioning systems.

%\subsubsection{Test of GEF predictions} \label{4-3-7}
% 
%\paragraph{SOFIA data:}
%
%\paragraph{VAMOS data:}
%
%\paragraph{Multi-nucleon transfer-induced fission in direct kinematics:}
   
\subsection{Other models} \label{4-5}

We have seen in the previous sections that the semi-empirical GEF model covers almost all
fission quantities, starting from the fission probabilities, multi-chance fission,
pre-scission neutron emission, and fission-fragment isotopic yields and kinetic energies, to the 
statistical de-excitation of the fission fragments,
and, finally, to the delayed processes after beta decay with all their correlations and interdependencies.
This assures that the different fission quantities are consistent, which can have important
impact on the results.  
The self-consistent and the stochastic models are much more restricted, but they are 
indispensable for exploring specific fundamental problems for example about the 
fission times or the shell effects in fission observables of systems in unexplored regions. 
One may state that GEF covers all fission phenomena that can be generalized: If one knows
the connected quantities for a certain number of systems, one can conclude on their values for
other systems without the need for additional specific experimental information.

However, this method has some limitations. In particular, the resonances in the fission
cross section induced by low-energy neutrons and the sub-barrier structure due to the
coupling between states in the first and the second well of the potential as a function
of elongation cannot be generalized due to their complexity. 
GEF can only provide a smooth average result.
Also for the accurate calculation of the prompt-gamma spectrum, in particular
the contribution from non-statistical gammas, the exact
experimental knowledge of the spectroscopy of the fission fragments is required.
In this case, GEF offers an approximate solution with its improved VMI 
(variable-moment-of-inertia)
model, which, however,
has the advantage to provide rather good results also for cases, where spectroscopic data
are missing. 
In the following, some dedicated models for these problems will be described that may be applied,
if the relevant experimental information is available. Their main importance lies in the
application for nuclear-reactor technology.  

\subsubsection{Fission cross section} \label{4-5-1}

The nuclear fission cross section is highly fluctuating and finely structured. 
At energies below the fission threshold, resonances appear due to the coupling
between states in the first and the second, and eventually in the third, minimum.
In neutron-induced reactions, which are most important for nuclear-reactor technology,
neutron resonances in the neutron-capture cross section reflect the levels in the compound system. They create a dense pattern of
resonances, which are separated up to a certain excitation energy (in the resolved-resonance region). They also influence the fission cross section in fissile nuclei, in which the
fission barrier is lower than the neutron separation energy.  
At higher energies, in the unresolved-resonance region, one still observes fluctuations. Only at even higher energies, in
the continuum region, where the levels overlap,
the capture cross section varies smoothly as a function of neutron energy.  Moreover, the transition states above the saddle points create a pattern in the fission cross section
at energies slightly above the saddle energies.
Structures in the fission cross sections appear again around the onset of first- and higher-chance
fission, due to the fissioning nuclei that proceed to fission at energies close to  
their fission barrier. 

Much effort has been invested into the accurate description of the nuclear fission cross section 
with all its complexity
\cite{Bjornholm80,JEFF18,Younes03,Sin06,Capote09,Goriely09,Kawano09,Goriely11,Bouland13,Romain16,Sin16}.
In any case, these models rely on a tremendous number of parameters that have to be deduced from 
experimental data, to a great part for each single system independently. 
These are very accurate parameters that define (i) the heights of the fission saddles, the
depths of the minima and the corresponding potential curvatures to deduce the transmission, 
considering the energies of the states in the second (and eventually also in the third) minimum, 
(ii) the parameters of the R-Matrix theory \cite{Lane58} that define the resonance energies, and 
(iii) the energies and other 
properties of the transition states above the saddle points, and several other parameters.
Several model codes contain very elaborate model descriptions that are able to reproduce the
complex data very well, for example the GNASH code \cite{Young92}, the
CCONE code \cite{Iwamoto07}, the EMPIRE code \cite{Sin07}, the CHO code \cite{Kawano10},
the TALYS code \cite{Koning12}, and the CONRAD code \cite{Archier14}.
Some of the recent developments for the description of the fission cross section are 
improved phenomenological optical-model potentials for neutrons and protons 
with incident energies from 1 keV up to 200 MeV \cite{Koning03},
the development of a model for the transmission through a triple-humped fission
barrier with absorption \cite{Sin07},
the parametrization of the threshold behavior near the fission barrier by an effective
barrier distribution, the extraction of an effective purely experimental fission barrier,
and an heuristic model of resonant transmission in \cite{Romain16}, and
the extraction of one-dimensional fission barriers from potential-energy surfaces calculated 
with macroscopic-microscopic models to obtain fission transmission coefficients that can be 
used in a Hauser-Feshbach model in \cite{Archier14}.

%fission barriers, nuclear level densities at the fission saddle points, In this field,the coupled-channels model, the dispersive optical potential, and an heuristic model for resonant transmission are of eminent importance.

\subsubsection{Fragment de-excitation} \label{4-5-2} 

Most part of the excitation energy of the fission fragments is liberated by the evaporation of
light particles. Because the fragments emerging from the fission of nuclei close to 
beta-stability are very neutron-rich with respect to beta stability, these are almost 
exclusively neutrons. 
The energy spectra of these neutrons as well as their multiplicity belong to the most
important nuclear data for the design and the operation of nuclear-fission reactors.
The modeling of the energy spectra, and to some extent, also the modeling
of the prompt-neutron multiplicities,
starting from an empirically deduced information on the excitation-energies of the
fragments and some additional ad-hoc assumptions, e.g. about the division of the
excitation energy between the two fragments, has a long tradition.
%\footnote{To our knowledge, the GEF model is the first one that includes the statistical de-excitation of the fission fragments in a full modeling of the fission process.}

One of the first widespread descriptions of the prompt-neutron spectrum was 
introduced by Watt \cite{Watt52}. He proposed a closed
formula, deduced from a Maxwell-type energy spectrum
from one or two average fragments and the transformation 
into the frame of the fissioning system with at least
two adjustable parameters: the temperature and the velocity 
of the average fragment. The ”Los-Alamos model” \cite{Madland82}
extended this approach essentially by the use of a
triangular temperature distribution of the fragments to
a four-term closed expression for an average light and an
average heavy fragment. 
A similar two-fragment model
was also used by Kornilov et al. in \cite{Kornilov01}. In 1989, Madland 
et al.~\cite{Madland89} introduced the point-by-point model by
considering the emission from all individual fragments,
specified by $Z$ and $A$. This model was further developed 
e.g. by Lemaire et al.~\cite{Lemaire05}, Tudora et al.~\cite{Tudora10}
and Vogt et al.~\cite{Vogt12}. In Refs.~\cite{Svirin97,Trufanov01,Lovchikova04,Maslov07}, 
the spectral shape was parametrized by the Watt formula \cite{Watt52} or
an empirical shape function that had been introduced by
Mannhart \cite{Mannhart87} in order to better model the shape of
the neutron energy spectra in the fragment frame. Kornilov \cite{Kornilov99} 
proposed a phenomenological approach for
the parametrization of a model-independent shape of
the prompt-neutron spectrum. This approach was later
also used by Kodeli et al.~\cite{Kodeli09} and Maslov et al.~\cite{Maslov11}. 
Recent refinements of the Los-Alamos model were proposed in Ref.~\cite{Madland17}.
These are (i) separate contributions of the (average) light and the (average) heavy
fragment to the prompt-neutron emission, 
(ii) departure from statistical equilibrium at scission,
(iii) separate nuclear level densities for the fragments, and
(iv) center-of-mass anisotropy in the angular distribution.
   
The Watt model \cite{Watt52} and the Los-Alamos model 
\cite{Madland82} are directly fitted to the
measured prompt-neutron spectrum, while the point-by-
point model \cite{Madland89} is based on the measured $A$-TKE distribution. 
Manea et al.~\cite{Manea11} proposed a scission-point model
that provides the TKE($A$) distribution, in order to allow 
for calculations of prompt-neutron spectra with the
point-by-point method, if only the mass distribution is known.

Recently, several models are being developed that treat the statistical
de-excitation of the fission fragments by prompt neutrons and prompt-gamma radiation
in a more elaborate way, partially by the Hauser-Feshbach formalism. We mention here
 MCHF \cite{Stetcu14}, CGMF \cite{Talou14a}, FIFRELIN \cite{Litaize15}, 
and FREYA \cite{Verbekea15,Vogt17}. 
A special feature of these models is
the most accurate reproduction of the prompt-gamma spectrum, in particular the
peak structure at low gamma energies, mostly produced by the non-statistical
gammas emitted at the end of the gamma de-excitation cascade based on detailed
tables of experimental spectroscopic information of the fission fragments.

\section{Discussion and outlook} \label{5}

After the detailed report on the different activities in fission research,
we will try to give an assessment of the status and the most important achievements
that have promoted the understanding of nuclear fission during the last years. 
This will lead to an outlook on the developments to be expected in the near 
future and on the challenges to be tackled. 

\subsection{Status of microscopic theories} \label{5-1}

There is no doubt that microscopic theories are indispensable for
a deeper understanding of the fission process. 
But in spite of considerable 
progress and many important results, the theoretical
description of the fission process with dynamical microscopic models is still
very difficult, because the most advanced models in nuclear physics that have
been developed for stationary states in heavy nuclei \cite{Bacca16}, 
for example modern versions of the interacting shell model or effective field theories of quantum chromodynamics
\cite{Hebeler15}, are not readily applicable to the decay of
a meta-stable state. 
Intense efforts are presently made to develop suitable theoretical tools, 
see Ref.~\cite{Schunck16} and section \ref{4-1}. 
Another difficulty arises from technical limitations. Still, the application 
of the most advanced models that are based on classical stochastic or self-consistent 
quantum-mechanical methods is heavily restricted by their high demand on computer
resources.
In this section, we list some of the most important conceptual and technical
challenges to which these theories are confronted.

\subsubsection{Restrictions by limited computer resources} \label{5-1-1}

\paragraph{Number of relevant degrees of freedom:}

%The high demand on computer resources is behind many difficulties and 
%restrictions in the development of microscopic fission models.
For the microscopic quantum-mechanical approaches based on TDGCM and for classical
stochastic models, the number of relevant degrees of freedom that 
are presently explicitly treated is insufficient for a realistic dynamical calculation
of the fission process and for covering the full variety of fission
observables. In these microscopic models 
% that rely on a potential-energy surface 
and in the stochastic models, the number of relevant
degrees of freedom is presently limited to two, respectively four or five, in the most
advanced approaches. The success of the random-walk approach of
Randrup, M\"oller et al.~in a five-dimensional deformation space \cite{Randrup11a}
in reproducing the mass distributions of
a great number of fissioning systems seems to indicate that the
number of relevant degrees of freedom is very important for obtaining
realistic fission-fragment distributions.
Their model allows for
fully independent shapes of the two nascent fragments. 
This elaborate feature seems to be more important than the
restrictions to a comparably simple handling of the dynamics,
assuming over-damped motion and using Metropolis sampling, as long
as fission-fragment mass distributions are concerned.
However, the calculation of other fission quantities, like the
charge polarization (N/Z degree of freedom),  
fission-fragment excitation energies or their angular momenta
requires including an extendet set of relevant degrees of freedom.
Recently, the charge polarization degree of freedom has already been
included \cite{Moeller15}. 
Progress is expected to come gradually by the continuous development
of computer technology.

In the TDDFT approach (e.g. \cite{Bulgac16}), the need to construct collective variables, potential energy surfaces and the corresponding mass tensor as a pre-requisite for the dynamical calculations is obviated.
Only the starting point of the TDDFT calculation in terms of suitably chosen constraints on the potential-energy surface needs to be determined, as discussed in section \ref{4-1}. 
Thus, the above-mentioned restrictions are overcome. 
However, this advantage is payed by a considerably increased computational cost which limits the applicability presently essentially to the calculation of most probable fission quantities and a few very specific problems.

\paragraph{Neglect or approximate treatment of effects beyond mean field:}

Another class of difficulties arises from effects beyond mean field
in microscopic quantum-mechanical models. 
%Explicit handling of the direct interactions between the nucleons
%(many-body interactions) 
Handling of many-body dynamics
is very difficult, and, thus, there is put much effort in
developing suitable approximate algorithms that do not demand 
too much computer resources (see e.g. \cite{Lacroix10}). Questions, how well the physics
is still represented, must be answered.

\subsubsection{Problems in determining the potential-energy surface} \label{5-1-2}

For the self-consistent microscopic approaches (see section \ref{4-1}) and for the stochastic models (see section \ref{4-2}),
the calculated potential energy in the space defined by the relevant degrees of 
freedom is the basis for the dynamical calculation of full distributions of
fission observables. There are several difficulties
associated with the determination of this multi-dimensional potential-energy
surface. 
 
Usually, in stochastic models 
a shape parametrization is used, and the potential-energy is calculated
with the macroscopic-microscopic approach. In this case, only a restricted class of shapes
can be realized. This means that the calculated potential energy is 
an upper limit of the optimum potential that the nucleus can adopt. 
The deviations can be reduced by increasing the dimension of the
deformation space. However, as said before, the tractable number of relevant degrees of freedom,
in particular in a dynamical calculation,
is restricted due to the limited computing resources as seen in section \ref{5-1-1}. 

When the potential energy is determined self-consistently with constraints
on some degrees of freedom, this is a safe method to find the optimum shape.
However, the optimum shape is determined independently on the different 
grid points defined by the constraints. This can lead to discontinuities
in the shape evolution from one grid point to the next one, because one
or several of the degrees of freedom that do not belong to the ones 
explicitly considered may take very different values. This means that 
the "real" nuclear potential may have a ridge 
that is not visible in the calculation \cite{Dubray12,Regnier16}. Again here, a small number of
dimensions aggravates the problem.

A similar problem arises when in the case of a shape parametrization, 
generally used in stochastic models,
an optimization with respect to additional shape degrees of freedom is
performed individually on each grid point that do not belong to the
degrees of freedom explicitly considered in the dynamic calculation \cite{Pashkevich08}. 

It is important to control the effect of such problems on the result
of the dynamical calculation, if they cannot be avoided.
The appearance of such problems can be detected by local unphysical
fluctuations \cite{Dubray12} in the calculated potential-energy landscape. It can be
reduced by an optimum choice \cite{Younes12} and by increasing the number of relevant shape parameters. 

As mentioned before, this problem does not exist 
when the TDDFT is used to compute the dynamical evolution of the nucleus from
an initial point beyond the second fission barrier up to
scission \cite{Bulgac16,Tanimura17}, because the shape of the system develops freely
in a self-consistent way without any constraint. 
%However, as discussed in section \ref{4-4} these studies require
%calculating a potential-energy surface with constraints
%to define the initial condition of the dynamical evolution. 

\subsection{Aspects of statistical mechanics} \label{5-2}

Most of the models applied to nuclear fission consider aspects of
statistical mechanics only by global properties of the degrees of freedom
that are treated as an environment.
This is, firstly, the level density of the fissioning system and, 
secondly, dissipation by the coupling
between the relevant degrees of freedom and the environment.
Phenomena that are connected with the energy transfer between subsystems
of the environment as described in section \ref{4-3-6} are most often 
not considered.

At present, stochastic models are able to treat dissipation by global
descriptions of one-body \cite{Blocki78,Randrup80} and two-body \cite{Davies76} 
mechanisms and to include the
effects of pairing correlations and shell effects on the transport coefficients 
and the level density
for determining the heat capacity of the environment and for 
deriving the driving force of the fission dynamics \cite{Usang16}.   

Self-consistent quantum-mechanical models overcame the restriction to
adiabaticity only recently and started to develop methods 
that enable considering quasi-particle excitations 
on the fission path
%by level crossings 
\cite{Bernard11} and one-body dissipation by the fast 
shape changes at neck rupture, e.g. \cite{Simenel14}.  

Energy exchange between the nascent fragments \cite{Schmidt10} 
or even the competition between quasi-particle 
excitations in the neutron and proton 
subsystems \cite{Rejmund00} of the fragments were only considered in dedicated approaches. 
However, for an understanding of the division of excitation energy
between the fragments or the odd-even effect in fission-fragment
nuclide distributions, the explicit consideration of the two environments 
in the two nascent fragments and their interaction is indispensable. 

Preliminary results about these rather complex features of
statistical mechanics with simplifying assumptions were already obtained.
These concern the phenomenon of energy sorting \cite{Schmidt10} and global features of 
the odd-even staggering in fission-fragment $Z$ distributions \cite{Jurado15}.

The division of excitation energy between the fragments has recently attracted 
quite some attention.
The energy dissipated separately in the individual nascent fragments on the fission
path was estimated by Mirea \cite{Mirea11} and compared with the experimental data.
The division of excitation energy between the fragments induced at neck rupture 
was studied in the sudden approximation \cite{Carjan12}. 
An interesting attempt to study the energy partition between the fragments with
a microscopic self-consistent approach has been performed in Ref.~\cite{Younes11}
by considering spatially localized quasi-particles in a frozen scission configuration.
The dominant role of statistical mechanics, and particularly the assumption of 
statistical equilibrium in the 
division of heat between the nascent fragments before scission that is made 
in refs. \cite{Schmidt10,Schmidt16}, is
questioned \cite{Younes11} or criticized \cite{Bulgac16} by several authors. But the remarkable
experimental result of Ref.~\cite{Naqvi86} that an increased initial excitation energy of the fissioning
system is found in the heavy fragment, while the neutron multiplicity of the light
fragment stays unchanged, has not yet been addressed by 
microscopic models. This is also true for the complex features of the odd-even effect
in fission-fragment $Z$ distributions, which is also successfully described in the framework of
statistical mechanics in Ref.~\cite{Jurado15}.

\subsection{Systematics and regularities} \label{5-3}
 
In section \ref{4-3} we presented the combinations of several empirical observations
and powerful theoretical ideas as a basis of the GEF model. 
They go well beyond purely empirical descriptions, 
because they do not only reproduce experimental data with a high accuracy, but,
due to their theoretical basis and the relatively small number of adjustable parameters
that describe all systems with identical values, 
they are also expected to provide reliable predictions for a large variety of 
fission quantities for a wide range of nuclei for which no experimental data exist.
In particular, it was demonstrated in section 4.4 that the results of GEF agree very well with very different experimental data that have not been used to fix the parameters of the GEF model.
The GEF model code \cite{Schmidt16} that exploits these ideas pursues the 
tradition of former inventive ideas like the macroscopic-microscopic model 
\cite{Brack72} and the concept of fission channels \cite{Brosa90}
and, partly, makes directly use of those.

The relationship between GEF and microscopic fission 
models may best be illustrated by recalling the role
of the liquid-drop model in the development of nuclear
mass models, although the dynamical fission process is
much more complex than the static properties of a 
nucleus in its ground state. For a long time, purely 
microscopic models were not able to attain the accuracy of
the liquid-drop model in reproducing the macroscopic 
nuclear properties. Only very recently, the accuracy of fully
microscopic and self-consistent models became comparable 
with the accuracy of macroscopic-microscopic 
models \cite{Goriely13,Sobiczewskif14}. While the powerful basic relations of the
liquid-drop model follow directly from the theoretical 
assumption of a leptodermous system, the
parameter values were determined by an adjustment to
experimental masses and other nuclear properties. 
Only microscopic models were able to relate the 
values of these model parameters to the properties of the
nuclear force \cite{Reinhard06}. Remembering this analogy clarifies
that GEF is not intended to compete with microscopic
models, although it is presently better suited as far as the
use for applications is concerned. On the contrary, both
approaches may be considered to be complementary for
extracting physics.
In particular, the semi-empirical GEF model helps 
to uncover regularities
that are not directly evident from the fission observables and 
to recognize the physics content of some systematic trends in the data.

\subsection{Uncomprehended observations} \label{5-4}

Beyond the general inability of theory 
in explaining many facets of the nuclear-fission process,
there are some specific observations that seem to contradict 
well established knowledge. In the following, we will present one of 
those cases that we believe to be among the most striking ones.

%\subsubsection{Apparent insensitivity of the fission probability to angular momentum}

\subsubsection{Dominance of "magic" proton numbers in fission-fragment distributions} \label{5-4-1}

The success of the GEF model \cite{Schmidt16} in reproducing the fission-fragment mass distributions
from the fission of actinides by a statistical approach, assuming universal 
fragment shells superimposed on the macroscopic potential, is already
a remarkable result. Even more striking is the constancy of the 
mean number of protons in the heavy fragment of the contributions from 
the asymmetric fission channels standard 1 and standard 2
over all systems investigated until now \cite{Boeckstiegel08}. In particular for the standard 2
fission channel, this finding seems to be in contradiction to the results 
of Strutinsky-type calculations \cite{Strutinsky68}
within the macroscopic-microscopic approach 
performed by Wilkins, Steinberg and Chasman 
\cite{Wilkins76} who attributed this dominant asymmetric fission channel
in the actinides to a shell at large deformation in the neutron subsystem 
at $N = 88$. The deformation of about $\beta = 0.6$ that they found in their
calculations is consistent with the neutron multiplicity observed in the 
heavy fragment. These calculations did not provide any evidence for a
proton shell at $Z = 55$ that one might suspect to be responsible for the
fixed mean number of protons found in the experiment. Other
systematic Strutinsky-type calculations performed by Ragnarsson and Sheline 
\cite{Ragnarsson84} yielded similar results. 

Until present, Randrup, M\"oller et al. \cite{Randrup13,Ghys14} performed the most extended
systematic calculations of the fission-fragment mass distributions for a large number of fissioning systems with the Brownian Metropolis shape-motion treatment. 
In a recent study, the mean number of protons in the heavy fragment has been investigated in dedicated calculations \cite{Moeller17} for a series of uranium isotopes ($^{234}$U, $^{236}$U, $^{238}$U, and $^{240}$U), and the experimental results, available for $^{234}$U, $^{236}$U, and $^{238}$U, were rather well reproduced. 
This model has the potential to shed some light on the prevalence of proton or neutron shells, because it is able to explicitly calculate the neutron shell-plus-pairing corrections and the proton shell-plus-pairing corrections.
However, Ref.~\cite{Moeller17} does not give any detailed information on this question. 
Furthermore, it is difficult to establish a relation to the hypothetical role of the fragment shells assumed in the GEF model \cite{Schmidt16}, because the fission channels standard 1 and standard 2 are not distinguished in \cite{Moeller17}.

%It would be very interesting to check in detail, whether this feature is also reproduced by other models.  
One might imagine a number of reasons for the observation of a nearly constant number of protons in the heavy fragment of the asymmetric fission channels under the assumption of the decisive role of fragment shells.
Some possible explanations could be 
(i) that the relation between the size of a shell-stabilized pre-fragment
and the size of the final fragment is not so strict, for example by
a variable division of the number of nucleons in the neck at scission, or
(ii) that the Strutinsky-type calculations of Refs.~\cite{Wilkins76,Ragnarsson84} are not realistic and miss a proton shell near $Z$ = 55 at large deformation.
In this context, we would like to remind the observation of a mutual support
of magicities in the surrounding of spherical doubly-magic nuclei 
\cite{Zeldes83}, because it violates the independence of shell effects in the neutron
and proton subsystems. A similar effect in deformed nuclei may be expected. 
May be, the position of a shell in proton or neutron number even moves as a function
of the number of nucleons of the other kind.
However, such an effect is not observed in the case of some local
stabilizations by neutron shells at $N = 152$ (see e.g. \cite{Patyk89}) and $N = 162$ (see e.g. \cite{Hessberger17}), which are stable 
in neutron number over several elements.

In any case, the almost constant number of protons in the heavy fragment, found in 
the contribution of the most important asymmetric fission channels in the 
actinides, is a very intriguing observation that asks for a deeper understanding.
Interesting enough, the position of the shell in the light fragment to which the double-humped mass distributions in the lead region are tentatively attributed shows the same feature, see section \ref{4-3-5}.

%\subsubsection{Energy dependence of delayed-neutron yields} \label{5-4-2}
%
%The observed energy dependence of delayed-neutron yields in neutron-induced fission, discussed in section \ref{4-3-5}, shows a few features that are not understood. 
%While the plateau below 4 MeV incident neutron energy can be attributed to the compensation of opposite trends from the odd-even staggering and the prompt-neutron emission, the origin of the fall-off around 6 MeV, at least its absolute value, is not clear. 
%The almost constant yield above the fall-off is even more intriguing, although it is not well established by the available data. 
%The variation of the prompt-neutron multiplicity and the well established change in the fission-fragment nuclide distribution, which are well documented by the available experimental data, should provide enough information to estimate the change in the production of beta-decay precursors. 
%This is a real puzzle. The best way to solve this puzzle would be to directly measure the production of the precursors as a function of the incident-neutron energy. 
%May be, the surrogate technique using transfer reactions in inverse-kinematic experiments could provide this information with the necessary resolution in $Z$ and $A$.

\subsection{Modeling of fission dynamics} \label{5-5}

In section \ref{4}, the different approaches and their methods of modeling fission 
dynamics have been described separately. In this section, we would like
to make a survey on the strengths and the limitations of each of these methods
for treating fission dynamics in a comparative consideration.
We will discuss the time-dependent density-functional theory (TDDFT), the time-dependent 
generator-coordinate method with the Gaussian overlap approximation (TDGCM-GOA), and stochastic models.
Finally, we will also have a short view on the semi-empirical GEF model.

TDDFT provides the most elaborate modeling of the shape evolution of the fissioning system.
It allows for all kind of shapes without any constraint, and, thus, 
all collective degrees of freedom (CDOF) are included. 
All shapes are allowed, and the nucleus chooses dynamically the path in the shape space.
The forces acting on nucleons are
determined by the nucleon distributions and velocities, and the nuclear system naturally and smoothly evolves into separated fission fragments.
It was found that the
strong energy exchange between a large number of CDOF slows down the rolling down 
towards scission and leads to a very long fission time \cite{Bulgac16}. One could say that the
collective degrees of freedom are strongly coupled between each-other and form a kind
of special heat bath, not in equilibrium with intrinsic excitations.
As discussed in section \ref{4-1}, one-body dissipation and some quasi-particle excitations are already included in TDDFT.
The authors of Ref. \cite{Bulgac16} state that extensions of the present approach to two-body observables\footnote{According to \cite{Bulgac16}, these observables cannot be determined
properly by considering only one-body interactions (the interaction of independent nucleons with the common potential). 
Also two-body interactions (interactions between two nucleons) must be included.} 
(fission-fragment mass, charge, angular momenta, and excitation-energy distribution 
widths) are rather straightforward to implement, and, eventually, stochasticity
of the mean field could be introduced. However, those calculations are still very much
hampered by the immense demand on computer resources, severely aggravated by the fact that distributions of fission quantities are constructed by accumulating the results of many fission events (Monte-Carlo method) \cite{Tanimura17}.

In TDGCM-GOA, the dynamics is calculated on a potential-energy surface, presently
spanned by two collective parameters (for example two multipole moments). 
Shapes and binding energies are calculated independently for the different points of the potential-energy grid. 
This makes it unpractical to handle energy transfer between the different shape degrees of freedom or excitations of collective modes. 
Moreover, it is necessary to impose a scission configuration with a somewhat arbitrarily defined criterion. 
In addition, the presently available calculations assume adiabaticity. 
The inclusion of dissipation, leading to intrinsic excitations (single-particle or quasi-particle excitations) is just being developed.

An elaborate consideration on the prospects and the limitations of the different self-consistent approaches has been formulated by Bulgac \cite{Bulgac10}.

Advanced stochastic models calculate the dynamics on a potential-energy surface, 
spanned by presently up to five dimensions. Like in the TDGCM-GOA method, the binding
energies are calculated independently for the different points of the potential-energy grid.
Due to the shape parametrization, only few of the collective degrees of freedom are included. 
Therefore, it is again unpractical to handle excitations of all collective modes and direct
energy transfer between different CDOF. Intrinsic degrees of freedom are included via a
representative heat bath with a certain temperature. The heat bath introduces stochastic
fluctuations in the trajectory in the collective-coordinate space. It is also the most
effective mean for (indirect) transfer of energy from one CDOF to another one. Note that this
scenario does not agree with the direct coupling between collective degrees of freedom found in TDDFT \cite{Bulgac16}.

In the GEF model, excitations of collective and intrinsic degrees of freedom are included only
in an effective way by temperature values of the relevant environment. Depending on the coupling
strengths between intrinsic and collective degrees of freedom, or between collective degrees
of freedom alone, this could be the
temperature of the heat bath that represents the intrinsic degrees of freedom, like in the
stochastic models, or it could be the average energy per collective degrees of freedom, which
is suggested by TDDFT.
GEF may act as a bridge between microscopic models and empirical information, by the values of the
temperature parameters that describe the population of the normal oscillators or the 
widths of the distributions of the respective fission quantities, see equations (\ref{EQ-Yields-Osci}) and (\ref{EQ-Sigma-Osci}). 
GEF is presently the only model that explicitly considers the heat transfer between the
nascent fragments as a collective degree of freedom.

Obviously, the different models are presently either not yet applicable 
or conceptually unable to treat all aspects of fission dynamics.
$TDDFT$ is the only approach that includes all collective degrees of freedom. However, it is still under development in many aspects and generally very much hampered by the high demand on computer resources.
$TDGCM-GOA$ is unpractical in handling excitations of all collective degrees of freedom and lacks inclusion of all kind of intrinsic excitations.
$Stochastic$ $models$ handle the coupling between collective and intrinsic excitations in the most
simple and most practical way. 
However, the excitation of collective degrees of freedom is very restricted, and the assumption of thermal equilibrium between collective and intrinsic degrees of freedom is a strong and
probably wrong assumption.
The $GEF$ $model$ contributes with the idea that the heat storage should be considered 
separately for the two nascent fragments. This emphasizes that the aspects of statistical mechanics should be taken into account in a realistic modeling of the fission process.

\subsection{Theoretical needs} \label{5-6}

In this review, many restrictions of the different theoretical approaches to 
nuclear fission were quoted. In the following, we summarize the 
expected evolution of fission theory to satissfy the 
main identified needs. 

\subsubsection{Extending the relevant degrees of freedom} \label{5-6-1}
With the progress in computer technology, the corresponding
restrictions will gradually become less severe. First of all, this will
allow extending the number of relevant degrees of freedom in stochastic and the
microscopic quantum-mechanical models that explicitly use a potential-energy surface. 
For example, it will become
customary that the shapes of the two nascent fragments are allowed to 
evolve independently, and the charge polarization will routinely be considered. 
These developments are important for handling fission-fragment 
distributions in a more realistic way and fully specified in mass 
and atomic number \cite{Moeller15}. 

Moreover, progress is needed to clarify the processes that are responsible for the generation of angular momentum
of the fission fragments and the eventual generation orbital angular momentum \cite{Kadmensky11}, 
as well as the evolution of the projection of the total angular momentum onto the 
symmetry axis of the fissioning nucleus \cite{Kadmensky09}.
Also the excitation of other collective 
modes that are already subject of dedicated theoretical considerations and some
stochastic calculations may be studied by microscopic models.

There are strong arguments that the intrinsic and the collective degrees of freedom
of the environment form separate thermodynamical units with different 
temperatures \cite{Noerenberg74,Bulgac16}.
Furthermore, it appears to be
mandatory for describing the heat transport between the nascent fragments 
that the part of the environment, which comprises the intrinsic degrees of freedom,
is divided into two parts, belonging to one and the other fragment. 
The properties of these partial environments should be treated separately, and their interaction should be considered.

\subsubsection{Effects beyond mean field} \label{5-6-2}

More powerful computing resources may also allow to apply more realistic
treatments of effects beyond mean field than those that are manageable at present,
see section section \ref{4-1}.

\subsubsection{Dissipation} \label{5-6-3}

While dissipation, the coupling between the relevant degrees of freedom and
the environment, is an inherent part of stochastic approaches,
microscopic quantum-mechanical models are originally restricted
to adiabatic processes. Algorithms for considering dissipative processes are 
presently being developed, for example by explicitly including the first 
quasi-particle excitations \cite{Bernard11} or by representing the heat bath of stochastic models
by a large number of identical quantum oscillators \cite{Reinhard13} (independent oscillator model \cite{Ford88}).
%\cite{Reinhard13}. 
There is great interest to develop
more realistic and more complete representations of the environment that better 
reflect the complex nuclear properties.

\subsubsection{Evolution from the mononuclear to the di-nuclear regime} \label{5-6-4}

The gradual transition from the mononuclear to the di-nuclear regime of
two nascent fragments that are coupled by the neck manifests itself in
several ways: The early predominance of fragment shells seems to be well 
established \cite{Mosel71,Mosel71a}, but also the localization of quasi-particles in the two fragments 
\cite{Krappe01} and the increase of the congruence energy \cite{Myers97}
need to be better understood.
The increase of the pairing gap is also expected to evolve gradially during
the fission process.
The odd-even staggering of the fission barrier
% increased odd-even staggering already at the fission barrier that is
demonstrated in section \ref{4-3-1}, which is not correctly reproduced
by theoretical models, e.g. \cite{Moeller09},
can be understood as a sign for the shape dependence of the pairing gap.

\subsubsection{Neck rupture} \label{5-6-5}

The realistic modeling of the violent processes around scission is very
demanding for the treatment of collective dynamics and the induced intrinsic 
and collective excitations. In principle, TDDFT masters this problem to some extent, but the
calculation of the corresponding fission observables like the odd-even effect in fission fragment $Z$ distributions or the generation of fragment angular momentum, is not yet possible. 
Progress in self-consistent modeling of quantum 
localization and other processes around neck rupture is expected
to improve the understanding of the instabilities at neck rupture and their effect on
different observables like the odd-even effect in fission-fragment $Z$ distributions
and in the kinetic energies or the angular momenta of the fission fragments.

\subsubsection{Combination of different approaches} \label{5-6-6}
Progress in the modeling of nuclear fission may also evolve from exploiting
the strengths of methods from different approaches. For example,
elaborate stochastic models provide the technical tools for handling dissipation, 
%that is the coupling of the relevant degrees of freedom with the environment, 
%because they are
which is directly considered in the Langevin equations by the 
friction tensor and the fluctuating term. Even more, the driving force 
$T dS/dq$ is essentially determined by the derivative of the entropy, a 
quantity that is primarily a property of the environment. 
It was mentioned that the microscopic models do not treat fully quantum 
mechanically intrinsic and collective degrees of freedom and their coupling. 
But without considering in some way the environment that creates
driving force, friction and fluctuations, a realistic description of the dynamics 
is not possible.  
A solution could be to use quantum-mechanical considerations 
for estimating the mass tensor \cite{Tanimura15},
the full friction tensor and the dependence of the entropy 
on the relevant degrees of freedom, including the excitation energy, and 
to perform the dynamical calculation with a stochastic approach. 
A step in this direction has already been made \cite{Sadhukhan16}.
However, the Langevin equations are based on the assumption that all degrees of
freedom of the environment form a heat bath, that means that they are in 
statistical equilibrium at every moment. This is probably not always a realistic
assumption, and it might be necessary to take this into account.

The transformation of energy and the transport of heat between different subsystems
during the fission process are genuine problems of statistical mechanics. 
It would be a great progress if considerations of statistical mechanics could be introduced
into microscopic quantum-mechanical approaches in some manageable way.       

The observation of gradual systematic variations of the fission quantities 
along a series of neighboring nuclei may be exploited to increase the accuracy of
microscopic models that treat each fissioning system independently with its
own technical uncertainties that are inherent, for example in the shell effects 
determined with the Strutinsky procedure or in the potential energy determined
in a restricted deformation space, discussed in section \ref{5-1-2}. 

One may also extend the validity range of semi-empirical approaches like the
GEF model, which covers a considerable variety of fission quantities,
if one succeeds in deriving the justification for some
approximations of the GEF model as well as for
the values of the model parameters on a microscopic basis.
 
\subsection{Experimental needs} \label{5-7}

In the field of nuclear fission, a process that is still so far from being
fully understood, it is not possible to give a list of most important missing
experimental information. One should be prepared for surprises and for
new problems to emerge when new data come up. However, some general rules 
may be given.
It is certainly beneficial to cover a range as wide as possible in the
choice of the fissioning system in terms of nuclear composition ($Z$ and $A$),
excitation energy and angular momentum. 
The angular momentum dependence can be investigated by forming the same fissioning nucleus with different reactions.
Moreover, the coverage of fission observables 
should be as complete as possible, and they should be measured with a resolution
as good as possible with as many quantities as possible in coincidence.  

In the following, we will illustrate the relevance of the rules mentioned
above by reminding the progress brought about by some specific
experimental information or the open questions that could be answered by
new measurements.

\subsubsection{Wide coverage and precise definition of initial excitation energy} \label{5-7-1}

\paragraph{Distinguishing different fission chances:}

At energies that are above the threshold for multi-chance fission, the 
fission fragments are emitted from a wide range of excitation energy.
Up to third-chance fission (the fission after the emission of two neutrons), step-like structures can be observed in the fission cross
section. These carry some information on the different contributions.
However at higher intial excitation energies, these data do not provide
unambiguous information on the evolution of the 
fission-fragment mass distributions and the fission probabilities.
Theoretical estimates have a large margin of uncertainty
\cite{Madland82,Svirin10}.  
See in particular figure 27 of Ref.~\cite{Svirin10}, where the uncertainty of the relative weights of the different fission chances are demonstrated. 
Further development of suitable differential techniques \cite{Delagrange77} to distinguish the fission
events from different fission chances would improve the experimental knowledge
on the washing out of shell effects in the fission probability and in the
fission-fragment production with increasing excitation energy at fission.

Valuable information on multi-chance fission can also be deduced from the
relative number of neutrons emitted before and after fission, or more exactly
before and after passing the fission barrier. However, measuring the number of neutrons
emitted from the fully accelerated fragments, which is easily determined by 
a multi-source fit of the prompt-neutron angular distribution, is not 
significant, because neutrons emitted from the first minimum cannot be distinguished experimentally from those emitted between barrier and scission.

\paragraph{Energy dependence of odd-even effect in different observables:}

The experimental information available at present about the decrease of the
odd-even effect in fission-fragment $Z$ distributions with increasing initial
excitation energy has been obtained with 
relatively broad excitation-energy distributions, e.g. by bremsstrahlung-induced
fission \cite{Pomme93}, or by electromagnetic-induced fission at relativistic energies
\cite{Steinhaeuser98}. First results on the energy dependence of the odd-even effect in fission-fragment 
$N$ distributions has only been deduced very recently by comparing new data from 
electromagnetic-induced fission at 
relativistic energies \cite{Boutoux13} with results from thermal-neutron-induced fission,
see section \ref{3-3}.
The determination of the energy-dependent
odd-even staggering in fission-fragment $Z$ and $N$ distributions, total kinetic
energies and other observables would certainly help to better understand the
influence of pairing correlations on the fission process. 

\subsubsection{Extended systematic coverage of fissioning systems} \label{5-7-2}

\paragraph{Shell structure in fragment distributions around $A = 200$:}

The observation of a double-humped mass distribution in the fission of $^{180}$Hg
\cite{Andreyev13} drew the attention to the appearance of structural effects in 
the fission of lighter fissioning nuclei ($A < 210$). In spite of
intense experimental effort, a comprehensive overview on mass distributions
in low-energy fission of these nuclei could not yet be established. 
The most efficient method to provide a wide systematics of 
fission-fragment distribution for fission from energies in the
vicinity of the fission barrier of neutron-deficient systems over a large
mass range is the electromagnetic-induced fission of relativistic 
projectile fragments that is presently used by the SOFIA experiment \cite{Pellereau17}.
This approach will even provide nuclide distributions, fully separated in $A$ and $Z$, 
which allow a much better characterization of the fission process than mass 
distributions with limited resolution obtained from direct-kinematic experiments.
New experimental results are urgently awaited. 

\paragraph{Fragment distributions for long isotopic chains:}

The evolution of fragment mass distributions is of great interest for two
reasons: Firstly, it will help to better estimate the mass distributions
from the fission of very neutron-rich nuclei on the astrophysical r-process
path. This information is urgently needed for simulating the nuclide
production in the r-process. Secondly, the
position of the mean $Z$ in the heavy component of the asymmetric fission
channels could be followed over a larger range. This could help to better
understand the mechanism behind the surprisingly almost constant mean $Z$ values
in the heavy components of the fission-fragment distributions for actinides and 
the weak variation of the proton number of the light peak from fission in the lead region.

\subsubsection{Correlations of as many observables as possible} \label{5-7-3}

The few data on the variation of the mass-dependent prompt-neutron multiplicities 
as a function of initial excitation energy are presently the only rather direct experimental
evidence for the energy-sorting process (see section \ref{4-3-6}). This illustrates the importance
of multi-parameter experiments for discovering new features of the fission
process. This concerns for example identification of fission fragments in $A$ and $Z$ 
and measurement of their kinetic energies, multiplicities and energies of prompt neutrons and prompt gammas. 
In general, such data will provide important constraints on the 
modeling of fission. 

\subsection{Possible scenario of the fission process}

\subsubsection{Proposed scenario}

On the basis of a synthetic view of experimental information and of the different theoretical approaches and ideas, we propose a possible scenario of the nuclear-fission process.
The following description is rather comprehensive in covering the whole fission process, but not complete in all details.

We consider a system that starts from a configuration in the first minimum with a certain excitation energy $E^*$ above the ground state.
On the way to fission, the excited nucleus has first to cross the fission barrier, which has a height of at least 3 MeV for the nuclei considered in this work.
In this case, the passage of the barrier is governed by the following characteristics:
Let us first consider the evolution of the intrinsic excitation energy $E^*$ (or
the entropy $S$, which is strictly connected to $E^*$ in case of stastistical equilibrium among the intrinsic degrees of freedom). 
If the entropy is preserved, the system can proceed to fission only by tunneling through the fission 
barrier.
Tunneling through the whole barrier is strongly hindered by a factor of roughly $10^{-15}$ or more \cite{Patyk89}.
Therefore, it is very probable that the system converts part of its excitation energy into deformation energy in order to pass the barrier without tunneling, if the initial excitation energy allows it.
The density of transition states above the saddle (or above the inner saddle in the case of a multi-humped barrier) is lower by a factor of about $exp(-3/T) = 0.0025$ (using a typical value of $T = 0.5$ MeV for $A \approx 200$ \cite{Egidy09}) or smaller, compared to the density of states in the first minimum. 
According to the transition-state model \cite{Bohr39}, the probability is very high $(> 0.9975)$ that the single-particle configuration of the system changes, before the system reaches the saddle.
There is no preference in populating any state at the barrier with the suitable energy, angular momentum and parity. 
Therefore, the distribution of transition states accumulated over a large number of fission events corresponds to statistical equilibrium of the system at the given energy \cite{Wigner38}. 
This means that there is no memory on the initial state in which the system was formed in the first minimum, or on any state before reaching the outer barrier, except excitation energy, angular momentum and parity.
Some pecularities should be mentioned: 
When the transition states at the given energy are resolved, the equilibrium distribution of transition states eventually shrinks to a single state. 
When the initial excitation energy is lower than the fission-barrier height, the system can pass the barrier only by tunneling. 
This occurs most probably with intrinsic excitation energy zero both at the entrance point before and at the exit point beyond the barrier.
In case of a double-humped barrier, the passage of the second barrier proceeds essentially with the same characteristics.  
For nuclei with a double-humped (or eventually with a triple-humped) barrier, this means that it fully adapts in all its degrees of freedom to the phase space above the outer-most barrier. 

Beyond the outer-most barrier, the system is driven to scission and, finally, to disintegration, in most cases into two fragments by the dominant influence of the Coulomb force. 
On the way from the outer saddle to scission, the system experiences a very complex evolution.
In addition to the intrinsic excitations, the system has full freedom in its shape evolution. 
(In a simplified way, the intrinsic excitations are often represented by a heat bath, while the shape evolution is often represented by the elongation and a number of normal modes like the octupole moment that determines the mass asymmetry and neck radius of the fissioning system, quadrupole moments of the two nascent fragments etc. 
The normal modes may be represented by harmonic oscillators with their minima at the static fission path. 
The characteristics of these harmonic oscillators evolve with elongation. 
The different normal modes may be coupled to each-other. 
For example, the mass asymmetry couples strongly with the quadrupole moments of the two nascent fragments.)
In this dynamic process, the driving force along the fission path drives the system continuously out of equilibrium, and statistical equilibrium is not attained any more.
Several forces are acting on the system:
The shape evolution is described by a kind of transport equation with appropriate coupling to the intrinsic degrees of freedom, including many-body interactions. 
Consideration of quantum mechanics is mandatory. 
(In the simplified picture, mentioned above, the motion towards increasing elongation is coupled to the other shape degrees of freedom. 
Moreover, the coupling of the shape degrees of freedom with the heat bath of intrinsic degrees of freedom induces entropy-driven forces and dissipation.) 

A few quantum-mechanical features, which are connected with residual interactions (that go beyond the independent-particle model or the mean-field approximation), are known to be subject to a considerable change during the fission process. 
In particular, this concerns the total pairing condensation energy and the total congruence energy. 
It is expected that both quantities change their magnitude continuously during the evolution from a mono-nucleaus to a di-nuclear system \cite{Myers97,Krappe01}. 
It is also expected that other properties of the fissioning system like the shell effects \cite{Mosel71,Mosel71a} and the level density approach more and more the superposition of the properties of the nascent fragments with a gradually decreasing coupling between them.   
The latter leads to a transport of thermal energy between the nascent fragments and, to the extent to which statistical equilibrium is realized, to the phenomenon of energy sorting \cite{Schmidt10}.

After scission, the fragments attain the shapes of their respective ground-state, experience some transformation of energy from deformation energy and collective excitation energy into intrinsic excitation energy, eventually experience some Coulomb excitation in the field of the complementary fragment \cite{Hoffman64}, and are accelerated to their final velocities, while conserving their total energies, angular momenta and parities.

After scission, the excitation energy and angular momentum difference to the ground states of the separate fragments are carried away in a statistical de-excitation process. On a longer time scale, the $\beta$-unstable fragments are subject to delayed processes that are, among others, responsible for the emission of anti-neutrinos, delayed neutrons and the decay heat of used fuel. 

\subsubsection{Confrontation with theoretical models}

The theoretical approaches, described in section \ref{4}, do not fully cover the above-mentioned scenario and partly show more or less severe deviations. 
We will mention only a few of the most important ones, and we concentrate on the
evolution of the system from the barrier to scission, which is the main subject of
most theoretical models.

Among the microscopic self-consistent approaches, the TDGCM with GOA assumes adiabaticity during the evolution from saddle to scission, considers only a very limited number of collective modes, and cannot model the scission process. 

The proposed TDDFT approaches handle only some limited aspects of dissipation. 
Ref. \cite{Tanimura17}, based on TDHF-BCS, starts the dynamic calculation around mid-way between saddle and scission with an arbitrary distribution of initial conditions and treats the effects connected with pairing correlations in the BCS approximation that includes pairing effects essentially only on static nuclear properties. 
TDHFB \cite{Bulgac16} in its concept agrees with the proposed scenario to a great extent, although dissipation is only partly considered, and the effect of residual interactions beyond pairing and the heat transfer in the dinuclear regime are not considered, at least not explicitly. In addition, the predicted quantities are still very limited. 

The stochastic models do not include quantum-mechanical effects in the dynamic evolution and only consider the motion in a limited number of collective coordinates by entropy-driven forces and dissipation in the interaction with the heat bath. Furthermore, only aa limited number of normal collective modes is considered. 

The semi-empirical GEF model \cite{Schmidt16} is guided by the proposed scenario, although the physics is not always explicitly formulated and often expressed by effective descriptions. 
The importance of statistical mechanics and
the early localization of the nuclear wave functions in the two nascent fragments before scission are particularly emphasized.
These aspects are not explicitly considered by the other models mentioned before.
The model owes its wide coverage of fission quantities to the choice of fast algorithms and simplified descriptions and its good accuracy to the adjustment of a number of model parameters to empirical information.

\section{Summary} \label{6}
 
The experimental and the theoretical activities of the last 
years that have most strongly promoted the understanding of nuclear fission, 
and the prospects for future developments have been covered in this review. 

On the experimental side, the application
of inverse kinematics extended the experimental capabilities in several aspects.
At present, this is the only way that is able to identify all fission products
in $Z$ and $A$. In an approach developed at GSI, Darmstadt,
fragmentation products from a relativistic $^{238}$U beam were fully identified
in $Z$ and $A$ and brought to fission by electromagnetic excitation in a heavy target
material. With this technique, low-energy fission of
a large number of nuclei with $A \le 238$ that were not accessible before can be investigated.
The fission products can be identified in $Z$ and $A$ with an excellent resolution. 
This technique has already proven its potential by mapping the transition from
symmetric fission to asymmetric fission around $^{226}$Th. But first results
demonstrate that this technique also offers unique possibilities for systematic 
experiments on lighter neutron-deficient nuclei in an extended region around 
$^{180}$Hg \cite{Martin15}.  
In another approach, developed at GANIL, Caen,
transfer reactions of a $^{238}$U primary beam in a carbon target 
gave access to experiments on fission for a number of heavier nuclei with well defined 
excitation energy and separation of all fission products in $Z$ and $A$.
At CERN-ISOLDE, the progress in LASER ionization made it possible to study 
beta-delayed fission with fully identified ISOL beams and to discover the asymmetric 
fission of $^{180}$Hg and structural effects in the mass distributions of other 
neighboring nuclei.

However, these measurements are still essentially restricted to fission-fragment yields and total kinetic energies. Extensions to include measurements of prompt neutrons and prompt gammas in
coincidence with the fragments would be very important for a better understanding of the energetics of the fission process and of the generation of the fragment angular
momenta.

On the theoretical side, much effort is being invested in developing 
microscopic quantum-mechanical self-consistent descriptions 
of the fission process. In spite of the difficulties caused by the high
demand on computer power, the lack of suitable tools to handle non-equilibrium
processes and the difficulties of introducing phenomena of statistical 
mechanics into a quantum-mechanical description, progress is being made in
promoting a qualitative understanding of fundamental aspects of nuclear fission.
The dynamical TDDFT approach that avoids any constraint on the collective variables is 
among the most interesting recent developments. 

The stochastic description of the fission process by the numerical solution 
of the Langevin equations, after being successfully applied for many years 
for studying high-energy fission, has
recently also been applied to low-energy fission, where shell effects
and pairing correlations play an important role. The strength of this
approach is the inherent treatment of statistical mechanics,
the drawback is the classical character of the Langevin equations and the 
assumption of a uniform heat bath.
Systematic dynamical calculations of the fission quantities and their
variation as a function of the nuclear composition and the excitation energy
are possible. Unfortunately, the necessity for
Monte-Carlo sampling entails a limitation in the number of relevant degrees
of freedom that are explicitly considered and, thus, a restriction in the
coverage of fission quantities, or a strongly simplified treatment of the dynamics.
A gradual extension is expected in line with the progress of computer technology.

Another possibility of modeling fission consists in the combination of 
powerful theoretical ideas and empirical knowledge. A rather successful
example is the recently developed GEF model that is
based on a global view on experimental findings and the application of 
a few general rules and ideas of physics and mathematics.
It covers almost all fission observables and is able to
reproduce measured data with high accuracy while having a remarkable 
predictive power by establishing and exploiting unexpected
systematics and hidden regularities in the fission observables.
This model revealed features that are not covered by current microscopic and 
self-consistent models, in particular several manifestations of 
statistical mechanics. A highlight is the discovery of energy sorting.

\section*{Acknowledgement}

%One of the authors (K.-H. S.) acknowledges stimulating discussions with the participants of the program on Quantitative Large Amplitude Shape Dynamics: fission and heavy ion fusion that was held at the INT of the University of Washington in Seattle, Washington, from 23 September to 15 November, 2013. 
The contribution of Alexey Stankovskij is gratefully acknowledged, who performed the 
decay-heat calculations.
We thank Denis Lacroix and Aurel Bulgac for a critical reading of the section on
self-consistent microscopic approaches and Christelle Schmitt for fruitful discussions.
%The development of the GEF model has been supported by the European Commission within the EURATOM FP7 Framework Programm through CHANDA (project no. 605203) and by the Nuclear Energy Agency of the OECD.
K.-H. S. thanks the CENBG for warm hospitality.

\end{document}